\documentclass[english]{iopart}

\usepackage{graphicx}
\usepackage{amssymb}
\newcounter{fig}

\begin{document}

\title[]{\Large Plea for diagonals and telescopers of rational functions}

\author{S. Hassani$^\S$, J.M. Maillard$^\pounds$, N. Zenine$^\S$  
}

\address{\S  Centre de Recherche Nucl\'eaire d'Alger, 
2 Bd. Frantz Fanon, BP 399, 16000 Alger, Algeria}
\address{$^\pounds$ LPTMC, UMR 7600 CNRS, 
Universit\'e de Paris, Tour 23,  5\`eme \'etage, case 121, 
 4 Place Jussieu, 75252 Paris Cedex 05, France} 

\begin{abstract}

  This paper is a plea for diagonals and telescopers of rational, or algebraic, functions
  using creative telescoping, in a computer algebra experimental mathematics
  learn-by-examples approach.

  We show that diagonals of rational functions (and this is also the case
with diagonals of algebraic functions) are left invariant when one performs
an  infinite set of birational transformations on the rational functions. 
These invariance results generalize to telescopers. We cast light on the
almost systematic property of homomorphism to their adjoint of the telescopers of 
rational, or algebraic, functions. We shed some light on the reason why
the telescopers, annihilating the diagonals of rational functions of the form
$\, P/Q^k$ and $\, 1/Q$, are homomorphic. For telescopers with solutions
(periods) corresponding to integration over
non-vanishing cycles, we have a slight generalization of this result.
We introduce some challenging examples of generalization of diagonals
of rational functions, like diagonals of transcendental functions, yielding
simple $\, _2F_1$ hypergeometric functions associated with elliptic curves,
or (differentially algebraic) lambda-extension of correlation of the Ising model.

\end{abstract} 
  
\vskip .5cm

\noindent {\bf PACS}:
05.50.+q, 05.10.-a, 02.10.De, 02.10.Ox

\vskip .5cm

\noindent {\bf Key-words}:
Diagonals of rational or algebraic functions, creative telescoping, 
globally bounded series, modular forms, multi-Taylor expansions, 
multivariate series expansions, magnetic susceptibility of the Ising model,
lattice Green functions, Fuchsian linear differential equations, 
homomorphisms of differential operators, self-adjoint operators,
Poincar\'e duality, differential Galois groups

\vskip .5cm
\vskip .2cm

\today

\vskip .3cm

{\em Contribution to the themed Issue in honor of 
Professor Richard Kerner on the occasion of his 80th Birthday,
 "The Languages of Physics"}              

\vskip .3cm

\section{Introduction: plea for a computer algebra experimental mathematics learn by example approach}

A paper in the honor of Professor Richard Kerner must be a paper on theoretical physics, mathematical physics,
physical mathematics, applied mathematics, applicable mathematics or even experimental mathematics~\cite{experimental}.
These different domains have large overlaps and, quite often, their differences, or shades, are slightly
irrelevant, only corresponding to social membership to different ``mathematical tribes''.  
This  computer algebra paper will actually be a plea for
{\em diagonals and telescopers of rational (or algebraic) functions}
and for {\em creative telescoping}, with a computer algebra experimental mathematics
learn-by-examples approach.  

\vskip .3cm

\subsection{Honor, pride and prejudice}

The Journal of Mathematical Physics defines mathematical physics  as
"the application of mathematics to problems in physics and the development of 
mathematical methods suitable for such applications and for the formulation of 
physical theories". An alternative definition would also include those mathematics
that are inspired by physics (also known as physical mathematics). Mathematical physics
clearly raises the question of the watershed between mathematics and physics (especially
in France ...). Does ``Mirror Symmetry''~\cite{Mirror,Candelas,Kont,Elegant}
which is a relationship between geometric objects called Calabi–Yau manifolds,
belong to  algebraic geometry or theoretical physics?
Does  ``Special Relativity'' belong to physics or mathematics, Einstein or Poincar\'e?
``Einstein was reluctant to acknowledge that the Michelson-Morley experiment had a signiﬁcant
inﬂuence on his road to special relativity''~\cite{Moore}. In fact, ``once Maxwell’s equations are properly
understood mathematically, special relativity is an inevitable consequence''~\cite{Moore}.
Physical mathematics is sometimes viewed with suspicion by both physicists {\em and} mathematicians.
On the one hand, mathematicians regard it as deﬁcient, for lack of proper
mathematical rigor. In the years since this ``mathematical physics debate'' erupted~\cite{Atiyah} there have been
many spectacular successes scored by physical mathematics, thanks  to the ``unreasonable effectiveness''
of physics in the mathematical sciences.
Dyson famously proclaimed: ``As a working physicist, I am actualy aware of the fact that the marriage
between mathematics and physics, which was so enormously fruitful in past centuries, has recently
ended in divorce''. 
This ``divorce'' is particularly serious in France,  because of the overwhelmingly
leading figure of Alexander Grothendieck and the huge influence of the Bourbaki group, which raises
the question of rigor~\footnote[1]{``We should not confuse rigor with rigor mortis'', Isadore Singer, see~\cite{Moore}.}
versus creativity.
Recalling Pierre Cartier~\cite{Marjorie}, 
the Bourbaki group has  been criticized by several mathematicians, including its own former members, for
a variety of reasons. `` Criticisms have included the choice of presentation of certain topics within
the \'El\'ements~\cite{Elements} at the expense of others,
{\em dislike of the method of presentation} for given topics,
dislike of the group's working style, and a perceived elitist mentality around Bourbaki's project
and its books, especially during the collective's most productive years in the 1950s and 1960s.
There is essentially {\em no analysis} beyond the foundations: {\em nothing about partial differential equations},
nothing about probability. There is also {\em nothing about combinatorics},
nothing about algebraic topology\footnote[9]{This is not totally true, see~\cite{Topol}.},
{\em nothing about concrete geometry}. Anything connected with {\em mathematical physics is totally absent}
from Bourbaki's text.''
Dieudonn\'e (founding member), later, regretted that Bourbaki's success had contributed to a snobbery
for pure mathematics in France, {\em at the expense of applied mathematics}~\cite{Mashaal,Mashaal2}.
In an interview\footnote[5]{To Marian Schmidt in 1990, see
  Chapter ``Bourbaki's choice: Neither Logic nor Applied Math'' page 118 in~\cite{Mashaal}.},
he said: "It is possible to say that there was no serious applied mathematics in France 
for forty years after Poincar\'e. There was even a snobbery for pure mathematics. When one noticed
a talented student, one would tell him ``You should do pure math.'' On the other hand, one would advise
a mediocre student to do applied mathematics while thinking, ``It's all that he can do! ...''. 
Apart from french\footnote[5]{When in doubt, blame the french (citation).} mathematicians,
this snobbery for pure mathematics met with harsh criticism from 
Vladimir Arnold in his deliciously polemical paper~\cite{Arnold} ``Sur l'\'education math\'ematique''. 

Quantum groups emerged from one (Yang-Baxter integrable) explicit example,  namely Quantum Toda, and
{\em not} from an ex-nihilo abstract, formal, construction of a noncommutative algebra formalism, and
other $\, C^{\star}$-algebras, dressed with coassociative coproducts. 
In theoretical physics we get used to the emergence of {\em modular forms} and sometimes {\em Shimura forms}~\cite{Heun}. 
If a physicist asks a mathematician for more  information on these structures
he will probably only get the academical Poincar\'e upper half-plane definition
and formalism which is totally and utterly useless for him
and he will not recognize the representation of {\em modular forms} and  {\em Shimura forms}
which naturally emerges in physics~\cite{Heun,Maier}
in terms of pullbacked $\, _2F_1$ hypergeometric functions\footnote[2]{Even worth the fact that
automorphic forms~\cite{Ford} are holonomic functions is seen as a surprising fact by some mathematicians
(see, for instance, paragraph
  ``How to recognize modularity'' page 7 of Zagier's paper~\cite{Zagier}).}. 
In theoretical physics we are flooded by {\em elliptic curves}, K3 surfaces, Calabi-Yau
manifolds~\cite{Candelas,TablesCalabi-2010,Almk-Zud-2006,broadhurst-2009,batyrev-kreuzer-2010,Straten,diffalgCalabi}. If
a physicist tries to discuss with a mathematician of the elliptic curve he just discovered (he has even calculated the
j-invariant, or the Hauptmodul, of this  elliptic curve ...), he might be severely rebuked 
that he has absolutely no right to talk of an elliptic curve because an elliptic curve must have
a ``specified point'', or will be seen with suspicion because his elliptic curve does not correspond to the complete
intersection of quadrics~\cite{Quasi,determin}
framework mathematicians like to consider in their theorems. 
Along this (slightly polemical ...) line,  pure mathematicians will, often\footnote[4]{Quite often
 in France.}, refuse to provide
{\em representation} of their formalism, in particular they will refuse to provide {\em examples}. If a physicist, eager
to understand a mathematical concept, asks for an example of an algebraic variety, an example of holonomic function,
or an example of functor, some mathematicians will, maliciously, reply:
a point, the constant function and the oblivion functor. In such
a frustrating  ``dialogue of the deaf'' between physicists and mathematicians,
mathematical physics is probably the perfect place to be criticized by physicists to be too abstract, or too mathematical,
and also by mathematicians for a lack of rigor, a lack of mathematical proofs. 

At this step, one of us (JMM) would like to seize the opportunity of this experimental mathematics paper in honor
of Professor Richard Kerner, to express his deep regrets for his numerous fruitful conversations
with Jean-Louis Verdier\footnote[9]{Jean-Louis Verdier  performed his thesis under the direction of Alexandre Grothendieck.
He was a member of the Bourbaki group. He passed away in August 1989.}
and its very generous pedagogical explanations. A discussion with him was not
flooded with ``Derived Categories'' or ``p-adic cohomology'', but with simple examples and representations
of the mathematical concepts.  A really good mathematician can provide examples, he is not afraid, or ashamed,
to provide  examples and representations. For Jean-Louis Verdier mathematics was not an obfuscation contest.

\vskip .3cm

This paper is an experimental mathematics~\cite{experimental} paper with a learn-by-example approach: we get puzzling {\em exact}
results from computer algebra\footnote[8]{Maple, Mathematica.}, and we hope mathematicians will be
interested to provide proofs of these results, in a proper framework.
Furthermore, these exact results, useful for physics, raise a lot of fascinating new questions at the crossroad
of different domains of mathematics.

\vskip .3cm

\subsection{Diagonal of rational functions, creative telescoping, birational transformations and effective algebtraic geometry}

Diagonals of rational functions (or diagonals of algebraic functions)
have been shown to emerge naturally~\cite{2013-rationality-integrality-ising} for $\, n$-fold integrals in
physics\footnote[1]{Corresponding to solutions of linear differential operators of quite high order~\cite{2009-chi5,High}.},  
field theory, enumerative combinatorics~\cite{combinatorics,binomial}, seen
as ``Periods''~\cite{lairez-2014} of algebraic varieties
(corresponding to the denominators of these rational functions). The fact that diagonals of rational, or algebraic,
functions occur frequently in physics, explains many unexpected mathematical properties encountered in physics,
that are far more obvious from a physics viewpoint.
For instance, the linear differential operators, annihilating these ``Periods'',
are  {\em globally nilpotent}~\cite{nilpotent,SpecialGeom}, and, consequently,
the {\em critical exponents}\footnote[2]{Physicists are
clearly very interested to see if the critical exponents of the {\em three-dimensional} Ising model
are, or are not, {\em rational} numbers. In contrast, since many
lattice Green functions in  any dimension~\cite{LGFGuttmann}
are diagonals of rational functions, 
their critical exponents are necessarily rational numbers in {\em any} dimension.}
of {\em all} the (regular) singular points of these operators are {\em rational numbers}\footnote[4]{Katz theorem
 states that {\em globally nilpotent} linear differential operators are {\em fuchsian} with
{\em rational exponents} (see for instance~\cite{Lepetit}).}. These $\, n$-fold integrals are also
{\em globally bounded}~\cite{2013-rationality-integrality-ising,globbounded}
series, which means that they  can be recast into
(finite radius of convergence) series with {\em integer coefficients}. Furthermore, these  series, with integer coefficients,
reduce {\em modulo every prime} to {\em algebraic functions}. The calculation of the linear differential operators
annihilating these $\, n$-fold integrals of algebraic functions can be systematically performed
using the {\em creative telescoping method}~\cite{Creative1,Creative2,Creative3}
which corresponds, essentially, to successive {\em differential algebra eliminations which are
blind to the cycles} over which one performs the $\, n$-fold integrals.
At first sight one expects the analysis of these $\, n$-fold integrals
to require, as in the S-matrix theory~\cite{Smatrix}, a lot of
complex analysis of {\em several complex variables}, 
but one quickly discovers, with creative telescoping, that one  needs
{\em differential algebra}, possibly {\em algebraic geometry}~\cite{MDPI},
because of the crucial role of an algebraic variety and, surprisingly one finds out almost ``arithmetical'' 
properties\footnote[5]{Like in the Grothendieck–Katz p-curvature conjecture which is a local-global principle
for linear ordinary differential equations, related to differential Galois theory.}. 
More experimentally, this time, one finds out that {\em almost all} the diagonals of rational, or algebraic, functions, 
corresponding, or  not, to physics, are annihilated by linear differential operators which are
{\em homomorphic to their adjoint}, and consequently, their differential Galois groups are (or are a subgroup of)
selected $ \, Sp(n, \, \mathbb{C})$ symplectic or $\, SO(n, \, \mathbb{C})$ orthogonal
groups~\cite{SpecialGeom,selected,Canonical}. More generally, one finds out that
the {\em telescopers of almost all}
the rational, or algebraic, functions are {\em also} homomorphic to their adjoint~\cite{selected}.
A physicist, already surprised to see the emergence  of all these mathematical concepts in his backyard,
will have the prejudice that these  {\em selected  differential Galois groups} are probably a consequence
of some ``sampling bias'', these diagonals and telescopers  being, in fact, related to (Yang-Baxter)
integrable models, like the $\, \chi^{(n)}$ components of the susceptibility of the Ising model~\cite{2009-chi5,High},
or beyond, Calabi-Yau manifolds, Mirror Symmetries,  Picard-Fuchs systems, and  other theory ``integrable''
in some way (Yang-Mills ...). In contrast, a mathematician will have
the prejudice that this is nothing but the {\em Poincar\'e duality}~\cite{Poincare}
since we have a canonical algebraic
variety for all these diagonals or telescopers~\cite{MDPI}. Experimentally it is quite hard to find telescopers,
or linear differential operators, that are not homomorphic to their adjoint, i.e.
that do not have selected symplectic, or orthogonal,
differential Galois groups~\cite{SpecialGeom,selected,Canonical}. If one
considers Christol's conjecture\footnote[5]{Christol conjectured that every D-finite globally bounded series
is the diagonal of a rational function.}~\cite{christol-83,christol-85,Birkhauser,EurMath,conjecture}, one can
seek for $\, _nF_{n-1}$ hypergeometric series with integer coefficients
that are candidates to be counter-examples to Christol's
conjecture~\cite{christol-83,christol-85,Birkhauser,EurMath}. Among these candidates
a sub-set has actually been seen~\cite{conjecture}
to be diagonals of rational, or algebraic, functions\footnote[1]{The fact that
the others, like the original example of G. Christol, $\, _3F_2([1/9,4/9,5/9],[1/3,1], \, 3^6 \, x)$,
are, or are not,  diagonals of rational or algebraic functions remains an open question. Even
the fact that the corresponding series with integer coefficients should reduce to algebraic functions
modulo primes, remains a humiliatingly difficult task even for small primes ! Christol's
conjecture~\cite{conjecture} remains an open question. }, like for instance $\, _3F_2([2/9,5/9,8/9],[2/3,1],x)$,
or $\, _3F_2([1/9,4/9,7/9],[1/3,1],x)$. 
It turns out that the linear differential operators of these $\, _nF_{n-1}$ candidates precisely
provide such rare examples of linear differential operators (annihilating
diagonals of rational, or algebraic, functions), that are {\em not} homomorphic to their adjoint. 
The existence of such examples (curiously related to Christol's conjecture ...) shows that
seeing the emergence of such selected differential Galois groups~\cite{selected} for diagonals of rational,
or algebraic, functions cannot simply be seen as some consequence of the {\em Poincar\'e duality}.
The Poincar\'e duality works for  {\em any} algebraic variety: the diagonal of {\em any} rational, or algebraic, function
should always yield ``self-dual'' linear differential operators in the sense that they are homomorphic to their adjoint.
This is {\em not} the case. Could it be that the physicist's prejudice is right and that, trying to be generic
in our computer algebra experiments,
we were, in fact, just exploring diagonals of selected subsets of rational, or algebraic, functions related to some
kind of ``integrable'' physics?

Like Monsieur Jourdain\footnote[2]{Le Bourgeois Gentilhomme. Moli\`ere.}
speaking ``prose'' without noticing himself, physicists often perform some fundamental mathematics when they work
on their $\, n$-fold integrals without noticing these $\, n$-fold integrals are, in fact, diagonals of rational,
or algebraic, functions. In fact diagonals of rational, or algebraic, functions,
and more generally telescopers, are a perfect subject
of analysis in mathematical physics: they are, essentially, not well-known by mathematicians and by physicists
(even if physicists speak ``diagonal'' without noticing ...),
and even when these concepts are superficially known, they are not taken seriously by mathematicians, probably
because the definition is so simple, and the calculations are just ``computer algebra'' which is not highly regarded
in the ``mathematical food chain''. This is in contrast with the fact that almost every calculation of
a diagonal of rational, or algebraic, function, or  calculation of a telescoper, yields interesting, or remarkable,
sometimes even puzzling exact results, providing answers in physics and mathematics, but also raising
new interesting questions, that could be called ``speculative mathematics''.

In a learn-by-example approach we are going to address the previous questions of ``duality-breaking'' of
some telescopers of rational, or algebraic, functions, and we will also sketch some remarkable
{\em birational symmetries}~\cite{Quasi,determin,Noether,zeta} of the diagonals and telescopers
of rational, or algebraic, functions.

\vskip .3cm

\section{Definition of the diagonals of rational, or algebraic, functions. Definition of telescopers.}
\label{recalls}

The purpose of this paper is not to provide an introduction to
{\em creative telescoping}~\cite{Creative1,Creative2,Creative3,Chyzak,Koutschan1,Koutschan2,ABC}
but, rather, to provide many (non-trivial) pedagogical examples of telescopers using extensively
the ``HolonomicFunctions'' package~\cite{HolonomicFunctions}.
One can obtain these telescopers using Chyzak’s algorithm~\cite{ABC} or Koutschan’s
semi-algorithm\footnote[1]{The termination is not proven}~\cite{HolonomicFunctions}. For
the examples displayed in this paper, Koutschan’s
package~\cite{HolonomicFunctions} is more users-friendly or efficient.

\vskip .3cm   

Creative telescoping~\cite{Creative1,Creative2,Creative3,Chyzak,Koutschan1,Koutschan2,ABC}
is a methodology to deal with parametrized symbolic 
sums and integrals that yields differential/recurrence equations for such expressions. This 
methodology became popular in computer algebra in the past twenty five years. By ``telescoper''
of a rational function, say $\,R(x, y, z)$, we here refer to the output of the
creative telescoping program~\cite{HolonomicFunctions}.
The telescoper $\, T$ represents a linear differential operator that is satisfied by the diagonal
$\, Diag(R)$, and {\em also all the other ``periods''}.

The paper is essentially dedicated to solutions of telescopers of rational functions which 
are {\em not necessarily} diagonals of rational functions. These solutions
correspond to ``periods''~\cite{lairez-2014}
of algebraic varieties over some cycles which are
{\em not necessarily vanishing  cycles}~\cite{Deligne},
like in the case of {\em diagonals} of rational functions.

The reader interested in the connection 
between the process of taking diagonals, calculating telescopers, and the notion of ``Periods'', 
de Rham cohomology (i.e. differential forms) and other Picard-Fuchs equations can read  
the thesis of Pierre Lairez~\cite{lairez} (see also~\cite{lairez-2014}). 

\vskip .3cm

\subsection{Definition}
\label{def}

Let us recall that the diagonal 
of a rational function in (for example) three variables is obtained through its multi-Taylor 
expansion [19,20] 
\begin{eqnarray}
\label{defdiag}
  \hspace{-0.96in}&&  \quad \quad \quad \quad \quad \quad \quad \quad  \quad 
  R(x, y, z) \, \, \, = \, \, \,
  \sum_{m}  \sum_{n}  \sum_{l} \, a_{m,n,l} \cdot \, x^m \, y^n \, z^l, 
\end{eqnarray} 
by extracting the ``diagonal'' terms, i.e. the powers of the product $ \, p = \,  xyz$: 
\begin{eqnarray}
\label{defdiag2}
  \hspace{-0.96in}&&  \quad \quad \quad \quad \quad \quad \quad \quad  \quad 
  Diag\Bigl(R(x, y, z)\Bigr)  \, \, \, = \, \, \,   \sum_{m} \, a_{m,m,m} \cdot \, p^m. 
\end{eqnarray}
In order to avoid a proliferation of variables, the variable $\, p$,
the diagonal (\ref{defdiag2}) depends on,
 is, in the following,  simply denoted $\, x$ (see below (\ref{defdiag2residues})).
Extracting these diagonal terms essentially amounts to finding  constant terms~\cite{Cst}
in several complex variables expansions, i.e. amounts to performing a
residue calculation in several complex variables expansions
\begin{eqnarray}
\label{defdiag2residues}
  \hspace{-0.98in}&&  
  Diag\Bigl(R(x, y, z)\Bigr)  \, \, \, = \, \, \,
  \int_{{\cal C}}\,\, {{1} \over {y \, z}} \cdot \, R \Bigl({{x} \over {y \, z}}, \, y, \,  z\Bigr) \cdot dy  \,\, dz
  \\
  \hspace{-0.98in}&& 
   = \, \, \
   {{1} \over {2 \, i \, \pi}} \int \, {{1} \over {2 \, i \, \pi}} \int \,\,\,
   \sum_{m}  \sum_{n}  \sum_{l} \, a_{m,n,l} \cdot \, x^m \, y^{n-m} \, z^{l-m} \cdot {{dy} \over {y}} \, {{ dz} \over {z}}
     \,  = \, \,  \sum_{m} \, a_{m,m,m} \cdot \, x^m,
  \nonumber 
\end{eqnarray}
or equivalently
 \begin{eqnarray}
\label{defdiag2residues2}
  \hspace{-0.96in}&&  \quad  \quad  \quad   \quad    \quad  
  Diag\Bigl(R(x, y, z)\Bigr)  \, \, \, = \, \, \,
  \, \int_{{\cal C}}\,\, {{1} \over {y \, z}} \cdot \, R \Bigl({{x} \over {y}}, \, {{y} \over {z}}, \,  z\Bigr) \cdot dy \, dz, 
\end{eqnarray}
 where $\, {\cal C}$ denotes a vanishing cycle~\cite{Deligne}, where $\,  \int_{{\cal C}}$ is a symbolic notation
 for the $\, n$-fold integral with the well-suited pre-factors,
 and where the diagonal (\ref{defdiag2residues2}) is seen as a function of the remaining variable $\, x$.
 This is the very reason why
diagonals of rational, or algebraic, functions can be interpreted
as $\, n$-fold integrals~\cite{2013-rationality-integrality-ising}. 
More generally, with $\, n$ variables, one can write the diagonal of a rational function of $\, n$-variables
as the residue in $\, n-1 \, $ variables $\, x_2, \,  \, \cdots, \, x_n$:
\begin{eqnarray}
\label{defdiag2residuesnvar}
  \hspace{-0.96in}&&  \, \, 
  Diag\Bigl(R(x_1, \, x_2, \, \cdots,  \, x_n)\Bigr)
   \\
\hspace{-0.96in}&&  \quad \, \, 
\, \, \, = \, \, \,
   {{1} \over {2 \, i \, \pi}} \int \, \cdots {{1} \over {2 \, i \, \pi}}
   \int \,  {{1} \over {x_2 \, \cdots \, x_n}} \cdot \,
 R \Bigl({{x_1} \over {x_2 \,  \, \cdots \, x_n}}, \, x_2, \,  \cdots,  \, x_n\Bigr)\Bigr)
 \cdot \, dx_2 \, \cdots \, dx_n. \nonumber
\end{eqnarray}

If the definition of the diagonal of a rational or algebraic function is very simple, it
{\em does not mean} that calculating such a diagonal is simple ! By ``calculating'' we mean
finding that the series, corresponding to the diagonal, is the series expansion of
some known special function~\cite{Ince,Grad,Whittaker,Abramowitz}
(an algebraic function~\cite{Denef-lipschitz-1987},
a pullbacked $\, _2F_{1}$ hypergeometric function
which turns out to be a {\em modular form}~\cite{Heun,Diag2F1,Modular}, a $\, _nF_{n-1}$   
hypergeometric function, a Heun function~\cite{MaierHeunG}, etc\footnote[5]{
Since diagonals of rational, or algebraic, functions are holonomic functions solutions of
Fuchsian ODEs~\cite{Fuchsian} (i.e. with {\em regular} singularities), one cannot have, among
Special Functions, D-finite functions with {\em irregular}
singularities like, for instance, modified Bessel functions.}).
Most of the time, it means, since diagonals of rational, or algebraic, functions
are selected\footnote[2]{Fuchsian equations~\cite{2009-chi5,High,Fuchsian}, G-nilpotent operators,
  globally bounded series~\cite{globbounded} reducing to algebraic curves modulo every prime.} D-finite functions,
finding the linear differential operator annihilating the diagonal series,
even if we are not able to ``solve\footnote[9]{For the time being ...}'' this linear differential equation.
Finding this linear differential operator can be performed by first getting large series
expansion of the diagonal and then finding, by a ``guessing''  approach, the linear differential
operator, or getting the linear differential operator from a more global differential algebra approach,
called creative telescoping. 

\vskip .3cm

\subsection{ Telescopers}
\label{telescopers}

For pedagogical reason let us sketch
creative telescoping~\cite{Creative1,Creative2,Creative3,Chyzak,Koutschan1,Koutschan2,ABC}
in the case of a rational function of three variables. By ``telescoper''
of a rational function, say $\,R(x, y, z)$, we here refer to the output of the
creative telescoping program~\cite{HolonomicFunctions}, applied to the transformed
rational function $\, \hat{R}\,  = \,\,  R(x/y, y/z, z)/(yz)$. 
Such a telescoper is a linear differential operator $\, T$ in $\, x$ and $\,\partial x$,
such that 
\begin{eqnarray}
\label{telescopequation}
  \hspace{-0.96in}&&  \quad \quad \quad \quad \quad \quad \quad \quad  \quad 
  T \cdot \Bigl({{1} \over {y\,z}} \cdot \, R \Bigl({{x} \over {y}}, \, {{y} \over {z}}, \,  z\Bigr) \Bigr)
  \, \, \,  + {{\partial U} \over {\partial y}} \,  + {{\partial V} \over {\partial z}} \, \, = \, \, \, \, 0,
\end{eqnarray}
where the so-called “certificates” $\, U$, $\, V$ are
rational functions\footnote[1]{These rational functions are often quite large rational functions.} in $\, x, \,y, \,z$.
This equation is called the {\em telescoping equation}. Extracting the diagonal of a rational function
amounts to calculating residues in several complex variables, namely
\begin{eqnarray}
\label{diagtelescopequation}
\hspace{-0.96in}&&  \quad \quad \quad \quad \quad \quad \quad \quad  \quad
Diag\Bigl(R(x, \, y, \, z)\Bigr) \, \, = \, \, \,
\int_{{\cal C}} \,\,  {{1} \over {y \, z}} \cdot \, R \Bigl({{x} \over {y}}, \, {{y} \over {z}}, \,  z\Bigr),
\end{eqnarray} 
where the cycle $\, {\cal C}$ is a vanishing cycle\footnote[4]{``Cycle \'evanescent'' in french~\cite{Deligne}.}.
Performing the previous integration over a cycle  $\, {\cal C}$ on the LHS of the
telescoping equation (\ref{telescopequation}) one will get (with the reasonable assumption
that the linear differential operator $\, T$ commutes with the integration):
\begin{eqnarray}
\label{telescopequation2}
\hspace{-0.96in}&&  \quad \quad \quad \quad \quad \quad \, \, \,  
T \cdot \, Diag\Bigl(R(x, \, y, \, z)\Bigr)
\, \,\,  + \int_{{\cal C}} \Bigl({{\partial U} \over {\partial y}} \,  + {{\partial V} \over {\partial z}}\Bigr)
\, \, \, = \, \, \, \, 0.
\end{eqnarray} 
Again  (with reasonable assumptions) one can expect the second term in (\ref{telescopequation2})
to be equal to zero, thus yielding the equation:
\begin{eqnarray}
\label{telescopequation3}
\hspace{-0.96in}&&  \quad \quad \quad \quad \quad \quad  
T \cdot \, Diag\Bigl(R(x, \, y, \, z)\Bigr) \, \, \,\,  = \, \, \,\,   
T \cdot \, \int_{{\cal C}}\, \,{{1} \over {y \, z}} \cdot \, R \Bigl({{x} \over {y}}, \, {{y} \over {z}}, \,  z\Bigr)
\,  \, = \, \,  \, \, 0.
\end{eqnarray} 
In  other words, the telescoper $\, T$ represents a linear differential operator annihilating
the diagonal $\, Diag(R)$. For the calculation of a diagonal, the cycle  $\, {\cal C}$
has to be a {\em vanishing cycle} (residue calculation).
Note that the creative telescoping calculations giving as an output the telescoper $\, T$ and the two
``certificates'' $\, U$ and $\, V$, essentially amounts to performing differential algebra
calculations\footnote[2]{Similar to integration by part for several complex variables.}.
Since these creative telescoping calculations are differential algebra eliminations,
{\em they are totally and utterly blind to the cycle} $\, {\cal C}$. Consequently,
even if one performs an integration over a {\em non-vanishing cycle}, the telescoper $\, T$
will also be such that
\begin{eqnarray}
\label{telescopequation4}
\hspace{-0.96in}&&  \quad \quad \quad \quad 
T \cdot \,  {\cal P}   \, = \, \, \, 0
\quad\quad \, \, \, \, \, \hbox{where:} \quad \quad \quad \, \, \,
   {\cal P} \, \, = \, \, \,
   \int_{{\cal C}}\, \,{{1} \over {y \, z}} \cdot \,
   R \Bigl({{x} \over {y}}, \, {{y} \over {z}}, \,  z\Bigr),
\end{eqnarray} 
this integral being {\em not necessarily equal} to the diagonal
$\, Diag(R(x, \, y, \, z))$ (which could be,
for instance, equal to zero). Equation (\ref{telescopequation4}) means that the
telescoper annihilates {\em all the periods} $\, {\cal P}$. 

The paper is essentially dedicated to solutions of telescopers of rational functions which 
are not necessarily diagonals of rational functions. These solutions correspond to {\em periods}~\cite{lairez-2014} 
of algebraic varieties over some cycles {\em which are not necessarily vanishing cycles}
like in the case of diagonals of rational functions.

\vskip .3cm

{\bf To sum-up:} In order to calculate the diagonal of a rational function one can try, in a very
down-to-earth way,  to get large enough series expansions of this diagonal from multi-series expansion,
and then try some guessing approach to obtain the linear differential operator annihilating
the diagonal of a rational function, or one can perform the creative telescoping approach that
will provide this telescoper even if the diagonal is zero, or cannot be nicely defined
because the rational function does not have a multi-Taylor expansion: in that case the
telescoper annihilates periods corresponding to {\em all} the cycles, in particular
non-vanishing cycles.
       
\vskip .3cm

\subsection{Diagonals versus telescopers:  vanishing cycles versus  non-vanishing cycles}
\label{telescopers}

\vskip .2cm

\subsubsection{Diagonals versus telescopers: a first example \\}
\label{telescopersfirst}

Let us first consider the following rational function of three variables
\begin{eqnarray}
\label{ratx5yfirst}
  \hspace{-0.96in}&&  \quad \quad \quad \quad \quad \quad \quad \quad \quad 
  R(x, y, z) \, \, \, = \, \, \,\,\,
  {{1} \over { -x \, -y \,\, -z^2}}. 
\end{eqnarray}
This rational function does not have a multi-Taylor expansion, and thus we cannot
define the diagonal of the rational function. 
This rational function has, however, a telescoper which is a linear differential operator of order
one, namely $\,\, 5  \, \theta \, +2$, where $\, \theta \, = x \, D_x \, = x \, d/dx$
is the homogeneous derivative.
Let us now consider a slightly more general rational function:
\begin{eqnarray}
\label{ratx5yfirstalpha}
  \hspace{-0.96in}&&  \quad \quad \quad \quad \quad \quad \quad \quad \quad 
  R(x, y, z) \, \, \, = \, \, \,\,
  {{1} \over { \alpha \, \, \, -x \, -y \,\,\,  -z^2}}. 
\end{eqnarray}
This rational function (\ref{ratx5yfirstalpha}) has a multi-Taylor expansion, and one can, thus,
get the first terms of the diagonal of this rational function
(\ref{ratx5yfirstalpha}):
\begin{eqnarray}
\label{Diagratx5yfirstalpha}
  \hspace{-0.96in}&&   \quad \, \,     
  Diag\Bigl(R(x, y, z)\Bigr) \, \,   = \, \,  \,  \, \,
  {{1} \over {\alpha}} \,\,  \, +{{30} \over {\alpha^6}} \cdot \, x^2
  \, \, +{{3150} \over {\alpha^{11}}} \cdot \, x^4 \,\,
  +{{420420} \over {\alpha^{16}}} \cdot \, x^6 \, \,\,\, \,\, + \, \, \cdots
\end{eqnarray}
The $\, \alpha$-dependent rational function (\ref{ratx5yfirstalpha}) has an order-four
$\, \alpha$-dependent telescoper $\, L_4(\alpha)$
\begin{eqnarray}
  \label{telescratx5yfirstalphasol}
  \hspace{-0.96in}&&   \quad \quad \quad \quad \,
  x^2 \cdot \, L_4(\alpha) \, \, = \, \, \,
  -5\cdot \, x^2 \cdot \, 
  (5 \, \theta \, +2)\cdot \, (5\, \theta \, +4) \cdot \, (5\, \theta \, +6)  \cdot \, (5 \, \theta \, +8) 
  \nonumber \\
 \hspace{-0.96in}&&   \quad   \quad\quad   \quad \quad  \quad   \quad   \quad  
  \, \,   +16\cdot \, \alpha^5 \cdot \, \theta^2 \cdot \, (\theta\, -1)^2,  
 \end{eqnarray}
which has the following $\, _4F_3\, $ hypergeometric function solution:
\begin{eqnarray}
\label{Diagratx5yfirstalphasol}
\hspace{-0.96in}&&  \quad \quad \quad \quad \quad \quad \quad 
  {{1} \over {\alpha}} \cdot \,
 _4F_3\Bigl( [{{1} \over {5}}, \, {{2} \over {5}}, \,  {{3} \over {5}}, \, {{4} \over {5}}],
  \, \, [{{1} \over {2}}, \, {{1} \over {2}}, \,  1],
  \, \, {{3125} \over {16  \,\, \alpha^5}} \cdot \, x^2 \Bigr).
\end{eqnarray}
The series expansion of this $\, _4F_3$ hypergeometric function (\ref{Diagratx5yfirstalphasol}) is
in agreement with the series expansion (\ref{Diagratx5yfirstalpha}).
In the $\, \alpha\, \, \, \rightarrow \, \, 0 \, $ limit the order-four
$\, \alpha$-dependent telescoper $\, L_4(\alpha)$ becomes
the direct-sum:
\begin{eqnarray}
\label{Diagratx5yfirstalphasoltheta}
\hspace{-0.96in}&&  \quad \quad \quad \quad 
-5 \cdot \, x^4 \cdot \,
\Bigl((5 \, \theta \, +2) \oplus \, (5\, \theta \, +4)
\oplus  \, (5\, \theta \, +6)  \oplus  \, (5 \, \theta \, +8) \Bigr).
\end{eqnarray}
We thus see, in this  $\, \alpha\, \, \, \rightarrow \, \, 0 \,  \, $ limit,
that one recovers,
among the different factors in (\ref{Diagratx5yfirstalphasoltheta}), the order-one
telescoper of the rational function (\ref{ratx5yfirst}), namely $\, 5 \, \theta \, +2$.
This first example being a bit too simple, or degenerate, let us consider another example.

\vskip .3cm

\subsubsection{Diagonals versus telescopers: a second example \\}
\label{telescopersecond}

Let us now consider the rational function of three variables:
\begin{eqnarray}
\label{ratx5y}
  \hspace{-0.96in}&&  \quad \quad \quad \quad \quad \quad \quad  \quad 
  R(x, y, z) \, \, \, = \, \, \,\,
  {{1} \over { -x \, -y \, -z \, \,\, -x^5\, y}}. 
\end{eqnarray}
This rational function has a telescoper $\, L_4$,
which is a linear differential operator of order four,
which reads:
\begin{eqnarray}
\label{x5y}
\hspace{-0.96in}&&  \quad \quad\quad \quad 
L_4  \, \,  = \, \, \,
-(800000\,\, x^5 \, -27)\cdot \, x^4 \, D_x^4\,  \, \,
-(11200000\, x^5 \, +27)\cdot \, x^3 \, D_x^3
\nonumber \\
\hspace{-0.96in}&&  \quad \quad \quad \quad \quad \quad 
\, \, -15 \cdot \, (2800000\, x^5\, -1)\cdot \, x^2 \, D_x^2
\, \,  \,-60 \cdot \, (700000\, x^5 \, -1) \cdot \, x \, D_x
\nonumber \\
\hspace{-0.96in}&&  \quad \quad \quad \quad \quad \quad\quad \quad 
\,\,   -12\cdot \, (437500\, x^5\, +9), 
\end{eqnarray}
or, introducing the homogeneous derivative $\, \theta \, = \, x \, D_x$,
\begin{eqnarray}
\label{x5ybis}
\hspace{-0.96in}&&  \quad  \quad \quad  \quad \quad \quad 
L_4  \, \, \, = \, \, \,\,
-50000 \cdot \, \, x^5 \cdot \, (2\,\theta \, +7)\,  (2\,\theta\,  +5)\, (2\,\theta \, +3)\, (2\,\theta \,  +1)\, 
\nonumber \\
\hspace{-0.96in}&&  \quad \quad \quad  \quad  \quad \quad  \quad  \quad 
\, \, \,   +3 \cdot \,(3\, \theta \,  +1)\, (3\, \theta \, -4)\, (\theta \, -3)^2.
\end{eqnarray}
The rational function (\ref{ratx5y}) does {\em not} have a multi-Taylor expansion.
We have a problem to define the diagonal of the rational function (\ref{ratx5y}).
The analytic solutions of (\ref{x5y}), or   (\ref{x5ybis}), are thus
just ``Periods'' of the rational function (\ref{ratx5y}), i.e. integrals over
a non-vanishing cycle of the rational function (\ref{ratx5y}).
A solution of (\ref{x5y}), or  (\ref{x5ybis}),
is, for instance, the hypergeometric function:
\begin{eqnarray}
\label{x5ybishyper}
\hspace{-0.96in}&&  \quad  \quad \quad  \quad  \quad  \quad 
x^3 \cdot \, _4F_3\Bigl([ {{7} \over {10}}, \, {{9} \over {10}}, \, {{11} \over {10}}, \, {{13} \over {10}} ],
\,\,  [1, \, {{4} \over {3}}, \, {{5} \over {3}}], \,\,  {{ 800000} \over {27  }} \cdot \,  x^5  \Bigr). 
\end{eqnarray}

If one finds that the concept of diagonal is easier to understand, compared to``Periods''
over non-vanishing cycles that are not really defined (we just know they exist),
such a result may look a bit too abstract, and thus slightly frustrating.
In fact one can recover some contact with the easier concept of diagonals,
performing some kind of ``desingularization''. 
Let us consider the more general $\, \alpha$-dependent
rational function of three variables:
\begin{eqnarray}
\label{ratxy5alpha}
  \hspace{-0.96in}&&  \quad \quad \quad \quad \quad \quad \quad 
  R(x, y, z) \, \, \, = \, \, \,\,
  {{1} \over { \alpha \,  \, \, -x \, -y \, -z \, \,  \, -x^5\, y}}. 
\end{eqnarray}
It has a telescoper which is a linear differential operator of order four $\, M_4(\alpha)$.
The first terms of the diagonal of that  rational function (\ref{ratxy5alpha})
can easily be calculated. 
We have calculated this  order four  linear differential operator $\, M_4(\alpha)$. It is a bit
too large to be given here. However one remarks that this $\alpha$-dependent
order four linear differential operator  $\, M_4(\alpha)$, is
actually related to the previous order-four linear differential operator $\, L_4$,
in the $\, \alpha \, \rightarrow \, 0 \, $ limit: 
\begin{eqnarray}
\label{limit}
\hspace{-0.96in}&&  \quad \quad \quad \quad \quad \quad \quad \quad
 M_4(0) \, \, = \, \, \, -675000000\, x^{11} \cdot \, L_4.
\end{eqnarray}

\vskip .3cm

{\bf To sum-up:} The telescoper corresponding to ``Periods''  over
a non-vanishing cycles can be obtained from a
one-parameter telescoper having clear-cut diagonal solutions
(``Periods''  over a vanishing cycle). 

\vskip .3cm

\subsection{The Devil is in the detail: the number of variables}
\label{sec4varbut3}

Let us consider the diagonal of the following rational function of four variables:
\begin{eqnarray}
  \label{4varbut3}
  &&  \hspace{-0.98in}  \quad \quad  \quad \quad  \quad \quad  \quad \quad  \quad
      \,  {{1} \over { 1 \, \, - \alpha\,  x - y  - z  \,\, \, - \beta \cdot  \, x \, u}}. 
\end{eqnarray}
Its telescoper is, {\em for any value of} $\, \alpha$, and for $\, \beta \, \ne \, 0$,
the order-two linear differential operator
\begin{eqnarray}
\label{L2_4but3}
&&  \hspace{-0.98in}  \quad \quad  \quad \quad  \quad
 L_2 \, \, = \, \, \,\,
   (1  \, -27\,\beta \cdot \,  x) \cdot \, x  \, D_x^2 \, \, \,
   +(1  \, -54 \,\beta \cdot \,  x) \cdot  \, D_x \,\,  \, -6 \, \beta, 
\end{eqnarray}
which has the following hypergeometric $\, _2F_1$ solution:
\begin{eqnarray}
\label{solL2}
&&  \hspace{-0.98in}  \quad \quad  \quad \quad  \quad \quad  \quad \quad  \quad
 _2F_1\Bigl([{{1} \over {3}}, \, {{2} \over {3}}], \, [1], \,  \,  \, 27\, \beta \cdot \,  x\Bigr). 
\end{eqnarray}

Recalling the definition of the diagonal of a  rational function based on multi-Taylor expansion,
it is easy to see, on this almost trivial example, that the various
powers of the product $\, t \, = \, x\, y\, z \, u \, \, $ that the diagonal extracts, 
require the occurrence of the variable $\, u$ which only occurs, in the denominator
of (\ref{4varbut3}), through the product
$\, x \, u\, $ yielding automatically the  occurrence of the variable $\, x$. Consequently, any
further occurrence of  the variable $\, x$, from the  $\, - \alpha\,  x$ monomial in the denominator
of (\ref{4varbut3}), is excluded.  This explains why the diagonal of (\ref{4varbut3})
is actually blind to the  $\, - \alpha\,  x$ term. In other words, the diagonal
of the four variables rational function (\ref{4varbut3}) is, {\em in fact the diagonal of a rational
function of three variables}
$\, y$, $\, z$, and the product $\, x \, u$.

\vskip .3cm

{\bf Remark 2.1:} To take into account this problem, we will introduce the concept of
``effective number'' of variables.  In the previous example the number of variables
is four but the ``effective number'' of variables is three.

\vskip .3cm

\subsection{Understanding the complexity of the diagonal of a rational function}
\label{minimal}

\subsubsection{Order of the linear differential operator and number of variables \\}
\label{Ordernumber}

The simplest example of diagonal of rational function of $\, n$ variables, corresponds
to the diagonal of the rational function 
\begin{eqnarray}
\label{1surn}
  &&  \hspace{-0.98in}  \quad \quad \quad  \quad   \quad  \quad  \quad \quad \quad
     {{1} \over { 1 \,\,\, -x_1 \, -x_2 \, -x_3 \,\, \, \, \cdots \,\, -x_n }}.  
\end{eqnarray}
The  diagonal of (\ref{1surn}) is annihilated by an order-$(n-1)$
linear differential operator with a $\,\, _{n-1}F_{n-2} \, $
hypergeometric solution:
\begin{eqnarray}
\label{nFns}
  &&  \hspace{-0.98in}  \quad \quad \quad  \quad   \,   \quad  \, 
    _{n-1}F_{n-2}\Bigl([{{1} \over {n}}, \,  {{2} \over {n}}, {{3} \over {n}}, \cdots,  \,   {{n-1} \over {n}}],
     \, [1, \, 1, \, \cdots,  \, 1], \,\, \,n ^n \cdot \, x \Bigr).  
\end{eqnarray}
This  simple example may provide the  prejudice that, for a given globally bounded series (\ref{nFn}),
the  number of variables of the rational function is related to the (minimal) order of the
linear differential operator annihilating the series.
One should note, however, for the class of the above example, that the
corresponding linear differential operator has
the  Maximally Unipotent Monodromy property
(MUM)\footnote[1]{A Maximally Unipotent Monodromy linear differential
operator (MUM) is a linear differential operator such that
all its indicial exponents (at the origin) are equal (see for instance~\cite{Straten,LGFGuttmann}).} .

This result is reminiscent of the  well-known 
$\, _4F_3([1/5,2/5,3/5,4/5], [1,1,1], \, x)$ Candelas et al. hypergeometric series
emerging in~\cite{Candelas} for a particular Calabi-Yau manifold.
Let us recall that Calabi-operators~\cite{Straten}, annihilating Calabi-Yau series~\cite{TablesCalabi-2010},
are (self-adjoint)
order-four linear differential operators
which have the  Maximally Unipotent Monodromy property
(MUM)  at $\, x \, = \, 0$: if one considers their formal series expansions at $\, x \, = \, 0$,
among the four formal series expansions,
one is analytic (it actually corresponds to our diagonals of rational functions), another one is a
formal series with a $\, \ln(x)^{1}$, another one is a formal series with a $\, \ln(x)^{2}$,
and the last one is a formal series with a $\, \ln(x)^{3}$.
Along this line ((\ref{1surn}) yielding (\ref{nFns})), one would expect that the diagonal of
rational function representation of a Calabi-Yau series
(solution\footnote[5]{The simplest Calabi-Yau series   (see for instance~\cite{TablesCalabi-2010})  
  are $\, _4F_3$ hypergeometric series like $\, _4F_2([1/2,1/2,1/2,1/2], [1,1,1], \, x)$,
  or $\, _4F_2([1/5,2/5,3/5,4/5], [1,1,1], \, x)$ (see equation 3.11 in~\cite{Candelas}).} of an order-four linear
differential operator) should require, at least {\em five} variables for the rational function.


\vskip .3cm 

\subsubsection{Order of the linear differential operator and degree in the variables \\}
\label{Orderdegree}

Let us now consider the diagonal of the following rational function of three variables
\begin{eqnarray}
\label{1surz2}
  &&  \hspace{-0.98in} \quad  \quad \quad \quad \quad  \quad   \quad  \quad  \quad \quad  \quad
     {{1} \over { 1 \,\, \, -x \, -\alpha \, y \,\,  \, -z^2}},  
\end{eqnarray}
whose diagonal writes as a simple $\, _4F_3$ hypergeometric solution:
\begin{eqnarray}
 \label{4F3}
  &&  \hspace{-0.98in} \quad  \quad  \quad  \quad \quad  \quad  \quad 
    _4F_{3}\Bigl([{{1} \over {5}}, \,  {{2} \over {5}},  \, {{3} \over {5}}, \,   {{4} \over {5}}],
     \, [1, \, {{1} \over {2}}, \,  {{1} \over {2}}], \, {{5^5} \over {2^4 }} \cdot \, \alpha^2  \cdot \, x^2 \Bigr).  
\end{eqnarray}

In contrast with the example (\ref{1surn}), here, we just need, for the rational function,
{\em three} variables, instead of the expected {\em five}
variables.
Note however, that the order-four linear differential operator
$\, L_4$, annihilating this hypergeometric solution (\ref{4F3}),
 does {\em not} have MUM.
 As usual, this order-four  linear differential operator is
 homomorphic to it adjoint with a very simple order-two intertwiner:
\begin{eqnarray}
  \label{HomoL4}
  &&  \hspace{-0.98in}   \quad    \quad    \quad    \quad    \quad \quad  \quad \, \, 
  L_4 \cdot \, \Bigl(x \, D_x^2\, \, +D_x \Bigr)
  \, \, \, = \, \, \,  \Bigl(x \, D_x^2\, \, +D_x \Bigr) \cdot \, adjoint(L_4).
\end{eqnarray}
One thus expects~\cite{Canonical} this order-four  linear differential
operator $\, L_4$ to have a symplectic
differential Galois group included in $\, Sp(4, \, \mathbb{C})$.
Actually the exterior square of this o.rder-four operator $\, L_4\, $ has
a simple rational function solution~\cite{Canonical}, namely $\,1/x/(5^5 \cdot \, x^2 \, -2^4)$.

\vskip .3cm

Let us now consider the diagonal of the following rational function of three variables:
\begin{eqnarray}
\label{1surz3}
  &&  \hspace{-0.98in}  \quad \quad \quad  \quad   \quad  \quad \quad    \quad  \quad \quad  \quad
     {{1} \over { 1 \, \,\, -x \, -\alpha \,  y \, \,\, -z^3}}.  
\end{eqnarray}
The linear differential operator annihilating this diagonal is an order-six linear differential
operator with a quite simple $\, _6F_5$ hypergeometric solution:
\begin{eqnarray}
  \label{6F5}
 &&  \hspace{-0.98in}   \quad  \quad    \quad    \quad \quad  
  _6F_{5}\Bigl([{{1} \over {7}}, \,  {{2} \over {7}},\,   {{3} \over {7}},
    \,   {{4} \over {7}}, \,   {{5} \over {7}}, \,   {{6} \over {7}}],
  \, [1, \, {{1} \over {3}}, \,  {{1} \over {3}}, \, {{2} \over {3}},
    \,  {{2} \over {3}}], \, {{7^7} \over {3^6 }} \cdot \, \alpha^3  \cdot \, x^3 \Bigr).  
\end{eqnarray}
Let us restrict to $\, \alpha \, = \, 1$. 
The order-six linear differential operator, annihilating
the diagonal of (\ref{1surz3}), does  {\em not} have MUM.
One has {\em three} series analytic at $\, x\, = \, 0$,
one of the form $\, x \cdot (1\, +2377375/6561 \, x^3 \, + \cdots )$,
one of the form $\, x^2 \cdot (1\, +16509584/32805 \, x^3 \, + \cdots )$,
and the third one being the diagonal of the rational function
which is the expansion of (\ref{6F5}):
\begin{eqnarray}
  \label{series6F5}
  &&  \hspace{-0.98in}   \quad \, \,
  1 \,  \, +140\, x^3 \, +84084\, x^6 \, +64664600\, x^9
  \, +55367594100\, x^{12}\, +50356110752640\, x^{15}
     \nonumber \\
  &&  \hspace{-0.98in}   \quad \quad \,  \quad \,
     +47606217704845800\, x^{18} \, +46236665756994672960\, x^{21}
     \, \, \, \,  + \cdots 
\end{eqnarray}
One also has three other formal series solutions with a $\, \ln(x)^{1}$,
but no $\, \ln(x)^{2}$ or  $\, \ln(x)^{3}$.

\vskip .3cm

As usual, this order-six linear differential operator is
homomorphic to its adjoint with a very simple order-four intertwiner:
\begin{eqnarray}
  \label{HomoL6}
  &&  \hspace{-0.98in}   \quad   \quad  
  L_6 \cdot \, \Bigl(x^2 \, D_x^4  +4 \, x\, D_x^3  +2\, D_x^2\Bigr)
  \, = \,   \Bigl(x^2 \, D_x^4  +4 \, x\, D_x^3  +2\, D_x^2\Bigr) \cdot \, adjoint(L_6).
\end{eqnarray}
One expects~\cite{Canonical} this order-six linear differential operator $\, L_6$
to have a symplectic
differential Galois group included in $\, Sp(6, \, \mathbb{C})$.
Actually the exterior square of this order-six linear differential operator $\, L_6\, $ has
a simple rational function solution~\cite{Canonical}, namely $\, \, 1/x/(7^7 \cdot \, x^3 \, -3^6) $.

\vskip .3cm 

{\bf Remark 2.2:} This result can be generalised. Let us consider the rational function:
\begin{eqnarray}
\label{generalised}
  &&  \hspace{-0.98in}  \quad \quad \quad  \quad   \quad  \quad  \quad \quad  \quad \quad  \quad \quad
     {{1} \over { 1 \, \, \, -x \, - \,  y \,  \,\, -z^n}}.  
\end{eqnarray}
The linear differential operator $\, L_{2n}^{(1)}$, 
annihilating this diagonal, is an order-$(2\, n)$ linear differential
operator with a quite simple $\,  _{2 n}F_{2n \, -1}  \,$ hypergeometric solution:
\begin{eqnarray}
  \label{nFn}
 &&  \hspace{-0.98in}   \,    \quad  \quad   
  _{2 n}F_{2n \, -1}\Bigl([{{1} \over {2\, n \, +1}}, \, \,
    {{2} \over {2\, n \, +1}}, \, \, {{3 } \over {2\, n \, +1}}, \,\,\,
     \cdots,
     \, {{2\, n} \over {2\, n \, +1}}],
  \nonumber \\
 &&  \hspace{-0.98in}   \quad  \quad   \quad     \quad  \quad    \quad   
  \, [1, \, {{1} \over {n}}, \,  {{1} \over {n}}, \, {{2} \over {n}}, \,  {{2} \over {n}}, \, \cdots,
    {{n-1} \over {n}}, \,  {{n-1} \over {n}}],
  \,\,  {{(2\, n \, +1)^{(2\, n \, +1)} }\over {n^{2\, n} }}  \cdot \, x^n \Bigr).  
\end{eqnarray}
Let us also consider the linear differential operators $\, L_{2n}^{(m)} \, $
annihilating the diagonal of the rational function:
\begin{eqnarray}
\label{generalised}
  &&  \hspace{-0.98in}  \quad \quad \quad  \quad   \quad  \quad  \quad \quad  \quad  \quad
     \Bigl({{1} \over { 1 \, \, -x \, - \,  y \, \, -z^n}}\Bigr)^m. 
\end{eqnarray}
One finds the following homomorphisms\footnote[9]{We use the Homomorphisms
command in Maple (DEtools).} between successive linear differential operators
$\, L_{2n}^{(m)}$:
\begin{eqnarray}
\label{telescQnHomMORE}
  \hspace{-0.96in}&&  \quad \quad \quad 
  Homomorphisms\Bigl(L_{2n}^{(m)} , \, \, L_{2n}^{(m+1)}  \Bigr)  \,  \,\, = \, \, \, \,
  (2\, n \, +1) \cdot  \, x \cdot \, D_x \, \, + \,  m \cdot \, n.
\end{eqnarray}
In other words one has the relations:
\begin{eqnarray}
\label{telescQnRelMORE}
  \hspace{-0.96in}&&  \quad  \quad \quad  \quad \quad \quad
  L_{2n}^{(m+1)}  \cdot \, \Bigl( (2\, n \, +1) \cdot  \, \theta \, \,  + m \cdot  \, n \Bigr)
  \, \, \, \, = \, \, \, \,  Z_1(m) \cdot\, L_{2n}^{(m)},
\end{eqnarray}
where $\, Z_1(m)$ is an order-one linear differential operator.
The linear differential operator $\, L_{2n}^{(1)} \, $ is simply homomorphic to its adjoint:
\begin{eqnarray}
\label{telescQnHomMORE}
\hspace{-0.96in}&&  \quad \quad \quad 
Homomorphisms\Bigl( adjoint( L_{2n}^{(1)}) , \, \, L_{2n}^{(1)}  \Bigr)  \, \, = \, \, \, \,
\nonumber \\
\hspace{-0.96in}&&  \quad \quad   \quad \quad  \quad \quad  
  {{1} \over {x^{n-1}}} \cdot \, \theta^2 \cdot \, (\theta \, -1)^2  \cdot \, (\theta \, -2)^2 \, 
    \cdot \, (\theta \, -3)^2  \,   \cdots \, \Bigl(\theta \, -(n\, -2)\Bigr)^2. 
\end{eqnarray}

\vskip .3cm 

{\bf Remark 2.3:}  With the previous, rather simple, examples we see that
the order of the linear differential operator annihilating the diagonal of a rational function,
is {\em not} related to the number of variables of the rational function (or even to the number
of ``effective'' variables see section \ref{sec4varbut3}). Furthermore, a given {\em globally bounded}
series can be seen to be the diagonals of an infinite number of rational functions of
a certain number of variables, but also, in the same time, of other infinite number of rational
functions with a different number of variables.
For a given globally bounded series we can find the (minimal order) linear differential operator
annihilating this series. Having this (minimal order) linear differential operator, the question is:
can we find the
{\em minimal number of variables} necessary to see this globally bounded series as the diagonal of
a rational function of that number of variables?
We will address these questions in a forthcoming paper~\cite{Preparation}.

\vskip .3cm

\section{Diagonals of rational  functions: should we restrict to rational  functions
of the form $\, 1/Q$?}
\label{restriction}

With $\, P$ and $\, Q \, $ multivariate polynomials (with $\, Q(0) \ne \, 0$),
the diagonals of the rational functions $\, P/Q^k \, $ are, for fixed polynomial $\, Q$,
and for arbitrary integer $\, k$, a {\em finite dimensinal} vectorial space 
related\footnote[1]{We are thankful to  P. Lairez for having clarified this point.},
as shown by Christol~\cite{christol-83,christol-85}, to the de Rham cohomology.
For physicists, not familiar with de Rham cohomology\footnote[2]{They are so many cohomologies
in mathematics. For non-mathematicians let us just say that the introduction
of a cohomology often amounts to seeing that ``something'' you expect, at first sight,
to be infinite, for instance the number of solutions of a system of PDE's,
is in fact a {\em finite} set (for instance for D-finite systems of PDE's).},
let us just say that this can be
seen as a consequence of the fact that these  $\, P/Q^k$ rational functions
are solutions of {\em D-finite systems}, which means that these systems of PDE's 
(partial differential equations)
have
a {\em finite set of solutions} of the form  $\, P/Q^k$. Being in such a ``finite box''
will force\footnote[5]{This requires to find a ``cyclic vector'' in mathematicians wording.}
the telescopers of the diagonals of $\, P/Q^k$ and $\, 1/Q$, to be related
(by  homomorphisms). 

Experimentally, if one considers the (minimal order) linear differential operators
for the diagonal of $\, P/Q^k$ and for the  diagonal of $\, 1/Q$, these two linear
differential operators are {\em actually homomorphic}. Note that this experimental result,
valid for diagonals (i.e.  integrals over vanishing cycles),
is {\em no longer} valid for telescopers of rational functions with analytic solutions corresponding
to ``periods'', $\, n$-fold integrals, over non-vanishing cycles. In this case we have
a slight generalization of that homomorphism between telescopers $\, P/Q^k$ and telescopers $\, 1/Q$,
that will be described in the sequel (see section \ref{GaloisSUB} below).

It is true that the analysis of lattice Green
functions (LGF)~\cite{2015-LGF-fcc7,2016-LGF-fcc8to12,joyce-1998,koutschan-2013,guttmann-2009}
in physics naturally yields
to diagonals of rational functions in the form $\, R=\, 1/Q$, where $\, Q$ is a polynomial.
However, the other $\,n$-fold integrals, emerging in physics, are much more complex (for instance
the $\, \chi^{(n)}$ terms of the susceptibility of the two-dimensional Ising model~\cite{High}).
The lattice Green
functions~\cite{LGFGuttmann,2015-LGF-fcc7,2016-LGF-fcc8to12,joyce-1998,koutschan-2013,guttmann-2009,LGFGuttmann,guttmann-prellberg-1993}
and some Occam's razor simplicity argument\footnote[4]{Or should we say laziness argument?
We might study something simple and possibly slighty irrelevant, just because we do not want to work hard
facing the true problem.} are not sufficient to justify a bias of studying, quite systematically,
rational functions of the form $\, R=\, 1/Q$ (as we often do). In fact these
de Rham cohomology arguments are the reason why, for diagonals (and diagonals only), one can restrict
to rational functions in the form $\, R= \, 1/Q$, but since these arguments may look too esoteric
for physicists, let us, in a learn-by-example, pedagogical approach, provide examples
showing that telescopers of rational functions in the form $\, R=\, 1/Q^k$
are homomorphic to telescopers  of rational functions in the form $\, R=\, 1/Q$,
and then that  telescopers  of rational functions in the form $\, R=\, P/Q$
are homomorphic to telescopers  of rational functions in the form $\, R=\, 1/Q$.

\vskip .3cm

\subsection{Diagonals of rational  functions: $\, R=\, 1/Q^k$ reducing to $\, 1/Q$}
\label{1/Q^k}

Let us denote $\, Q$ the polynomial:
\begin{eqnarray}
  \label{usingdenote}
&&  \hspace{-0.98in} \quad  \quad  \quad  \quad  \quad \quad  \quad 
\, Q \, \, = \, \,\, \, 
   1 \,\, \,  + x\,y + y\,z + z\, x \,\,\,  + 3 \cdot \, (x^2 + y^2 + z^2).
   \end{eqnarray}
Let us denote  $\, L_4^{(n)}$ the telescopers of $\, Diag(1/Q^n)$:
\begin{eqnarray}
\label{telescQn}
  \hspace{-0.96in}&&  \quad \quad \quad \quad \quad \quad \quad \quad  \quad  \quad  \quad 
                 L_4^{(n)}  \cdot \,    Diag\Bigl( {{1} \over {Q^n}}  \Bigr) \,=\,0. 
\end{eqnarray}
One remarks that these telescopers are all of order four.
One actually finds the following homomorphisms between successive telescopers (\ref{telescQn}):
\begin{eqnarray}
\label{telescQnHom}
  \hspace{-0.96in}&&  \quad \quad \quad \quad \quad
  Homomorphisms\Bigl(L_4^{(n)}, \, \, L_4^{(n +1)} \Bigr)
  \, \, = \, \, \, 3 \, x \cdot \, D_x \, \, + \,  2 \, n.
\end{eqnarray}
In other words one has the relations:
\begin{eqnarray}
\label{telescQnRel}
  \hspace{-0.96in}&&  \quad \quad \quad \quad \quad  \quad \quad \quad
         L_4^{(n +1)} \cdot \, (3 \, \theta \, + 2 \, n) \, \, \, = \, \, \, \,  Z_1(n) \cdot\, L_4^{(n)},
\end{eqnarray}
where $\, Z_1(n)$ is an order-one linear differential operator,
the intertwining relation (\ref{telescQnRel}) yielding:
\begin{eqnarray}
\label{telescQnRelA}
  \hspace{-0.96in}&&  \quad \quad \quad \quad  \quad \quad 
  L_4^{(n +1)} \cdot \,  (3 \, \theta \, + 2 \, n)  \,
  \,  \cdots \,  \,  (3 \, \theta \, + 6) \cdot \, (3 \, \theta \, + 4) \cdot \, (3 \, \theta \, + 2)
  \nonumber \\ 
  \hspace{-0.96in}&&  \quad \quad \quad \quad \quad \quad \quad \quad\quad 
  \, \, = \, \, \, \,
  Z_1(n)  \, \,  \cdots \,  \,   Z_1(3)  \cdot \, Z_1(2) \cdot\, Z_1(1) \cdot \, L_4^{(1)}.
\end{eqnarray}
One deduces:
\begin{eqnarray}
  \hspace{-0.96in}&&  \quad \quad \quad
2^n \cdot \, n! \cdot \, Diag\Bigl( {{1} \over {Q^{n+1}}}  \Bigr)
\nonumber \\ 
  \hspace{-0.96in}&&  \quad \quad \quad  \quad \quad
 \, \, = \, \, \,
(3 \, \theta \, + 2 \, n)  \,   \,  \cdots \,  \,  (3 \, \theta \, + 6)
\cdot \, (3 \, \theta \, + 4) \cdot \, (3 \, \theta \, + 2) \cdot \,  Diag\Bigl( {{1} \over {Q}}  \Bigr). 
\end{eqnarray}
In other words the diagonal of $\, 1/Q^{n+1}\, $ can be simply deduced from the diagonal of $\, 1/Q$.

\vskip .2cm

{\bf Remark 3.1:} The product
$\,  (3 \, \theta \, + 2 \, n)  \,   \,  \cdots \,  \,  (3 \, \theta \, + 6)
\cdot \, (3 \, \theta \, + 4) \cdot \, (3 \, \theta \, + 2)$,
in the intertwining relation (\ref{telescQnRelA}), {\em is in fact a direct sum}:
\begin{eqnarray}
  \hspace{-0.98in}&&  \quad \quad \quad \quad           
                     (3 \, \theta \, + 6) \cdot \, (3 \, \theta \, + 4) \cdot \, (3 \, \theta \, + 2)
      \nonumber \\ 
  \hspace{-0.96in}&&  \quad \quad \quad      \quad  \quad      \quad \quad \quad                
  \, \, = \, \, \,
  27\, x^3 \cdot \, LCLM\Bigl(  (3 \, \theta \, + 6),  \,  \, (3 \, \theta \, + 4),  \,  \, (3 \, \theta \, + 2)\Bigr).  
\end{eqnarray}

\vskip .3cm

One has, for instance, the relations:
\begin{eqnarray}
\label{diagupto5}
  \hspace{-0.96in}&&  \quad \quad \quad
2 \cdot \,  Diag\Bigl( {{1} \over {Q^2}}  \Bigr) \,  \, = \, \,  \,
   (3 \, \theta \, + 2)    \cdot \,   Diag\Bigl( {{1} \over {Q}}  \Bigr)
                     \nonumber \\ 
  \hspace{-0.96in}&&  \quad \quad \quad
8 \cdot \,  Diag\Bigl( {{1} \over {Q^3}}  \Bigr) \,  \, = \, \,  \,
    (3 \, \theta \, + 4)    \cdot \,    (3 \, \theta \, + 2)    \cdot \,   Diag\Bigl( {{1} \over {Q}}  \Bigr)
                     \\ 
  \hspace{-0.96in}&&  \quad \quad \quad
48 \cdot \,  Diag\Bigl( {{1} \over {Q^4}}  \Bigr) \,  \, = \, \,  \,
(3 \, \theta \, + 6)    \cdot \,   (3 \, \theta \, + 4)
\cdot \,    (3 \, \theta \, + 2)    \cdot \,   Diag\Bigl( {{1} \over {Q}}  \Bigr)
  \nonumber \\
  \hspace{-0.96in}&&  \quad \quad \quad
 384 \cdot \,  Diag\Bigl( {{1} \over {Q^5}}  \Bigr) \,  \, = \, \,  \,
 (3 \, \theta \, + 8)    \cdot \,  (3 \, \theta \, + 6)    \cdot \,   (3 \, \theta \, + 4)
 \cdot \,    (3 \, \theta \, + 2)    \cdot \,   Diag\Bigl( {{1} \over {Q}}  \Bigr).
   \nonumber 
\end{eqnarray}
Of course, since the telescoper of $\, Diag\Bigl( {{1} \over {Q}}  \Bigr)$
is an order four linear differential operator,
the  order-$(k-1)$ product in front of $\, Diag\Bigl( {{1} \over {Q}}  \Bigr)$ in
(\ref{diagupto5}) can be, for $\, Diag\Bigl( {{1} \over {Q^k}}  \Bigr)$,  reduced to an order-three
linear differential operator (the simple products
$\, (3 \, \theta \, + 2 \cdot \, (k-1)) \cdots  (3 \, \theta \, + 4) \cdot \,    (3 \, \theta \, + 2) $
in (\ref{diagupto5}) being taken ``modulo'' $\, L_4$, 
for $\, k\, \ge 5$). 

\vskip .3cm

\subsection{Diagonals of rational  functions: $\, R=\, P/Q \, $ reducing to $\, 1/Q$}
\label{P/Q}

Experimentally one finds, quite often, that the telescoper of a rational  function of the form
$\, R=\, P/Q \, $ and  the telescoper of the simple rational  function $\, 1/Q$
with its numerator normalized to $\, 1$, are homomorphic. The intertwiner $\, M$
occurring in the homomorphisms of these two telescopers yields a relation of the form
\begin{eqnarray}
  \label{oftheform}
  \hspace{-0.7in} \quad  \quad  \quad \quad  \quad \quad 
  Diag\Bigl( {{P} \over {Q}}\Bigr)
  \, \, = \, \, \, M \cdot \, Diag\Bigl( {{1} \over {Q}}\Bigr),
\end{eqnarray}
yielding the prejudice that the diagonals of the  rational  functions of the form
$\, P/Q \, $ should reduce to the ``simplest\footnote[1]{Simplest in some sense. In fact
  one is looking for a {\em cyclic vector}, and the cyclic vector is not
  necessarily  $\, Diag(1/Q)$ 
(see relation (\ref{lastQ})   and (\ref{lastQxy}) below).}'' diagonal, namely $\, Diag(1/Q)$.
In fact things are slightly more subtle, as will be seen below. 

\vskip 0.1cm

Sticking with the polynomial (\ref{usingdenote}), one has
\begin{eqnarray}
  \label{L4first}
  \hspace{-0.7in} \quad  \quad  \quad \quad  \quad \quad 
 L_4^{(1)} \cdot \, Diag\Bigl( {{1} \over {Q}}\Bigr)  \, \, = \, \, \, 0, 
\end{eqnarray}
and considering the diagonal of $x\,y/Q$, one obtains an
order-five differential operator with unique factorization:
\begin{eqnarray}
  \label{L4xyDxdiag}
  \hspace{-0.7in} \quad  \quad  \quad \quad  \quad \quad 
 L_{4}^{(xy)} \cdot \, D_x \,  \cdot \,  Diag\Bigl( {{x \, y} \over {Q}}\Bigr)  \, \, = \, \, \, 0.
\end{eqnarray}

The homomorphisms between $L_4^{(1)}$ and $L_{4}^{(xy)}$ amounts to seeking for linear
differential operators that map the solutions of one differential operator into the other. These relations are 
\begin{eqnarray}
  \label{Q3}
  \hspace{-0.7in} \quad  \quad  \quad \quad  \quad \quad 
  L_{4}^{(xy)} \cdot \,  Q_3 \, \, = \, \, \,  K_3 \cdot \, L_4^{(1)}, 
\end{eqnarray} 
and 
\begin{eqnarray}
 \label{alsohas}
  \hspace{-0.7in} \quad  \quad  \quad \quad  \quad \quad 
L_4^{(1)} \cdot \, J_3   \, \, = \, \, \,   P_3 \cdot \,  L_{4}^{(xy)},
\end{eqnarray}
where the intertwiners $Q_3$, $K_3$, $J_3$ and $P_3$ are linear differential operators of order three.

\vskip 0.1cm

Note that the above two relations show \cite{diffalgCalabi} that
the linear differential operator $ \, J_3 \cdot Q_3$ (resp. $ \, Q_3 \cdot J_3$)
leaves the solutions of $\, L_4^{(1)}$ (resp.   $ \, L_{4}^{(xy)}$)
unchanged\footnote[4]{Equivalently, the adjoint of $ \, P_3 \cdot K_3$
  (resp. the adjoint of $K_3 \cdot P_3$) leaves the solutions o
  f the adjoint of $L_4$ (resp. the adjoint of $L_4^{(xy)}$) unchanged.},
\begin{eqnarray}
  \label{gives}
  \hspace{-0.7in}  \quad  \quad \quad  
  J_3 \cdot \, Q_3  \cdot \,  Diag\Bigl( {{1} \over {Q}}\Bigr)
  \, \, = \, \, \,  Diag\Bigl( {{1} \over {Q}}\Bigr)
  \\
  \hspace{-0.98in}  \quad \quad \quad  \quad \quad  \quad
  \, \, = \, \, \,  \,
  1 \, \,\, -195\, x^2 \, +135225\, x^4  \, -143647728\, x^6  \, +182699446545\, x^8 \,
  \nonumber \\
  \hspace{-0.98in}  \quad \quad \quad  \quad \quad \quad \quad  \quad \quad \quad
  \, -252437965534755\,\,  x^{10}
  \, +364803972334074000 \, \, x^{12} \,\, \, \, + \, \, \, \cdots
 \nonumber 
\end{eqnarray} 
and
\begin{eqnarray}
  \label{givestogether}
  &&  \hspace{-0.98in}  \quad \quad  \quad
 Q_3 \,  \cdot \,  J_3  \cdot \, D_x  \cdot \,  Diag\Bigl( {{x \, y} \over {Q}}\Bigr)
  \, \, = \, \, \,  D_x  \cdot \, Diag\Bigl( {{x \, y} \over {Q}}\Bigr) \\
  &&  \hspace{-0.98in} \quad  \quad  \quad \quad  \quad
  \, \, = \, \, \, \, \,
  16\, x \,\, \, -38400 \,x^3\,\, +71593536 \,x^5\, -126120445440 \,x^7\, 
  \nonumber \\
&&  \hspace{-0.98in} \quad  \quad  \quad  \quad \quad  \quad \quad  
  \, +218901889206000 \,x^9\, -378463218115207680\,x^{11} \, \,\, + \, \, \cdots 
\end{eqnarray}

\vskip 0.1cm

Introducing the differential operator $\, D_x$ on both sides of  (\ref{alsohas}),
and using  (\ref{L4xyDxdiag}), one obtains:
\begin{eqnarray}
\label{lasequ}
\fl \qquad \quad
L_4^{(1)} \cdot \, J_3 \cdot \,D_x  \cdot \,Diag\Bigl( {{x \, y} \over {Q}}\Bigr)
\, \, = \, \,
P_3 \cdot \, ( L_{4}^{(xy)} \cdot \,D_x ) \cdot Diag\Bigl( {{x \, y} \over {Q}}\Bigr). 
\end{eqnarray}
The RHS of (\ref{lasequ}) cancels and therefore, the LHS of (\ref{lasequ}),
according to  (\ref{L4first}), leads to 
\begin{eqnarray}
\label{lastQ}
Diag\Bigl( {{1} \over {Q}}\Bigr)
\,\, = \,\, J_3 \cdot \,D_x  \cdot \,Diag\Bigl( {{x \, y} \over {Q}}\Bigr).
\end{eqnarray}
Also, acting by both sides of  (\ref{Q3}) on $\, Diag(1/Q)$, using  (\ref{L4first}),
and (\ref{L4xyDxdiag}) in mind  leads to:
\begin{eqnarray}
\label{lastQxy}
D_x  \cdot \,Diag\Bigl( {{x \, y} \over {Q}}\Bigr)
\,\, = \,\, Q_3 \cdot Diag\Bigl( {{1} \over {Q}}\Bigr).
\end{eqnarray}

With these relations we see that the derivative of the
diagonal of $\, xy/Q$ simply reduces to the diagonal of $\, 1/Q$, but
the diagonal of $\, xy/Q$ does not simply
reduce\footnote[5]{Here $\, 1/Q$ is not the ``cyclic vector''.}
to the diagonal of $\, 1/Q$.

\vskip .3cm

\section{Diagonals of algebraic functions}
\label{diagalg}

\subsection{Diagonals of algebraic functions: a first example}
\label{diagalgfirst}

Let us consider the algebraic functions: 
\begin{eqnarray}
\label{defdiagexample}
  \hspace{-0.96in}&&  \quad \quad \quad \quad \quad  \quad \, \,  
  A(x, y) \, \, \, = \, \, \,\,
  {{1} \over { \Bigl(1 \, -\alpha \cdot \, (x\,+y) \Bigr)^{1/n} }}
  \quad \quad \quad \, \, \,
  n \, = \, \, 2, \, 3, \, \,\, \cdots 
\end{eqnarray} 
The telescopers of these algebraic functions are order-two linear differential operators
with the simple $\, _2F_1$ hypergeometric solution:
\begin{eqnarray}
\label{defdiagexamplesolugen_n}
\hspace{-0.96in}&&  \quad \quad 
_2F_1\Bigl([{{1} \over {2 \, n }}, \, {{n\, +1} \over {2\, n }}], \, [1], \,
4 \cdot \, \alpha^2  \cdot \, x\Bigr)
\nonumber \\
\hspace{-0.96in}&&  \quad \quad  \quad
\, \, = \, \, \,
1 \,\,\, + {{n+1} \over {n^2}}  \, \alpha^2 \,  x \, \,\,
+  {{(1 \, +n) \cdot \,(1\, +2\, n) \cdot \,(1 \, +3 \,n)  } \over {4 \cdot n^4}}  \, \alpha^4 \,  x^2
\,\,\, \,\, \, + \, \,\, \cdots  
\end{eqnarray}
Note that, among these $\, _2F_1$  hypergeometric functions,
the $\, n=\, 2$,  $\, n=\, 3$, $\, n=\, 4$, $\, n=\, 6$ cases
correspond to {\em modular forms} (see Appendix B in~\cite{Heun}).   

These  hypergeometric series can be seen to be, as it should, the diagonals of the algebraic functions
(\ref{defdiagexample}).
In particular, for $\, n= \, 2$, one gets:
\begin{eqnarray}
\label{defdiagexamplesolu}
\hspace{-0.96in}&&   \quad \,  \,  
_2F_1\Bigl([{{1} \over {4 }}, \, {{3} \over {4 }}], \, [1], \,  \, 4 \cdot \, \alpha^2  \cdot \, x\Bigr)
\nonumber \\
\hspace{-0.96in}&&  \quad \quad \, \,  
\, \, = \, \, \, 
\Bigl( {{1} \over {1 \, -3 \, \alpha^2  \, x }} \Bigr)^{1/4}  \cdot \,
_2F_1\Bigl([{{1} \over {12}}, \, {{5} \over {12}}], \, [1], \, 
     {{27} \over {4}} \cdot \,
     {{\alpha^4  \cdot \, x^2 \cdot \, (1 \, -4 \, \alpha^2 \, x)} \over {(1\, -3 \, \alpha^2 \, x)^3}} \Bigr)
   \\
\hspace{-0.96in}&&   \quad \quad  \, \,  
\, \, = \, \, \,\, \,
1 \,\,\, \, + {{3} \over {4}}  \, \alpha^2 \,  x \,\, \,+  {{105} \over {64}}  \, \alpha^4 \,  x^2
\, +  {{1155} \over {256}}  \, \alpha^6 \, \, x^3 \, +  {{225225} \over {16384}}  \, \alpha^8 \, \, x^4
\,\, \, \,\, + \, \,\, \cdots   \nonumber 
 \end{eqnarray}
For $\, n\, = \, 2\, $ it is natural  to associate  the denominator
of (\ref{defdiagexample}), with the algebraic surface
\begin{eqnarray}
  \label{surface}
  \hspace{-0.96in}&&  \quad \quad \quad \quad\quad \quad \quad \quad \quad \quad \quad
  z^2 \, \,  = \, \,  \, \, 1 \,\, \,  -\alpha \cdot \, (x\,+y),  
\end{eqnarray}
and, following ideas developped in~\cite{MDPI}, since calculating the diagonal of the function
(\ref{defdiagexample}) for $\, n= \, 2$, 
amounts, in the multi-Taylor expansion, to extracting the terms depending
only on the product $\, p \, = \, x \, y$, take the intersection of the algebraic surface (\ref{surface})
with the surface  $\, p \, = \, x \, y$. This amounts, for instance, to eliminating
$\, y \, = \, p/x$ in (\ref{surface}),
thus getting the algebraic curve 
\begin{eqnarray}
\label{algcurve}
  \hspace{-0.96in}&&  \quad \quad \quad \quad \quad \quad \quad
 - \alpha \cdot \, x^2 \, \, -x\, z^2 \, \, -\alpha \cdot \, p \, \, +x \, \,  \, = \,  \, \, \, 0, 
\end{eqnarray}
which turns out to be an {\em elliptic curve}  (genus-one). Calculating the j-invariant of the
elliptic curve (\ref{algcurve}), one deduces the following Hauptmodul
\begin{eqnarray}
\label{algcurveHau}
\hspace{-0.96in}&&  \quad \quad \quad \quad\quad \quad \quad
  {\cal H} \, \, = \, \, \, {{1728 } \over {j}} \, \, = \, \, \,\,
  {{27} \over {4}} \cdot \,
  {{\alpha^4 \cdot \, p^2 \cdot \, (1 \, -4 \, \alpha^2 \, p)} \over {(1\, -3 \, \alpha^2 \, p)^3}}, 
\end{eqnarray}
which is actually the Hauptmodul pullback in (\ref{defdiagexamplesolu}).
This example gives some hope that the effective algebraic geometry approach
of diagonals of rational functions,
detailed in~\cite{MDPI}, could also work with  diagonals of algebraic  functions.

For $\, n\, \ne \, 2\, $  it is tempting to associate  the denominator
of (\ref{defdiagexample}), with the algebraic surface
\begin{eqnarray}
  \label{surfacen}
  \hspace{-0.96in}&&  \quad \quad \quad \quad\quad \quad \quad \quad \quad \, \, 
  z^n \, \,  = \, \,  \, 1 \, -\alpha \cdot \, (x\,+y),  
\end{eqnarray}
and after the  elimination $\, y \, = \, p/x$ in (\ref{surface}), the algebraic curve 
\begin{eqnarray}
\label{algcurven}
  \hspace{-0.96in}&&  \quad \quad \quad \quad\quad \quad \quad \quad \quad
 - \alpha \cdot \, x^2 \, \, -x\, z^n \, \, -\alpha \cdot \, p \, \,\,  +x \, \, = \, \, \, 0, 
\end{eqnarray}
but such algebraic curves turn out to be of genus $\, g = \, n-1$.
Understanding the emergence of {\em modular forms} for the $\, n=\, 3$, $\, n=\, 4$, $\, n=\, 6$
subcases of (\ref{defdiagexamplesolugen_n})
from (respectively) genus $\, 2$, $\, 3$, and $\, 5$
algebraic curves, is an open (and challenging) problem.

\vskip .3cm

{\bf Remark 4.1:} From the definition of the diagonals of a rational, or algebraic, functions it is
straightforward to see that the  diagonals of the algebraic functions (\ref{defdiagexample})
are  series of the variable $\, \alpha^2 \, x$. Consequently, the previous calculations for
a particular value of $\, \alpha$, are sufficient to recover the previous results valid
for arbitrary $\, \alpha$. For that reason we will, in the next example, take
specific values of the parameters. 

\vskip .3cm

\subsection{Diagonals of algebraic functions: a second example}
\label{diagalgsec}

Let us consider the algebraic functions: 
\begin{eqnarray}
\label{defdiagexamplesec}
  \hspace{-0.96in}&&  \quad \, \,  \, \, 
  A(x, y) \, \, \, = \, \, \,\,
  {{1} \over { \Bigl(1 \, -3 \cdot \, (x\,+y) \, +5 \cdot \, (x^2\,+y^2) \Bigr)^{1/n} }}, 
  \quad \quad \, \,  \, \, 
  n \, = \, \, 2, \, 3, \, \, \, \cdots 
\end{eqnarray} 
For $\, n=\, 2$ the telescoper of the algebraic function (\ref{defdiagexamplesec})
is an order-two linear differential operator
with the pullbacked $\, _2F_1$ hypergeometric solution:
\begin{eqnarray}
\label{defdiagexamplesolusec}
\hspace{-0.96in}&&  \quad \quad  
       {{1} \over { (1 \, -30 \, x)^{1/2} }}
       \cdot \, _2F_1\Bigl([{{1} \over {4 }}, \, {{3} \over {4 }}], \, [1], \,  \,
       - {{ 4 \cdot \, (11 \, -200 \, x)  \cdot \, x} \over { (1 \, -30 \, x)^{2} }} \Bigr)
  \nonumber \\
\hspace{-0.96in}&& \quad  \quad  
\, \, = \, \, \, 
 {{1} \over {(1 \, -27  \, x \, +300 \, x^2)^{1/4} }} 
 \\
\hspace{-0.96in}&&  \quad  \quad  \quad    \quad  \quad \quad \,  \times \, 
_2F_1\Bigl([{{1} \over {12}}, \, {{5} \over {12}}], \, [1], \, \, 
     {{27} \over {4}} \cdot \,
     {{  x^2 \cdot \,  (11 \, -200 \, x)^2  \cdot \,(1 \, -16 \, x \,+100 \, x^2)  } \over {
         (1\, -27\, x \, +300 \, x^2)^3}} \Bigr)
     \nonumber
   \\
\hspace{-0.96in}&&  \quad \quad \quad  
\, \, = \, \, \, \, \,
1 \,\,\,  \, + {{27} \over {4}}   \,  x \,\,  \,+  {{4305} \over {64}}  \,  x^2
\, +  {{199395} \over {256}} \, \, x^3 \, +  {{167040825} \over {16384}}  \, \, x^4
\,\, \,\, \, \, + \, \,\, \cdots
\nonumber 
\end{eqnarray}
From multi-Taylor series expansion, it is straightforward to see that the hypergeometric series
is actually the diagonal of the algebraic function (\ref{defdiagexamplesec}) for $\, n= \, 2$.

As in the previous subsection we introduce the algebraic surface
\begin{eqnarray}
  \label{surfacenew}
  \hspace{-0.96in}&&  \quad \quad \quad \quad\quad \quad \quad\quad \quad
  z^2 \, \,  = \, \,  \, 1 \, \, - 3 \cdot \, (x\,+y) \, +5  \cdot \, (x^2\,+y^2),   
\end{eqnarray}
and, again, eliminate $\, y \, = \, p/x \, $ in (\ref{surfacenew}),
thus getting the algebraic curve 
\begin{eqnarray}
\label{algcurvenew}
  \hspace{-0.96in}&&  \quad \quad \quad \quad\quad \quad \quad\, \, 
 5\,x^4 \, \, -x^2\, z^2 \, -3\, x^3 \, +5\, p^2 \, -3\, p\, x \,\,  +x^2 \,\,  \, = \,\,  \, \, 0, 
\end{eqnarray}
which turns out to be an {\em elliptic curve}  (genus-one). Calculating the j-invariant of the
elliptic curve (\ref{algcurvenew}), one deduces the following Hauptmodul
\begin{eqnarray}
\label{algcurveHaupt}
\hspace{-0.96in}&&  \quad \quad  \, \, \, 
  {\cal H} \, \, = \, \, \, {{1728 } \over {j}} \, \, = \, \, \,\,
    {{27} \over {4}} \cdot \,
     {{  p^2 \cdot \,  (11 \, -200 \, p)^2  \cdot \,(1 \, -16 \, p \,+100 \, p^2)  } \over {
         (1\, -27\, p \, +300 \, p^2)^3}}, 
\end{eqnarray}
{\em which is actually the Hauptmodul} pullback in (\ref{defdiagexamplesolusec}).
Again, this last example gives some hope that the effective algebraic geometry approach of diagonals
of rational functions, detailed in~\cite{MDPI}, could also work with
diagonals of algebraic functions.
For $\, n\, \ne 2$, it is tempting to introduce   the algebraic surface
\begin{eqnarray}
  \label{surfacenewnew}
  \hspace{-0.96in}&&  \quad \quad \quad \quad\quad \quad \quad\quad \quad
  z^n \, \,  = \, \, \, \, 1 \,\, - 3 \cdot \, (x\,+y) \,\,  +5  \cdot \, (x^2\,+y^2),   
\end{eqnarray}
and, again, eliminate
$\, y \, = \, p/x$ in (\ref{surfacenew}),
thus getting the algebraic curve 
\begin{eqnarray}
\label{algcurvenew }
  \hspace{-0.96in}&&  \quad \quad \quad \quad\quad \quad \quad  
  5\,x^4 \,\, -x^2\, z^n \, \, -3\, x^3 \, +5\, p^2 \, -3\, p\, x \,\, +x^2
  \, \,\, = \, \,\, \, 0, 
\end{eqnarray}
which is an algebraic curve of genus $\, g \, = \, 2 \, n \, -3\,$ for $\, n$ even,
and  $\, g \, = \, 2 \, n \, -2\,$ for $\, n$ odd.
For $\, n\, = \, 3$ (genus $\,4$) the telescoper of the algebraic function
(\ref{defdiagexamplesec}) is an (irreducible) {\em order-three} linear differential  operator
{\em which is not homomorphic to its adjoint}.
The interpretation of such non-self-dual order-three linear differential operators from
these higher genus algebraic curves is a totally open problem. 

\vskip .3cm

\section{Understanding the emergence of selected differential Galois groups for diagonals of rational functions}
\label{Galois}

Experimentally one finds that almost all the linear differential operators annihilating
the diagonal of a rational, or algebraic, function are homomorphic to their adjoint~\cite{selected}. 
For instance, recalling an example in~\cite{selected}
\begin{eqnarray}
  \label{3F2ChristolBISbis}
  \hspace{-0.98in}&& \quad \quad
  _4F_3\Bigl([{{1} \over {3}},\, {{1} \over {3}}, \, {{2} \over {3}}, \, {{2} \over {3}}],
  \, [{{1} \over {2}}, \, 1, \, 1],
  \, \, {{729 } \over {4}} \cdot \,  x\Bigr)
  \, \, = \, \, \, Diag\Bigl( {{  1}  \over { 1 \, \, -(1\, +u) \cdot \, (x \,+y\, +z)}} \Bigr)
  \nonumber \\
 \hspace{-0.98in}&& \quad\quad\quad\quad\quad  
  \, \, = \, \, \, \, 1 \, \,  \, + \, 18\, x \,\, \,  +1350 \, x^2 \,\, \, \, + \, \cdots 
\end{eqnarray}
we find the corresponding order-four linear differential operator
\begin{eqnarray}
\label{orderfourL4}
  \hspace{-0.98in}&& \quad \quad \quad \quad \, \,
  x \cdot \, L_4 \, \,\, \, = \, \, \,\, \,
  2 \cdot \, x \cdot \, (3\, \theta \, +2)^2 \cdot \, (3\, \theta \, +1)^2
  \, \, \,\,   -81\cdot \, \theta^3 \cdot \, (2\, \theta \, -1),
\end{eqnarray}
which can be seen to be non-trivially homomorphic to its adjoint:
\begin{eqnarray}
\label{homoorderfourL4}
  \hspace{-0.98in}&& \quad \quad \quad \quad \quad \quad \quad 
  \, L_4 \cdot \, \Bigl(\theta \, + {{1} \over {2}} \Bigr)
  \, \, = \, \, \,   \,      \Bigl(\theta \, + {{1} \over {2}} \Bigr)\cdot \, adjoint(L_4).
\end{eqnarray}
Beyond diagonals of a rational, or algebraic, functions, one also finds experimentally, that
almost all the telescopers of rational or algebraic functions are homomorphic to their adjoint.
This homomorphism to the adjoint property is so systematic that, following a mathematician's
prejudice one can imagine that this is nothing but the {\em Poincar\'e duality}.  
The Poincar\'e duality~\cite{Poincare} works for  {\em any} algebraic variety: the diagonal
of {\em any} rational, or algebraic, function
should yield self-dual linear differential operators in the sense
that they are homomorophic to their adjoint.
This is {\em not} the case. It turns out that the linear differential operators of some $\, _nF_{n-1}$,
candidates to rule-out Christol's conjecture~\cite{christol-83,christol-85,conjecture},
precisely provide such rare examples of linear differential operators annihilating
diagonal of rational or algebraic functions that are {\em not} homomorphic to their adjoint. 
Among these candidates  a large set has been seen to actually be diagonals of rational, or
algebraic, functions~\cite{conjecture,Yurkevich}. 

\vskip .3cm 

\subsection{A recall on Christol's conjecture}
\label{recallChristol}

Let us recall one of the $\, _3F_2$  hypergeometric candidates introduced to
rule out Christol's  conjecture:
\begin{eqnarray}
  \label{3F2Christol}
  \hspace{-0.98in}&& \quad \quad
_3F_2\Bigl([{{2} \over {9}},\, {{5} \over {9}}, \, {{8} \over {9}}], \, [{{2} \over {3}}, \, 1],
\, \,  27 \cdot \,  x\Bigr)
  \\
\hspace{-0.98in}&& \quad   \quad \quad    
\, \, = \, \, \, \, 1  \, \,  \, + \, {{40} \over {9}} \cdot \,  x \, 
\, \, +  \, {{5236} \over {81}} \cdot \,  x^2
 \, \, +  \, {{7827820} \over {6561}} \cdot \,  x^3
\, \, +  \, {{1444588600} \over {59049}} \cdot \,  x^4  \, \,   \,  \, \, +  \, \cdots
 \nonumber 
\end{eqnarray}
It is a globally bounded series (change $\, x \, \rightarrow \, 3^3 \cdot \, x \, \, $
to get a series with integer coefficients).
In fact it actually corresponds~\cite{conjecture} to the diagonal of the algebraic function:
\begin{eqnarray}
  \label{Alg3F2Christol}
  \hspace{-0.98in}&& \quad \quad \quad \quad \quad \quad \quad \quad \quad \quad \quad
      {{  (1 \, \, -y \, -z)^{1/3}}  \over { 1 \, -x \, - \, y  \, -z}}.    
\end{eqnarray}
The telescoper of the algebraic function (\ref{Alg3F2Christol}) is the order-three linear differential
operator which has (\ref{3F2Christol}) as a solution. This order-three linear differential
operator is {\em not} homomorphic to its adjoint.  We have a $\, SL(3, \, \mathbb{C})$
differential Galois group. 

\vskip .2cm

Other similar examples are, for instance: 
\begin{eqnarray}
  \label{3F2ChristolBIS}
  \hspace{-0.98in}&& \quad \quad\quad\quad\quad
_3F_2\Bigl([{{1} \over {9}},\, {{4} \over {9}}, \, {{7} \over {9}}], \, [{{2} \over {3}}, \, 1],
  \, \,  27 \cdot \,  x\Bigr)
  \, \, = \,\, \, \, Diag\Bigl( {{  (1 \, \, -y \, -2\, z)^{2/3}}  \over { 1 \, -x \, - \, y  \, -z}} \Bigr), 
\end{eqnarray}
or
\begin{eqnarray}
  \label{3F2ChristolTER}
  \hspace{-0.98in}&& \quad\quad\quad\quad\quad
_3F_2\Bigl([{{2} \over {9}},\, {{5} \over {9}}, \, {{8} \over {9}}], \, [{{5} \over {6}}, \, 1],
  \, \,  27 \cdot \,  x\Bigr)
  \, \, = \, \, \,\,
  Diag\Bigl( {{  (1 \, \, -y \, -2\, z)^{1/3}}  \over { 1 \, -x \, - \, y  \, -z}} \Bigr),
\end{eqnarray}
or even the $\, _4F_3 \, $ hypergeometric function:
\begin{eqnarray}
  \label{3F2ChristolTERTER}
  \hspace{-0.98in}&& \quad\quad\quad\quad\quad
  _4F_3\Bigl([{{2} \over {9}},\, {{5} \over {9}}, \, {{8} \over {9}}, \,  {{1} \over {2}}],
  \, [ {{1} \over {3}}, \, {{5} \over {6}}, \, 1],
  \, \,  27 \cdot \,  x\Bigr)
  \, \, = \, \,
  Diag\Bigl( {{  (1 \, \, -x)^{1/3}  } \over {1\, -x \, -y \, -z}} \Bigr).
  \end{eqnarray}
Again these three diagonals (\ref{3F2ChristolBIS}), (\ref{3F2ChristolTER}) and  (\ref{3F2ChristolTERTER})
are solutions of telescopers that {\em are not} homomorphic to their adjoint.

These examples are taken in a list of 116 potential counter-examples
constructed in 2011 by Bostan et al.~\cite{2013-rationality-integrality-ising}. Note that, more recently,
38 cases in that list of 116, have actually been found to be diagonals of algebraic functions~\cite{Yurkevich}.
The two relations (\ref{3F2ChristolBIS}) and (\ref{3F2ChristolTER}) can be generalized~\cite{Yurkevich,class}
as follows:
\begin{eqnarray}
  \label{3F2ChristolMORE}
  \hspace{-0.98in}&& \quad\quad\quad\quad\quad
  _4F_3\Bigl([{{1\, -(R+S)} \over {3}},\, {{2\, -(R+S)} \over {3}}, \, {{3\, -(R+S)} \over {3}}, \,{{1 \, -S} \over {2}}],
  \nonumber \\
\hspace{-0.98in}&& \quad\quad\quad\quad\quad \quad \quad\quad \quad  \,
         [{{1\, -(R+S)} \over {2}},\,\, {{2\, -(R+S)} \over {2}}, \, 1],
         \, \,  27 \cdot \,  x\Bigr)
         \nonumber \\
\hspace{-0.98in}&& \quad\quad\quad\quad\quad \quad\quad 
\, \, = \, \, \,
Diag\Bigl( {{  (1 \, \, -x)^R \cdot \, (1 \, \, -x \, -2 \, y)^S } \over {1\, -x \, -y \, -z}} \Bigr), 
\end{eqnarray}
where $\, R$ and $\, S$ are rational numbers.
These diagonals are annihilated by the order-four linear differential operator:
\begin{eqnarray}
  \label{3F2ChristolMOREoperator}
  \hspace{-0.98in}&& \quad \quad \quad 
  2\cdot \, x \cdot \, (S -1 \, -2\, \theta)  \cdot \, (S+R \, \,  -3\, \theta)
    \cdot \, (S+R-1 \, \,  -3\, \theta) \cdot \, (S+R-2 \,\,   -3\, \theta)
  \nonumber \\
  \hspace{-0.98in}&& \quad\quad\quad \quad\quad \quad\quad 
  \, \,
  -\theta^2 \cdot \, (S+R+1 \,\,  -2\, \theta) \cdot \, (S+R\, \, -2\, \theta).
 \end{eqnarray} 
This order-four linear differential operator is {\em not} homomorphic to its adjoint.
Other more involved similar relations can be found in section 2.1 of chapter 2 of~\cite{Yurkevich}.

\vskip .2cm

Experimentally we found, after quite systematic calculations of thousands of telescopers
of rational, or algebraic, functions, that the telescopers are (almost always)
homomorphic to their adjoint,
or if they are not irreducible, that each of the factors of these telescopers
are homomorphic to their adjoint.
Such  previous examples like (\ref{3F2Christol}), (\ref{Alg3F2Christol}), or (\ref{3F2ChristolBIS})
and  (\ref{3F2ChristolTER}),
curiously related
to Christol's conjecture,
provide  the {\em rare} examples of diagonals of algebraic functions such that their corresponding
telescopers are {\em not} homomorphic to their adjoint. 
We have similar results with the algebraic function:
\begin{eqnarray}
  \label{Alg3F2Christolother}
  \hspace{-0.98in}&& \quad \quad \quad \quad \quad \quad \quad \quad \quad\quad \quad
      {{  x^{1/3}}  \over { 1 \,\,  -x \, - \, y  \, -z}}.    
\end{eqnarray}
In order to understand this ``duality-breaking'' (the telescoper is not self-adjoint up to homomorphisms),
it is tempting to introduce the (algebraic) function:
\begin{eqnarray}
  \label{Alg3F2Christolotheralpha}
  \hspace{-0.98in}&& \quad \quad \quad \quad \quad \quad \quad \quad \quad 
      {{  1}  \over { 1 \,\, \,  -x \, - \, y  \, -z \, \, \,\,  -\alpha \cdot \, x^{1/3} }}.    
\end{eqnarray}
However, in order to avoid the introduction of rational functions of $\, n$-th roots of
variables, we will (changing $\, x,\, y, \, z \, $ into $\, x^3,\, y^3, \, z^3$)
rather introduce the diagonal of the following rational function: 
\begin{eqnarray}
  \label{christola}
  &&  \hspace{-0.98in}  \quad \quad  \quad \quad \quad \quad  \quad \quad  \quad 
      \,  {{1} \over { 1 \, \,\, \,  -  x^3 - y^3  - z^3  \, \,\, \,  - \alpha \cdot  \, x}}. 
\end{eqnarray}

\vskip .3cm

\subsection{Understanding the emergence of selected differential Galois groups for almost all the diagonal of rational functions}
\label{GaloisSUB}

The linear differential operator annihilating the diagonal of the rational function (\ref{christola})
is a (quite large) order-eight linear differential operator $\, L_8(\alpha)$,
depending on the parameter $\, \alpha$,
which is homomorphic to its adjoint with an order-six intertwiner. This order-eight
linear differential operator $\, L_8(\alpha)$ is irreducible except at $\, \alpha \, = \, 0$.
For $\,  \alpha \, = \, 1$,  $\,  \alpha \, = \, 2$,  $\,  \alpha \, = \, 3$ the  order-eight
linear differential operator $\, L_8(\alpha)$ is homomorphic
to its adjoint with an order-six intertwiner\footnote[1]{For $\,  \alpha \, = \, 0$ the  order-eight
linear differential operator $\, L_8(\alpha)$ is homomorphic
to its adjoint with {\em several} order-six intertwiners.
Its exterior square has the rational solution $\,1/x/(1 \, -27 \, x^3)$, which is
solution of the exterior square $\, L_2$. }.
The differential Galois group should, thus, be included in $\, Sp(8, \, \mathbb{C})$.
This is confirmed when calculating~\cite{Canonical} the exterior square of  $\, L_8(\alpha)$.
This exterior square has a rational function solution $\, P_a/x/Q_a$, where the polynomials $\, P_a$
and $\, Q_a$ read:
\begin{eqnarray}
\label{ratsolu}
&&  \hspace{-0.98in} 
P_a \, \, = \, \,
(4\,\alpha^3 -27)\cdot  \, (20\,\alpha^3 -81) \,
+18 \cdot \, (-6561 \, -891\, \alpha^3 \, +500\, \alpha^6) \cdot \, x^3 \, +1594323\, x^6,
     \nonumber \\
&&  \hspace{-0.98in}   
 Q_a \, \, = \, \,  387420489\, x^9  \, -531441 \cdot \, (81+100\, \alpha^3)\cdot \, x^6
      \\
&&  \hspace{-0.98in}  \quad  \quad  \, 
      \, +(1594323 \, -2972133\, \alpha^3 \, +729000\, \alpha^6 \, -50000\, \alpha^9) \cdot \, x^3
      \,  \, \,  -27 \cdot \, (4\, \alpha^3 -27)^2.
       \nonumber 
\end{eqnarray}

Let us now take the $\, \alpha \, \rightarrow \, 0 \, $
limit of the order-eight linear differential operator $\, L_8(\alpha)$.
In this limit the order-eight linear differential operator just becomes the direct-sum
\begin{eqnarray}
\label{limitchristol}
&&  \hspace{-0.98in}  \quad \quad  \quad \quad \quad \quad \quad \quad \quad \quad \quad
       L_2 \, \oplus \, L_3  \, \oplus \, M_3, 
\end{eqnarray}
where the order-two linear differential operator $\, L_2 $ has the $\, _2F_1$ hypergeometric solution
\begin{eqnarray}
\label{limitchristol}
  &&  \hspace{-0.98in}  \quad \quad \quad \quad \quad \quad \quad \quad \quad
     _2F_1\Bigl([{{1} \over {3}}, \, {{2} \over {3}}], \, [1], \, 27 \, x^3\Bigr), 
\end{eqnarray}
where the order-three linear differential operator $\, L_3$ has the $\, _3F_2$
hypergeometric function solution
\begin{eqnarray}
\label{solchristolL3}
  &&  \hspace{-0.98in}  \quad \quad \quad \quad \quad \quad \quad \quad \quad
_3F_2\Bigl([{{5} \over {9}}, \, {{8} \over {9}}, \, {{11} \over {9}}],
\,\, [{{2} \over {3}}, 1], \, 27 \, x^3\Bigr), 
\end{eqnarray}
and where the order-three linear differential operator $\, M_3$ has the $\, _3F_2$
hypergeometric function solution:
\begin{eqnarray}
\label{solchristolM3}
  &&  \hspace{-0.98in}  \quad \quad \quad \quad \quad \quad \quad \quad \quad
_3F_2\Bigl([{{7} \over {9}}, \, {{10} \over {9}}, \, {{13} \over {9}}],
\,\, [{{1} \over {3}}, 1], \, 27 \, x^3\Bigr).
\end{eqnarray}

These two order-three linear differential operators, similarly
to the previous example (\ref{3F2Christol}),
are {\em not} homomorphic to their adjoint: they
{\em break the self-adjoint duality}\footnote[2]{Up to homomorphisms
of operators.}, and thus  
have a $\, SL(3, \, \mathbb{C})$ differential Galois group.

These two hypergeometric series are exactly on the same footing as (\ref{3F2Christol}):
they are {\em globally bounded series} (just change $\, x^3 \, \rightarrow \, 3^3 \, x^3 \, $
in order to get a series with integer coefficients), and their respective order-three
linear differential operators are {\em not} homomorphic to their adjoint, their differential Galois group
being $\, SL(3, \, \mathbb{C})$.
Let us note, however, that the  order-three linear differential operator $\, L_3$ 
{\em is actually homomorphic to the adjoint of} $\, M_3$, and of course  the
order-three linear differential operators $\, M_3$ is  homomorphic to the adjoint of $\, L_3$.

If, in an algebraic geometry perspective~\cite{MDPI}, one sees the fact that all our linear differential operators,
annihilating diagonals of rational functions, are homomorphic to their adjoint
as the differential algebra expression of
the Poincar\'e duality on the algebraic varieties corresponding to the denominators
of our rational functions~\cite{MDPI}, the fact that this Poincar\'e duality is broken for $\, L_3$
or $\, M_3$ {\em is, in fact, restored in the bigger picture} (\ref{christola})
with the linear differential order-eight operator.
In the $\, \alpha \, \rightarrow \, 0 \, $ limit we see that these two linear differential
operators breaking the duality,
actually emerge in a dual pair, thus {\em restoring the duality}.
For instance, if one focuses on   $\, L_6 \, \, = \, \, \, L_3  \, \oplus \, M_3 \, $
in (\ref{limitchristol}), one finds easily that this order-six linear differential 
operator is homomorphic to its adjoint. Its exterior square has the
following rational function solution:
\begin{eqnarray}
\label{ratsolL6}
  &&  \hspace{-0.98in}  \quad \quad \quad \quad \quad \quad \quad \quad \quad \quad \quad \quad
    {{ 4 \,\, +621\,\, x^3  } \over {(1 \,\, -27 \,\, x^3)^3 \cdot \, x }}. 
\end{eqnarray}

\vskip .3cm

Since these calculations are in the $\, \alpha \, \rightarrow \, 0 \, $ limit, let us expand, in $\, \alpha$,
the rational function (\ref{christola}):
\begin{eqnarray}
  \label{christolaterexpandfirst}
  &&  \hspace{-0.98in}  \quad \quad 
   {{1} \over { 1 \, \, -  x^3 - y^3  - z^3  \, \, - \alpha \cdot  \, x}}
  \,  \, \, \, = \,  \, \, \, \, {{1} \over { 1 \, \, -  x^3 - y^3  - z^3}}
  \, \, \,  \, \,  +{{x} \over { (1 \, \, -  x^3 - y^3  - z^3)^2 }} \cdot \, \alpha
  \nonumber \\
  &&  \hspace{-0.98in}  \quad \quad \quad \quad  \quad  \quad 
  \, \, +{{x^2} \over { (1 \, \, -  x^3 - y^3  - z^3)^3 }} \cdot \, \alpha^2 \, \, \, \, \,
  +{{x^3} \over { (1 \, \, -  x^3 - y^3  - z^3)^4 }} \cdot \, \alpha^3
  \nonumber \\
  &&  \hspace{-0.98in}  \quad \quad \quad \quad \quad \quad  \quad  \quad  \quad \, \,
  +{{x^4} \over { (1 \, \, -  x^3 - y^3  - z^3)^5 }} \cdot \, \alpha^4 \,  \, \,
   \, \, \, + \, \, \cdots 
\end{eqnarray}
The diagonal of a sum is clearly the sum of the diagonals. Thus the diagonal of the LHS
of (\ref{christolaterexpandfirst}) will be the sum of the various rational function terms
in $\, \alpha^n$ in the RHS of (\ref{christolaterexpandfirst}). The diagonal of the $\, \alpha^1$ term
in the $\, \alpha$-expansion (\ref{christolaterexpandfirst}) 
\begin{eqnarray}
  \label{alpha1}
  &&  \hspace{-0.98in}  \quad \quad  \quad \quad  \quad \quad \quad  \quad \quad  \quad 
   {{x} \over { (1 \, \, -  x^3 - y^3  - z^3)^2 }},  
\end{eqnarray}
is clearly equal to zero, since the diagonal extracts, in the multi-Taylor series, the terms in
the product $\, p \, = \, \, x \, y \, z$, or, in this case, the terms in
the product $\,  \, \, x^3 \, y^3 \, z^3$. Similarly the diagonal of the $\, \alpha^2$ term
in the $\, \alpha$-expansion (\ref{christolaterexpandfirst}) 
\begin{eqnarray}
  \label{alpha2}
  &&  \hspace{-0.98in}  \quad \quad  \quad \quad  \quad \quad \quad \quad  \quad  \quad  
   {{x^2} \over { (1 \, \, -  x^3 - y^3  - z^3)^3 }}, 
\end{eqnarray}
is also zero, but the diagonal of the $\, \alpha^3$ term
\begin{eqnarray}
  \label{alpha3}
  &&  \hspace{-0.98in}  \quad \quad  \quad \quad  \quad \quad \quad  \quad  \quad \quad 
   {{x^3} \over { (1 \, \, -  x^3 - y^3  - z^3)^4 }},  
\end{eqnarray}
is not zero.  Actually this last diagonal reads:
\begin{eqnarray}
  \label{diagalpha2}
  &&  \hspace{-0.98in}  \quad \quad\quad  \quad 
  -{{1} \over {9}} \cdot \, {{ 1 \, +216\,x^3 } \over {(1 \, -27 \, x^3)^3 }}
  \cdot \, x \cdot \, {{d} \over {dx}} \, _2F_1\Bigl([{{1} \over {3}}, \, {{2} \over {3}}], \, [1], \, 27 \, x^3\Bigr)
  \nonumber \\
   &&  \hspace{-0.98in}  \quad \quad \quad \quad \quad \quad \quad \quad  \quad \quad \quad
  -18 \cdot \, {{x^3} \over {(1 \, -27 \, x^3)^2  }}  \cdot \,
  _2F_1\Bigl([{{1} \over {3}}, \, {{2} \over {3}}], \, [1], \, 27 \, x^3\Bigr). 
  \\
   &&  \hspace{-0.98in}  \quad \quad  \quad 
  \, \, = \, \, \, \,  -20\, x^3 \, \,  -1680\, x^6 \, -92400\, x^9 \, -4204200\, x^{12}
  \, \,  -171531360\, x^{15} \, \, \, \, + \, \, \, \cdots  \nonumber 
\end{eqnarray}
It is annihilated by an order-two operator $\, M_2$.

We have a different story with telescopers.
Since the telescoper of a sum of rational functions is the direct sum (LCLM) of the telescopers
of these rational functions\footnote[1]{In fact one expects the telecoper of a sum of rational functions to
  be equal, or to be a rightdivisor, of the LCLM of the telescopers of these rational functions. 
  In contrast the diagonal of a sum of rational functions is equal
  to the sum of the diagonals of these rational functions,
as long as each rational function, in the sum, depends on all the variables.},
let us consider the telescopers of the first five terms in the RHS of
(\ref{christolaterexpandfirst}).  The telescoper of the first term is, of course,
the order-two linear differential operator
$\, L_2$ annihilating the diagonal of this rational function. The telescoper of the second term (in $\, \alpha^{1}$),
is  the previous order-{\em three} linear differential operator $\, L_3$. The telescoper
of  the third term  (in $\, \alpha^{2}$)
is exactly  the previous $\, M_3$.  The telescoper of the fourth term (in $\, \alpha^{3}$),
is the order-two linear differential operator $\, M_2$. 
The telescoper of the sum of the first orders in $\alpha$ in the expansion (\ref{christolaterexpandfirst})
\begin{eqnarray}
  \label{truncchristolaterexpand}
  &&  \hspace{-0.98in}  \, \, \, \, 
  {{1} \over { 1 \, \, -  x^3 - y^3  - z^3}}
  \, \,   +{{x} \over { (1 \, \, -  x^3 - y^3  - z^3)^2 }} \cdot \, \alpha
   \, \, +{{x^2} \over { (1 \, \, -  x^3 - y^3  - z^3)^3 }} \cdot \, \alpha^2, \,  \, \,\, 
 \end{eqnarray}
is actually the LCLM of the three telescopers $\, L_2$,  $\, L_3$ and $\, M_3$
{\em which is precisely the $\, \alpha \, \rightarrow \, 0 \, $
  limit of the order-eight linear differential operator !}

\vskip .3cm

\subsection{Revisiting $\, 1/Q \, \rightarrow \, P/Q^k$ for telescopers}
\label{revisiting}
The next terms in the $\alpha$-expansion (\ref{christolaterexpandfirst}), namely the terms
in $\, \alpha^{4\, +3\, n} \, $ with $\, n=\, 0, \, 1, \, \cdots$ 
\begin{eqnarray}
  \label{alpha4n}
  &&  \hspace{-0.98in}  \quad \quad  \quad \quad  \quad \quad  \quad \quad  \quad \quad  \quad 
   {{x^{4\, +3\, n}} \over { (1 \, \,  \, -  x^3 - y^3  - z^3)^{5\, +3\, n} }},  
\end{eqnarray}
have  telescopers {\em actually homomorphic} to the telescoper $\, L_3$ for (\ref{alpha1}).
Similarly, considering in the $\alpha$-expansion (\ref{christolaterexpandfirst}), namely the terms
in $\alpha^{5\, +3\, n}$ with $\, n=0, \, 1, \, \cdots$ 
\begin{eqnarray}
  \label{alpha5n}
  &&  \hspace{-0.98in}  \quad \quad  \quad \quad  \quad \quad  \quad \quad  \quad \quad 
   {{x^{5\, +3\, n}} \over { (1 \, \,  \, -  x^3 - y^3  - z^3)^{6 \, +3\, n} }},  
\end{eqnarray}
 have  telescopers {\em actually homomorphic} to the telescoper $\, M_3$ for (\ref{alpha2}).
Finally,  the terms
in $\alpha^{3\, +3\, n} \, $ with $\, n=0, \, 1, \, \cdots$ 
\begin{eqnarray}
  \label{alpha3n}
  &&  \hspace{-0.98in}  \quad \quad  \quad \quad  \quad \quad  \quad \quad  \quad \quad 
   {{x^{3 \, +3\, n}} \over { (1 \,  \, \, -  x^3 - y^3  - z^3)^{4\, +3\, n}}},  
 \end{eqnarray}
have  telescopers {\em  homomorphic} to the telescoper $\, L_2$,
generalizing the result (\ref{diagalpha2}) for $\, n=\, 0$. 
This last sequence of telescopers can be understood from the ideas sketched in subsections (\ref{1/Q^k}) 
and (\ref{P/Q}) for diagonals (changing for instance $\, (x,y,z)$ into $\, (x^3,y^3,z^3)$). However,
we see that these ideas {\em do not work anymore} when we compare the telescopers
for (\ref{alpha4n}) (resp.  the telescopers for (\ref{alpha5n})) with
the telescopers for (\ref{alpha3n}). These different telescopers are {\em not homomorphic}.
They correspond to {\em three different sequences} of telescopers of different nature,
corresponding to three hypergeometric function of {\em quite different nature}:
\begin{eqnarray}
\label{threedifferent}
  &&  \hspace{-0.98in} 
     _2F_1\Bigl([{{1} \over {3}}, \, {{2} \over {3}}], \, [1], \, 27 \, x^3\Bigr), 
_3F_2\Bigl([{{7} \over {9}}, \, {{10} \over {9}}, \, {{13} \over {9}}], \, [{{1} \over {3}}, 1], \, 27 \, x^3\Bigr), 
_3F_2\Bigl([{{5} \over {9}}, \, {{8} \over {9}}, \, {{11} \over {9}}], \, [{{2} \over {3}}, 1], \, 27 \, x^3\Bigr).
\nonumber 
\end{eqnarray}
Along this line similar $\alpha$-dependent examples are sketched in \ref{General}.
     
\vskip .3cm
  
{\bf To sum-up:}  The ideas sketched in subsections (\ref{1/Q^k})  and (\ref{P/Q}) for diagonals, can be
generalized to telescopers (which may  correspond to vanishing cycles i.e. diagonals), with the caveat that
the unique ``root'' rational function $\, 1/Q$, has to be replaced by a finite set of rational functions
($1/Q_1$, $\, 1/Q_2$, $\, 1/Q_3$ in our previous example).

\vskip .3cm

\section{An infinite number of birational symmetries of the diagonals and telescopers}
\label{infinite}

Let us consider the simplest example of non-trivial diagonal of rational function, namely
the diagonal of the rational function of three variables:
\begin{eqnarray}
\label{simplest}
  \hspace{-0.96in}&&  \quad \quad \quad \quad \quad \quad \quad \quad \, \,   
  R(x, y, z) \, \, \, = \, \, \,\,
  {{1} \over { 1 \, \, -x \, -y \, -z }}. 
 \end{eqnarray} 
Let us consider the {\em birational transformation} $\, B$:
\begin{eqnarray}
\label{simplestbirat}
  \hspace{-0.96in}&&  \quad \quad 
  B: \quad  (x, \, y, \, z) \, \, \, \, \, \longrightarrow \, \quad \, \, 
  \Bigl(x, \,\,\, y \cdot \, (1\, +3\, x\, +7\, x^2), \, \,\, {{z} \over {1\, +3\, x\, +7\, x^2 }}  \Bigr).
 \end{eqnarray} 
It is birational because its compositional inverse is also a rational function:
\begin{eqnarray}
\label{simplestinverse}
  \hspace{-0.96in}&&  \quad \quad  \quad \quad 
  (x, \, y, \, z) \, \, \, \longrightarrow \, \quad
  \Bigl(x, \, \, {{y} \over {1\, +3\, x\, +7\, x^2 }},\,  \, z \cdot \, (1\, +3\, x\, +7\, x^2) \Bigr).
\end{eqnarray}
Note that this birational transformation preserves the product $\, p \, = \, \, x \, y\, z$, as well as
the neighbourhood of the point $\,  (x, \, y, \, z) \, = \, \, (0, \, 0, \, 0)$.
This birational transformation is an {\em infinite order} transformation. The composition
of this transformation $\, n$ times gives:
\begin{eqnarray}
\label{simplestn}
  \hspace{-0.96in}&&  \quad \quad  \quad 
  (x, \, y, \, z) \, \, \, \, \, \longrightarrow \, \quad \, \, 
  \Bigl(x, \, \,\,y \cdot \, (1\, +3\, x\, +7\, x^2)^n,\, \,\, {{z} \over {(1\, +3\, x\, +7\, x^2)^n }}  \Bigr).
\end{eqnarray}
The rational function (\ref{simplest}),
transformed by the (infinite order) birational transformation (\ref{simplestbirat}),
reads:
\begin{eqnarray}
\label{simplestB}
\hspace{-0.96in}&&  \quad \quad  \quad  \quad  \, \, 
R_B(x, y, z) \, \, \, = \, \, \,\,
R\Bigl(x, \,  \, \,  y \cdot \, (1\, +3\, x\, +7\, x^2), \,  \, \, {{z} \over {1\, +3\, x\, +7\, x^2 }}  \Bigr)
  \nonumber \\
\hspace{-0.96in}&&  \quad \quad  \quad  \quad    \quad  \quad 
  \, \, \, = \, \, \,\, 
  {{1} \over { 1 \,\,\,  -x \, \, -y\cdot \, (1\, +3\, x\, +7\, x^2) \,\,  -z/(1\, +3\, x\, +7\, x^2) }}. 
 \end{eqnarray} 
On the multi-Taylor expansion of (\ref{simplestB}) one finds easily that the diagonal
of (\ref{simplest}) and  (\ref{simplestB}) are {\em actually identical}.

\vskip .3cm 

More generally, let us consider 
\begin{eqnarray}
\label{simplestbiratQ}
  \hspace{-0.96in}&&  \quad \quad  \quad \quad \quad \quad 
  B_x: \quad \,  (x, \, y, \, z) \,\,  \, \, \, \longrightarrow \, \quad \, \, 
  \Bigl(x, \, \, \, y \cdot \, Q_1(x), \, \, \, {{z} \over {Q_1(x) }}  \Bigr), 
 \end{eqnarray} 
where $\, Q_1(x)$ is a rational function\footnote[9]{See however section (\ref{transcendental}).}
with a Taylor expansion such that  $\, Q_1(0) \, \ne \, 0$.
One also finds for any such rational function $\, Q_1(x)$, that the diagonal of (\ref{simplest}) and  (\ref{simplestB})
are actually identical. This can be seen from the  multi-Taylor expansion of (\ref{simplestB}):
\begin{eqnarray}
\label{defdiagBB}
  \hspace{-0.96in}&&  \quad
  R_B(x, y, z) \, \, \, = \, \, \, \,
  \sum_{m}  \sum_{n}  \sum_{l} \,
  a_{m,n,l} \cdot \, x^m  \cdot \, y^n \cdot \, Q_1(x)^n  \cdot \, z^l \cdot \, Q_1(x)^{-l}
   \\
 \hspace{-0.96in}&&  \quad \quad 
  \, \, \, = \, \, \,  \sum_{m}  \, a_{m,m,m} \cdot \, (x\, y\, z)^m
  \, \, \, \,\,  + \sum_{(m,n,l) \ne \, (m,m,m)}
  \, a_{m,n,l} \cdot \, x^m  \cdot \, y^n   \cdot \, z^l \cdot Q_1(x)^{n-l}.
  \nonumber 
\end{eqnarray}
The second triple sum can be decomposed into the terms such that  $\, n \ne \, l$,
which cannot contribute to the diagonal (which extracts terms in $\, p \, = \, x \, y \, z$
and thus terms in the product $ \, y \, z$), and the $\, n = \, l\, $ terms (such that
the $\, Q_1(x)^{n-l}$ factor in (\ref{defdiagBB}) disappear):
\begin{eqnarray}
\label{defdiagBBoftheform}
  \hspace{-0.96in}&&  \quad \quad \quad \quad \quad \quad \quad \quad
 \sum_{m \ne \, n} \, a_{m,n,n} \cdot \, x^m  \cdot \, y^n   \cdot \, z^n.
\end{eqnarray}
This last sum (\ref{defdiagBBoftheform}), which excludes the power of $\, x$ to be equal
to the power of the product $\, y\, z$,  cannot contribute to the diagonal. 
We have thus proved that the diagonal of (\ref{simplest}) and  (\ref{simplestB}) are equal.

Of course there is nothing particular with the variable $\, x$.
We can also introduce other birational transformations which single out respectively $\, y$ and $\, z$:
\begin{eqnarray}
\label{simplestbiratQ2}
  \hspace{-0.96in}&&  \quad \quad  \quad \quad  \quad 
  B_y: \quad \,  (x, \, y, \, z) \, \, \, \,\, \longrightarrow \,\,  \quad\,
  \Bigl(x \cdot \, Q_2(y), \, \, \, \, y, \, \, \, \,  {{z} \over {Q_2(y) }}  \Bigr), 
\end{eqnarray}
and 
\begin{eqnarray}
\label{simplestbiratQ3}
  \hspace{-0.96in}&&  \quad \quad  \quad \quad  \quad 
  B_z: \quad \,  (x, \, y, \, z) \, \,\,  \, \,\longrightarrow \, \quad\,\, 
  \Bigl(x \cdot \, Q_3(z), \, \, \, \, \, \, {{y} \over {Q_3(z) }}, \, \, \, z \Bigr), 
 \end{eqnarray} 
for any rational functions $\, Q_2(x)$ and  $\, Q_3(x)$ with a Taylor expansion
and such that $\, Q_2(0) \, \ne \, 0$ and  $\, Q_3(0) \, \ne \, 0$.
We can compose these birational transformations
(\ref{simplestbiratQ}), (\ref{simplestbiratQ2}) and  (\ref{simplestbiratQ3}),
in any order and changing the various  $\, Q_1(x)$, $\, Q_2(x)$  and  $\, Q_3(x)$ at each step.
We get that way a quite large {\em infinite} set of birational transformations
preserving the product $\, p \, = \, \, x\, y\, z$
and the neighbourhood of the point $\,  (x, \, y, \, z) \, = \, \, (0, \, 0, \, 0)$.
Since the product $\, p \, = \, \, x\, y\, z$ is preserved,
let us eliminate (for instance) the variable $\, z \, = \, p/x/y$.
The three previous birational transformations (\ref{simplestbiratQ}), (\ref{simplestbiratQ2}) and  (\ref{simplestbiratQ3}),
on the three variables $\, x, \, y, \, z$, become
birational transformations depending on a parameter $\, p$, of only two variables $\, x, \, y$:
\begin{eqnarray}
\label{simplestbiratQp}
  \hspace{-0.96in}&&  \quad \quad  \quad \quad  \quad  \quad  \quad 
  \tilde{B}_x: \quad\,   (x, \, y) \, \,  \, \, \, \longrightarrow \, \, \quad \,  \, 
  \Bigl(x, \, \, \, y \cdot \, Q_1(x)  \Bigr), 
 \end{eqnarray} 
\begin{eqnarray}
\label{simplestbiratQ2p}
  \hspace{-0.96in}&&  \quad \quad  \quad \quad  \quad  \quad  \quad 
  \tilde{B}_y: \quad \,  (x, \, y) \, \, \, \,  \, \longrightarrow \,\,  \quad \,  \, 
  \Bigl(x \cdot \, Q_2(y), \, \, \, y  \Bigr), 
\end{eqnarray}
and 
\begin{eqnarray}
\label{simplestbiratQ3p}
  \hspace{-0.96in}&&  \quad \quad  \quad \quad  \quad  \quad  \quad 
  \tilde{B}_z: \quad  \,  (x, \, y) \,  \, \,\, \, \longrightarrow \, \quad \,\, \, 
  \Bigl(x \cdot \, Q_3\Bigl({{p} \over {x\, y}} \Bigr),
  \, \, \, \, \, \, y/Q_3\Bigl({{p} \over {x\, y}} \Bigr) \Bigr), 
 \end{eqnarray}
Composing these birational transformations of two variables
(\ref{simplestbiratQp}), (\ref{simplestbiratQ2p}) and  (\ref{simplestbiratQ3p}),
in any order and changing the various  $\, Q_1(x)$, $\, Q_2(x)$  and  $\, Q_3(x)$
at each step, one gets that way
a quite large subset of the (huge set of) {\em Cremona transformations}~\cite{Noether,Cremona}.

\vskip .3cm

{\bf Remark 6.1:} Of course there is nothing specific with the particularly
simple example (\ref{simplest})
of rational function. The previous birational transformations
(\ref{simplestbiratQp}), (\ref{simplestbiratQ2p}) and  (\ref{simplestbiratQ3p}),
are symmetries of the
diagonals of any rational function of three variables. Furthermore, there is nothing specific with
rational function of three variables. We can generalize such  birational transformations
for diagonal of rational function of $\, n$ variables, for any integer $\, n$.

\vskip .3cm

\subsection{Non birational symmetries for diagonals}
\label{notpreserving}

\subsubsection{Monomial transformation\\}
\label{notpreservingmo,omial}

Let us consider the  (non-birational) {\em monomial} transformation:
\begin{eqnarray}
\label{monomial}
  \hspace{-0.96in}&&  \quad \quad  \quad \quad   \quad \quad  \quad \quad 
  M: \quad \, \, (x, \, y, \, z) \, \,\,\,\, \, \, \longrightarrow \,\, \,  \quad
  \Bigl( x, \,  x^2\, y^2, \,  y\, z^3\Bigr). 
\end{eqnarray}
Let us perform this  monomial transformation (\ref{monomial}) on
the rational function (\ref{simplest}), one gets the new  rational function:
\begin{eqnarray}
\label{involutrationalm}
\hspace{-0.96in}&&   \quad \quad   \quad \quad
R_M(x, \, y, \, z) \, = \, \,\, R\Bigl( x, \,  x^2\, y^2, \,  y\, z^3\Bigr)
\, = \, \, \, \, {{1} \over { 1 \,\,\, -x \,\, -x^2\, y^2  \,\,\, - y\, z^3}}.
\end{eqnarray}
The calculation of the telescoper of (\ref{involutrationalm}) gives an order-two linear differentizal
operator which has the $\, _2F_1$ hypergeometric series solution:
\begin{eqnarray}
\label{nonbiratseries}
  \hspace{-0.96in}&&  \quad \quad  \,\,\,
  _2F_1\Bigl([{{1} \over {3}}, \, {{2} \over {3}}], \, [1], \, \, 27 \,  x^3  \Bigr)
  \, \,\, = \, \, \,\,  1 \, \,+6 \, x^3 \, +90\, x^6 \, +1680\, x^9 \, +34650\, x^{12}
  \nonumber \\
 \hspace{-0.96in}&&  \quad \quad  \quad \quad \quad   \quad \quad   \quad  
 \,
 \, +756756\, x^{15} \,\, +17153136\, x^{18} \,\, \,\, + \, \, \cdots 
\end{eqnarray}
One verifies easily, on the multi-Taylor expansion of (\ref{involutrationalm}), that
its diagonal is actually the $\, _2F_1$ hypergeometric series (\ref{nonbiratseries}).
The fact that the diagonal is the diagonal of (\ref{simplest}), where
$\, x$ is changed into $\, x^3$, is a consequence of the fact that the product
$\, p \, = \, \, x\, y\, z$ is  changed into $\, p \, = \, \, x^3\, y^3 \, z^3 \, $
by the monomial transformation (\ref{monomial}).

\subsubsection{Non-birational transformation \\}
\label{notpreservingmonomial}

Let us now consider the  non-birational ``monomial-like'' transformation
\begin{eqnarray}
\label{nonbirat2}
  \hspace{-0.96in}&&  \quad \quad  \quad \quad   \quad \quad 
  B: \quad  \,  (x, \, y, \, z) \,  \, \, \, \longrightarrow \, \quad \, 
  \Bigl( x, \, \,  \, x^2\, y^2 \cdot \, (1 \,+3\,x), \,  \, \, {{y\, z^3} \over {1 \,+3\,x}}\Bigr). 
\end{eqnarray}
Let us perform this non-birational monomial transformation (\ref{nonbirat2}) on
the rational function (\ref{simplest}), one gets the new  rational function
\begin{eqnarray}
  \quad 
\hspace{-0.96in}&&  \quad \quad  \quad   \quad   \quad   \quad 
R_B(x, \, y, \, z) \, \,  = \, \, \, 
R\Bigl( x, \, \,\, x^2\, y^2 \cdot \, (1 \,+3\,x), \, \,\, {{y\, z^3} \over {1 \,+3\,x}}\Bigr)
\nonumber \\
\label{involutrationalmbis}
\hspace{-0.96in}&&  \quad \quad  \quad \quad   \quad   \quad   \quad  \quad 
\, = \, \, \,
   {{1} \over { 1  \, \,\, -x \,\, \, \, -x^2\, y^2 \cdot \, (1 \,+3\,x) \, \, \, - y\, z^3/ (1 \,+3\,x)}}.
\end{eqnarray}
The calculation of the telescoper of (\ref{involutrationalmbis}) gives an order-two linear differential
operator which has, again, the $\, _2F_1$ hypergeometric series solution:
\begin{eqnarray}
\label{nonbiratseries2}
  \hspace{-0.96in}&&  \quad \quad \quad \quad  \,  \, 
  _2F_1\Bigl([{{1} \over {3}}, \, {{2} \over {3}}], \, [1]\, \, 27 \,  x^3  \Bigr)
  \, \,\,  = \, \, \,  \, \,  \, 1 \, \, \,  +6 \, x^3 \, \,  +90\, x^6 \, +1680\, x^9
 \nonumber  \\
 \hspace{-0.96in}&&  \quad\quad \quad  \quad  \quad \quad   \quad \quad  
 \,  \, +34650\, x^{12}
 \, +756756\, x^{15} \, +17153136\, x^{18}\,  \, \,  \, \, + \, \, \cdots 
\end{eqnarray}
One verifies easily on the multi-Taylor expansion of (\ref{involutrationalmbis}) that
its diagonal is the $\, _2F_1$ hypergeometric series (\ref{nonbiratseries2}).
This result can be understood from the results on (\ref{involutrationalm})
and the diagonal-preservation results on the birational transformations
(\ref{simplestbiratQ}), (\ref{simplestbiratQ2}) and  (\ref{simplestbiratQ3}).

Consequently we have another infinite set of (non-birational) transformations such that
the diagonal of a rational function is changed into the diagonal of that rational function
where $\, x$ is changed into $\, x^N$.

\vskip .3cm

\subsection{ Birational symmetries for telescopers}
\label{birat}
Recalling the creative telescoping equation (\ref{telescopequation}) and (\ref{telescopequation3}),
we have verified experimentally, on thousands of examples, that the previous birational transformations
generated by (\ref{simplestbiratQ}), (\ref{simplestbiratQ2}) and  (\ref{simplestbiratQ3}),
are actually compatible with the creative telescoping equations (\ref{telescopequation}) and (\ref{telescopequation3}).
Note however, in the birationally transformed creative telescoping equations,  that if the
telescoper does remain invariant ({\em even if we are not in a context where the rational function has a multi-Taylor
  expansion}), the two ``certificates'' $\, U$ and $\, V$ are transformed in a very involved way (they become quite
large rational functions).

\vskip .3cm

\subsubsection{ Birational symmetries not preserving $\,  (x, \, y, \, z) \, = \, \, (0, \, 0, \, 0)$\\}
\label{notpreserving}

Let us consider the involutive birational transformation:
\begin{eqnarray}
\label{involut}
  \hspace{-0.96in}&&  \quad \quad  \quad \quad   \quad \quad 
  I: \quad \,   (x, \, y, \, z) \, \,  \, \,  \, \, \longrightarrow \, \quad \,  \, \, 
  \Bigl( {{1} \over {x}}, \,  \,  {{1} \over {y}}, \,  \,\,  x^2 \, y^2 \, z\Bigr). 
 \end{eqnarray}
This involutive birational transformation transforms the rational function (\ref{simplest}) into:
\begin{eqnarray}
\label{involutrationalmf}
  \hspace{-0.96in}&&  \quad \quad  \quad \quad  \quad \quad  \quad \quad 
  R_I(x, \, y, \, z) \, \, = \, \, \, \,
  -\, {{x \, y} \over {  x^2\, y^3\, z \, \,  -x\, y \,\,\, +x \, +y }}. 
 \end{eqnarray}
The calculation of the telescoper of (\ref{involutrationalmf}) gives the same telescoper
as  the telescoper of (\ref{simplest}), whose diagonal  is the hypergeometric series:
\begin{eqnarray}
\label{diagsimplestfirst}
  \hspace{-0.96in}&&  \quad  \quad   
  _2F_1\Bigl([{{1} \over {3}}, \, {{2} \over {3}}], \, [1], \, 27\, x\Bigr)
 \nonumber  \\
 \hspace{-0.96in}&&  \quad \quad   \,\, 
  \, \, = \, \, \, (1\, -24\, x)^{-1/4}  \cdot \,  _2F_1\Bigl([{{1} \over {12}}, \, {{5} \over {12}}], \, [1],
  \, \, {{ 1728 \, x^3 \cdot \, (1 \, - 327\, x ) } \over {(1 \, -24 \, x)^3 }} \Bigr)
   \\
 \hspace{-0.96in}&&  \quad \quad   \, \,  
  \, \, = \, \, \,
  1 \, +6\,x \, +90\, x^2 \, +1680\, x^3 \, +34650\, x^4 \, +756756\, x^5 \, +17153136\, x^6 \, \,\,  + \, \, \cdots
  \nonumber
\end{eqnarray}
The hypergeometric series (\ref{diagsimplestfirst}) (which is equal to the diagonal of (\ref{simplest})),
 is, here, just an analytical solution of  the telescoper of (\ref{involutrationalmf}),
that is, a ``Period'' of  (\ref{involutrationalmf}) but corresponding to a non-vanishing cycle,
since  (\ref{involutrationalmf}) does not have a multi-Taylor expansion.

\vskip .3cm

\subsubsection{Birational symmetries from collineations\\}
\label{col}
Let us recall Noether's theorem~\cite{Noether1,Noether2,Noether} on the decomposition~\cite{Noether3} of Cremona
transformations\footnote[5]{Noether's theorem~\cite{Noether1} on the generation of the Cremona group
 by quadratic transformations, like many theorems
in mathematics, is not a constructive theorem. }. Noether's theorem shows that any  Cremona
transformation can be seen as the composition~\cite{Noether3,Noether} of {\em collineation transformations}
and of the Hadamard inverse transformation\footnote[1]{Called, in a projective $\, \mathbb{C}P_2$ approach,
  the quadratic transformation because it reads in the three homogeneous variables
$\, (x, \, y, \, t) \,  \, \rightarrow \,  \, (y \, t, \, x \, t, \, x\, y)$.}
\begin{eqnarray}
\label{quadratic}
\hspace{-0.96in}&&  \quad \quad  \quad \quad  \quad \quad \quad \quad
(x, \, y)  \,\, \quad  \longrightarrow \, \,\quad  \quad  \Bigl({{1} \over{x}}, \,\,  {{1} \over{y}} \Bigr). 
\end{eqnarray}

Let us consider Cremona transformations preserving $\, (x, \, y) \, = \, \, (0, \, 0)$:
\begin{eqnarray}
\label{collin}
\hspace{-0.96in}&&  \quad \quad  \quad \quad  \quad \quad \quad \quad
(x, \, y)  \, \quad  \longrightarrow \, \quad  \quad
\Bigl({{x} \over{ 1\, -\, x \, +2 \, y}}, \, \, \, \,  {{y} \over{ 1\, -\, x \, +2\, y}} \Bigr). 
\end{eqnarray}
With this theorem in mind, since we have already considered the involutive transformation (\ref{involut})
corresponding to the Hadamard inverse (\ref{quadratic}), let us just introduce the following
birational transformation associated with the {\em collineation} (\ref{collin}):
\begin{eqnarray}
\label{collinxyz}
\hspace{-0.96in}&&  \, \, \,  \quad 
(x,  \, y, \, z)  \, \, \quad  \longrightarrow \, \quad  \quad
\Bigl({{x} \over{ 1\, -\, x \, +2 \, y}}, \,  \,\,  \, {{y} \over{ 1\, -\, x \, +2\, y}} , \, \,  \,
 z \cdot \, (1 \, -x \, +2\,y)^2 \Bigr).
\end{eqnarray}
Such  a birational transformation (associated with collineations) is an (infinite order)  transformation.
It preserves $\, (x, \, y, \, z) \, = \,  (0, \, 0, \, 0)$ and the product $\, p \, = \, x \, y\, z$.
Let us perform this birational transformation (\ref{collinxyz}) on
the rational function (\ref{simplest}). One gets a new  rational function whose telescoper
is an order-four linear differential operator $\, L_4$ which is the product of two
order-two linear differential operator $\, M_2$ and $\, N_2$:  $\, L_4 \, = \, \, M_2 \cdot \, N_2$. 
The order-two linear differential operator $\, M_2$ is (non-trivially) homomorphic to
the order-two telescoper of the rational function (\ref{simplest}). The second order-two
linear differential operator $\, N_2$
corresponds to algebraic functions.
For such transformations, associated with collineations, we see that
{\em the telescoper is not preserved:
we just have a  (non-trivial) homomorphism property}.

More examples of birational symmetries for telescopers, associated with collineations,
are given\footnote[9]{The example (\ref{collinxyz}) is revisited in detail in \ref{collineevenanothersimplerfirst}.}   
in \ref{colline}.
These examples illustrate the complexity of the homomorphism. 

\vskip .3cm

\subsection{Algebraic geometry comments on these birational symmetries}
\label{comments}

The diagonal of the rational function (\ref{simplest}) is the hypergeometric series:
\begin{eqnarray}
\label{diagsimplest2F1}
  \hspace{-0.96in}&&  \quad  \quad  \quad  \quad 
  _2F_1\Bigl([{{1} \over {3}}, \, {{2} \over {3}}], \, [1], \, 27\, x\Bigr)
 \nonumber  \\
 \hspace{-0.96in}&&  \quad \quad   \quad 
  \, \, = \,\, \, (1\, -24\, x)^{-1/4}  \cdot \,  _2F_1\Bigl([{{1} \over {12}}, \, {{5} \over {12}}], \, [1],
  \, \, {{ 1728 \, x^3 \cdot \, (1 \, - 327\, x ) } \over {(1 \, -24 \, x)^3 }} \Bigr)
   \\
 \hspace{-0.96in}&&  \quad \quad   \quad 
  \, \, = \, \, \,
  1 \, \, +6\,x \, +90\, x^2 \, +1680\, x^3 \, +34650\, x^4
    \, +756756\, x^5 \, +17153136\, x^6 \, \, + \, \, \cdots
  \nonumber
\end{eqnarray}

The algebraic curve, associated with the denominator of the
rational function (\ref{simplest}), is the {\em genus-one} algebraic curve (elliptic curve): 
\begin{eqnarray}
\label{diagsimplest}
\hspace{-0.96in}&&  \quad \quad  \, \, 
1 \, -x \, -y \, -{{p} \over {x \, y}} \, \,\, = \, \,\, \, 0
\quad \quad  \, \hbox{or:} \quad \quad
-x^2\, y \, -x\, y^2 \, +x\, y \, -p \, \, \,= \, \,\, \, 0.  
\end{eqnarray}
The calculation of its j-invariant gives the following Hauptmodul
\begin{eqnarray}
\label{Hauptsimplest}
\hspace{-0.96in}&&  \quad \quad  \quad \quad \quad \quad \quad \quad
       {\cal H} \, \, = \, \, \, {{1728 } \over {j}} \, \, = \, \, \,
       {{ 1728 \, p^3 \cdot \, (1 \, - 27\, p ) } \over {(1 \, -24 \, p)^3 }}, 
\end{eqnarray}
which is exactly the Hauptmodul pullback in (\ref{diagsimplest2F1}). 

Let us consider the rational function (\ref{simplestB}), the algebraic curve corresponding
to eliminate $\, z \, = \, p/x/y \, $ in the denominator of  (\ref{simplestB}) reads:
\begin{eqnarray}
\label{diagsimplestBIS}
\hspace{-0.96in}&&  \quad \quad \quad \quad  \quad \quad
-49 \, x^5\, y^2 \,\,  -42\, x^4\, y^2 \, -7\, x^4\, y\, -23\, x^3\, y^2 \,
+4\, x^3\, y \, \, -6\, x^2\, y^2 \,
\nonumber  \\
\hspace{-0.96in}&&  \quad \quad   \quad \quad  \quad\quad \quad \quad \quad \quad
+2\, x^2\, y \, -x\, y^2 \, +x\, y \, \, -p
\, \,\,\,  = \, \, \,\, \, 0.
\end{eqnarray}
This  algebraic curve is a {\em genus-one} algebraic curve (elliptic curve) and
the calculation of its j-invariant gives the same Hauptmodul pullback in (\ref{diagsimplest2F1})
as the  Hauptmodul (\ref{Hauptsimplest}) for (\ref{diagsimplest}). This is in agreement
with the fact that  the diagonal of
(\ref{simplest}) and (\ref{simplestB}) are equal. At first sight, the fact
that (\ref{diagsimplestBIS}) is an elliptic curve is not totally obvious, however
it is a consequence of the fact that (\ref{diagsimplest}) and (\ref{diagsimplestBIS}) are
{\em birationally equivalent elliptic curves} (since one gets one from the other one
from a birational transformation). {\em Consequently they should have the same j-invariant}.

This kind of remark will be seen as obvious, or slightly tautological, for an algebraic geometer,
however, as far as down-to-earth computer algebra calculations of diagonals of rational functions
or telescopers of rational functions are concerned, it becomes more and more spectacular for more complicated
birational transformations generated by the composition of birational transformations
like (\ref{simplestbiratQ}), (\ref{simplestbiratQ2}) and  (\ref{simplestbiratQ3}).

More generally, the previous birational transformations preserving the product
$\, p \, = \, \, x\, y\, z$, $\, p \, = \, \, x\, y\, z \, u$,  ... occurring in the diagonals,
will preserve the algebraic geometry description of the diagonal of rational functions~\cite{MDPI}.
For instance the genus-two curves associated with split
Jacobians\footnote[1]{Which corresponds to products
of elliptic curves~\cite{MDPI}.} situation we have encountered
in~\cite{MDPI},  will be preserved by such  birational transformations.

\vskip .3cm

\subsection{Diagonal of transcendental functions }
\label{transcendental}

Generalizing the rationals functions  
\begin{eqnarray}
\label{simplestBb}
\hspace{-0.96in}&&  \,  
R_B(x, y, z) \,  =  \,\,
R\Bigl(x,  \, \,  y \cdot \, Q_1(x),  \, \, {{z} \over {Q_1(x) }} \Bigr)
   \, \, =  \, \,\, 
  {{1} \over { 1 \, -x \, -y\cdot \, Q_1(x)\, -z/Q_1(x) }}, 
\end{eqnarray}
deduced from (\ref{simplestB}), using birational
transformations like (\ref{simplestbiratQ}), 
one can consider, beyond, {\em transcendental} functions like
\begin{eqnarray}
\label{simplestBbtrans}
\hspace{-0.96in}&&  \,   
  R_T(x, y, z) \,  =  \, R\Bigl(x,  \, \,  y \cdot \, \cos(x),  \, \, {{z} \over {\cos(x)}} \Bigr)
   \,  =  \,  
  {{1} \over { 1 \,\, -x \, \, -y\cdot \, \cos(x) \,\, -z/\cos(x) }}. 
 \end{eqnarray} 
One can easily verify, from the multi-Taylor expansion of the (simple)
transcendental function (\ref{simplestBbtrans}), that its diagonal
{\em is actually the same as the one}
of (\ref{simplest}), namely (\ref{diagsimplest2F1}). 
This is not a surprise since the
demonstration of the invariance of the diagonal by birational transformation
sketched in section \ref{infinite} (see (\ref{defdiagBB})), just requires that 
$\, Q_1(0) \ne 0$ with $\, Q_1(x)$ behaving at the origin as a polynomial.

\vskip .3cm

\section{Conclusion}
\label{Conclusion}

Diagonals of rational functions have been shown to emerge naturally for $\, n$-fold integrals in
physics, field theory, enumerative combinatorics, seen as ``Periods'' of algebraic varieties
(corresponding to the denominators of these rational functions). On the thousands of examples
we have analyzed, corresponding to  $\, n$-fold integrals of theoretical physics
(in particular the $\, \chi^{(n)}$'s of the susceptibility of the Ising model, ...), or 
corresponding to rather academical diagonal of rational functions, we have seen the
emergence of many striking properties, and we want to understand if these remarkable
properties are inherited from the ``physics'', and, more precisely, the rather ``integrable'' 
framework of these examples (Yang-Baxter integrability, 2D Ising models, Calabi-Yau
and other mirror symmetries, ...) or, on the contrary, are a consequence of the remarkable
nature of diagonals  of rational functions in the most general framework.

\vskip .3cm

This paper is a plea for diagonals of rational, or algebraic, functions and more generally
telescopers of rational or algebraic functions.

\vskip .3cm

$\bullet$ We show that ``periods'' corresponding to non-vanishing cycles, obtained as solutions of telescopers
of rational  functions can sometimes be recovered from  diagonals of rational
functions corresponding to vanishing cycles, 
introducing an extra parameter. These two concepts are not that compartmentalized.

\vskip .3cm

$\bullet$ When considering diagonals of rational functions we have shown that the number of variables
of a rational function must, from time to time, be replaced by a notion of ``effective number'' of variables.

\vskip .3cm

$\bullet$ We have shown that the ``complexity'' of the  diagonals of a rational  function, and for instance the order
of the (minimal order) linear differential operator annihilating this diagonal, is not  related
to the number of variables, or ``effective number'' of variables of the rational  function. In a forthcoming
publication~\cite{Preparation} we will try to understand what is
the minimal number of variables necessary to represent
a given D-finite globally bounded series as a diagonal of a rational function.

\vskip .3cm

$\bullet$ We have shown that the algebraic geometry approach of the diagonals of rational  functions,
or of the telescopers of these  rational  functions,
described in~\cite{MDPI}, can, probably, be generalized to  diagonals of algebraic functions,
or telescoper of algebraic  functions.  These are just preliminary studies and
almost everything remains to be done.

\vskip .3cm

$\bullet$ When studying diagonals of rational functions, our explicit examples enable to understand why one can 
actually restrict  to rational functions of the form $\, 1/Q$ provided the polynomial at the denominator is
irreducible. The situation where the denominator $\, Q$ factorizes clearly needs further
analysis that will be displayed in a forthcoming paper~\cite{Preparation}. The case
of the calculations of telescopers
is slightly different: one can (probably), again, reduce to  rational functions of the form $\, 1/Q$
but with a {\em finite set} of polynomials $\, Q$.

\vskip .3cm

$\bullet$ We have shown that diagonals of rational functions (and this is also the case
with diagonals of algebraic functions) are left invariant when one performs
an {\em infinite set of birational transformations} on the rational functions.  This
remarkable result can, in fact,
be generalized to {\em infinite set of rational transformations}, the diagonals of the transformed 
rational functions becoming the  diagonal of the original rational function
where the variable $\, x$ is changed into $\, x^n$.
These invariance results generalize to telescopers.  More general (infinite) set
of birational transformations are shown to correspond to more convoluted
``covariance'' property of the telescopers (see \ref{colline}).

\vskip .3cm

$\bullet$ We provide some examples of diagonals of transcendental functions which can also yield
simple $\, _2F_1$ hypergeometric functions associated with elliptic curves. The analysis
of diagonal of transcendental functions is clearly an interesting new domain to study.

\vskip .3cm

$\bullet$ Finally, when trying to understand the puzzling fact that telescopers of rational functions
are almost always homomorphic to their adjoint, and thus have selected symplectic or orthogonal
differential Galois groups, 
we understand a bit better the emergence of curious examples of telescopers that are
{\em not} homomorphic to their adjoint, this  (up to homomorphisms) {\em self-duality-breaking}
ruling out a Poincar\'e duality
interpretation of this quite systematic emergence of operators  homomorphic to their adjoint.
A ``desingularization'' of such puzzling cases, corresponding to the introduction of
an extra parameter, shows that such operators now occur in dual (adjoint) pairs, thus restoring the
duality (homomorphism to the adjoint). The limit when the extra parameter goes to zero,
is the {\em direct sum} of different telescopers corresponding to the first rational function terms of
the expansion of the extended rational function in term of this extra parameter. With subsection \ref{GaloisSUB}
we see that the puzzling (non self-adjoint up to homomorphism)
order-three linear differential operator $\, L_3$ with $\, SL(3, \, \mathbb{C})$ differential Galois group,
is better understood as a member of a triplet of three ``quarks''
(\ref{limitchristol}), (\ref{solchristolL3}), and (\ref{solchristolM3}), which restores the
duality. This may suggest that the quite strange $\, _3F_2$ hypergeometric functions (\ref{solchristolL3})
or (\ref{solchristolM3}), could be related to (\ref{limitchristol}) which has a clear elliptic curve
origin. After all, these functions are three periods of the same algebraic variety. The existence
of such a relation between  hypergeometric functions of totally and utterly different nature, is
a challenging open question.

\vskip .3cm

$\bullet$ In \ref{colline} the calculations of telescopers of
rational functions, associated with very simple collineations, 
yield quite massive linear differential operators which factor
into an order-two operator associated with an elliptic curve,
and a ``dressing'' of products of factors which turn out to be direct sums of operators with algebraic function
solutions. This occurrence of this ``mix'' between products and direct sums of a large number
of operators\footnote[2]{Occurring, for instance,
  for the linear differential operators annihilating the $\, \chi^{(n)}\, $
  components of the susceptibility of the Ising model~\cite{experimental,2009-chi5,High}.}
will be revisited in a forthcoming paper~\cite{Preparation}.  

\vskip .3cm

Instead of pursuing one specific mathematical problem this paper can be seen as
a journey into the amazing world of integer sequences, and differential equations.
With all the examples displayed in this paper we provide some answers, sometimes some plausible scenarii,
to many important questions naturally emerging when working on diagonals of rational or algebraic  functions,
or on telescopers of rational, or algebraic, functions, related, or not related,
to problems of physics or enumerative combinatorics.
Like any fruitful concept, everything answered questions
does not ``close'' the subject but, on the contrary, often
raises more new questions than the number of answered questions.

\vskip .3cm

Diagonals of rational, or algebraic, functions, correspond to (globally bounded) series that can be recast
into series with integer coefficients which are solutions of {\em linear} differential operators. When studying
the two dimensional Ising model and its related {\em Painlev\'e equations},
one finds that the $\, \lambda$-extensions
of the correlation functions~\cite{Painleve1,Painleve2} can also produce series with
integer coefficients which are solutions
of {\em non-linear} differential equations\footnote[5]{Differentially algebraic functions~\cite{diffalg}.}
of the Painlev\'e type, these series being also such that
their reduction modulo primes give algebraic functions, just like
diagonals of rational or algebraic functions\footnote[9]{For other examples of differentially algebraic
series with integer coefficients see for instance~\cite{Tutte}.}.

This paper tries to show that the concept of diagonals of  rational, or algebraic, functions is a remarkably
rich and fruitful concept providing tools for physics but also bridging, in a quite fascinating way, different
domains of mathematics. The case of diagonal of transcendental functions, or of these $\, \lambda$-extensions
seems to show that the ``unreasonable richness'' of diagonals and telescopers, may just be the top of an
even more fascinating mathematical ``iceberg'' of mathematical physics.

\vskip .3cm
\vskip .3cm
\vskip .3cm

{\bf First acknowledgment.} One of us (JMM) would like  to thank Prof. Richard Kerner  
for decades of courteous and rich discussions in our laboratory
giving the comforting, and certainly illusory, feeling to belong to a privileged group
of educated  people, blind to the planned disappearance of
mathematical physics in France\footnote[1]{``All those moments will be lost in time,
like ... tears in rain'',  Rutger Hauer in ``Blade Runner''.}.

\vskip .3cm

{\bf More acknowledgments.}
One of us (JMM)  would like to thank P. Lairez for generous de Rham cohomology explanations.
He also thanks C. Koutschan for help in a telescoper calculation. He also thanks Prof. R.J. Baxter
for a kind invitation at the Royal Society in London, where part of this work has been completed.
We also thank A. Bostan, G. Christol, J-A. Weil  and S. Yurkevich
for so many discussions on diagonals of rational functions.

\vskip 1cm

\appendix

\section{Other $\, \alpha$-dependent example}
\label{General}

\subsection{A first very simple example\\}
\label{firstsimplex}

Another example,  similar to the rational function (\ref{christola})
studied in section \ref{GaloisSUB}, is
\begin{eqnarray}
  \label{christolater1}
  &&  \hspace{-0.98in}  \quad \quad  \quad \quad  \quad \quad  \quad\quad  \quad
      \,  {{1} \over { 1 \, \,  \,  \, -  x^2 - y^2  - z^2  \,\, \, \, - \alpha \cdot  \, x \, y^2}}. 
\end{eqnarray}
Its telescoper is an order-four linear differential operator which becomes in the
$\, \alpha \, \rightarrow \, 0$ limit the LCLM of two order two linear differential operators,
one, $\, L_2$,  corresponding to the hypergeometric
solution (which is actually the $\alpha \, = \, 0$ diagonal)
\begin{eqnarray}
  \label{christolabissol1bis}
  &&  \hspace{-0.98in}  \quad \quad  \quad \quad  \quad \quad \quad  \quad \quad
     _2F_1\Bigl( [{{ 1} \over { 3}}, {{2 } \over {3 }} ], \, [1], \, \, 27 \, x^2\Bigr), 
\end{eqnarray}
and an order-two linear differential operator $\, M_2$ having the solution
\begin{eqnarray}
  \label{christolabissol1bisbis}
  &&  \hspace{-0.98in}  \quad \quad  \quad \quad  \quad \quad \quad  \quad \quad
    {{d} \over { dx}} \,  _2F_1\Bigl( [{{ 1} \over { 6}}, {{5 } \over {6 }} ], \, [1], \, \, 27 \, x^2\Bigr), 
\end{eqnarray}
This order-two operator $\, M_2$ is not homomorphic to the  order-two operator $\, L_2$.
Let us consider the $\, \alpha$ expansion of (\ref{christolater1})
\begin{eqnarray}
  \label{expandchristolater}
  &&  \hspace{-0.98in}  \quad 
  \,  {{1} \over { 1 \, \, -  x^2 - y^2  - z^2  \, \, - \alpha \cdot  \, x \, y^2}}
  \, \, \, = \, \, \,  \,\, {{1} \over { 1 \, \, -  x^2 - y^2  - z^2}} \, \, \, \,
  +{{x \, y^2} \over { (1 \, \, -  x^2 - y^2  - z^2)^2}} \cdot \, \alpha
  \nonumber \\
 &&  \hspace{-0.98in}  \quad  \quad  \quad  \quad   \quad  \quad  \quad  \quad  
  +{{x^2 \, y^4} \over { (1 \, \, -  x^2 - y^2  - z^2)^3}} \cdot \, \alpha^2
  \,\, \, \, \, \, + \, \, \cdots 
\end{eqnarray}
The diagonal of the term in $\, \alpha^{1}$ in (\ref{expandchristolater}) is trivial: it is equal to zero. 
In contrast, the telescoper of the term in $\, \alpha^{1}$ in (\ref{expandchristolater}) is
actually nothing but the order-two linear differential operator $\, M_2$.
The telescoper of the term in $\, \alpha^{2}$ in (\ref{expandchristolater}) is
an order-two linear differential operator homomorphic to the previous
order-two linear differential operator $\, L_2$. Similarly to the calculations
displayed in  (\ref{christola}), the telescopers for the terms in $\, \alpha^{2 \, n}$
in the expansion (\ref{expandchristolater})
yield order-two linear differential operators, homomorphic to $\, L_2$,
when the telescopers for the terms
in $\, \alpha^{2 \, n\, +1}$ yield order-two operators, homomorphic to $\, M_2$.

\vskip .3cm

\subsection{Christol: breaking the duality symmetry}
\label{Christol2}

These results can be compared with ones for the diagonal of the rational function
\begin{eqnarray}
  \label{christolater}
  &&  \hspace{-0.98in}  \quad \quad  \quad \quad  \quad \quad \quad \, \, \quad  \quad
      \,  {{1} \over { 1 \, \,  \,-  x^4 - y^4  - z^4  \, \, \,  \, - \alpha \cdot  \, x}}. 
\end{eqnarray}
The linear differential operator annihilating the diagonal of the rational function (\ref{christolater})
is an  order-ten linear differential operator $\, L_{10}(\alpha)$ depending on the parameter $\, \alpha$,
which is homomorphic to its adjoint with an order-eight intertwiner. 
Consequently its differential Galois
group is included in $\, Sp(10, \, \mathbb{C})$. 
This order-ten
linear differential operator $\, L_{10}(\alpha)$ is irreducible except at $\, \alpha \, = \, 0$.

At $\, \alpha \, = \, 0$ it is the direct sum
$\, LCLM(L_2, \, M_2, \, L_3, \, M_3)$, of two order-three linear differential operators and
two order-two linear differential operators, namely
$\, L_2$ corresponding to the solution
\begin{eqnarray}
  \label{christolabissol1}
  &&  \hspace{-0.98in}  \quad \quad \quad \quad  \, \,
  _2F_1\Bigl( [{{ 1} \over { 3}}, {{2 } \over {3 }} ], \, [1], \, \, 27 \, x^4\Bigr)
  \\
  &&  \hspace{-0.98in}  \quad \quad  \quad \quad  \quad \quad 
  = \, \, \, \, \, 1 \,\, \, +6\, x^4 \, \, +90\, x^8 \, +1680\, x^{12} \, +34650\, x^{16}
  \, +756756\, x^{20} \,\,\,  \, + \, \,  \cdots
  \nonumber 
\end{eqnarray}
as it should (this is the diagonal of (\ref{christolater}) at $\, \alpha \, = \, 0$),
and the other one, $\, M_2$, corresponding to the globally bounded  series
solution expressed in terms of HeunG functions\footnote[1]{Use Table page 24 of~\cite{MaierHeunG}.}:    
\begin{eqnarray}
  \label{christolabissol21}
  &&  \hspace{-0.98in}  \quad \quad  \quad \quad  \quad \quad  \quad
     {{ (1 \, -24\, x^4)^2} \over { (1 \, -27\, x^4)^2 }} \cdot \,
     HeunG\Bigl( {{9} \over {8}}, \,  {{97} \over {32}}, \,  {{7} \over {6}}, \,  {{5} \over {6}}, \, 1, \, -1;
       \, \, 27 \cdot \, x^4 \Bigr).  
\end{eqnarray}
This linear differential operator $\, M_2$ is homomorphic to the order-two linear differential
operator corresponding to the modular form
(see Appendix B in~\cite{Heun}):
\begin{eqnarray}
  \label{christolabissol2new}
  &&  \hspace{-0.98in}  \quad \quad   \quad \quad   \quad \quad  \quad \quad  \quad
     _2F_1\Bigl( [{{ 1} \over { 6}}, \, {{5 } \over {6 }} ], \, [1], \, \, 27 \, x^4\Bigr). 
\end{eqnarray}
Using the identity
\begin{eqnarray}
  \label{HeunG}
  &&  \hspace{-0.98in}  \quad \quad    \quad \quad  
  HeunG\Bigl( {{9} \over {8}}, \,  {{97} \over {32}}, \,  {{7} \over {6}}, \,  {{5} \over {6}}, \, 1, \, -1;
  \, \, 27 \cdot \, x \Bigr)
     \, \, = \, \, \,
     \nonumber \\
 &&  \hspace{-0.98in}  \quad \quad    \quad \quad   \quad   \quad   \quad \,  
     4\cdot \, (1 \, -27\, x)\cdot \, {{(27\,x \, +2)} \over {(1\, -24\,x)^2 }}
     \cdot \,  x \cdot \,
           {{d } \over {dx}} \, _2F_1\Bigl( [{{ 1} \over { 6}}, \, {{5 } \over {6 }} ], \, [1],
     \, \, 27 \, x\Bigr)
    \nonumber \\
 &&  \hspace{-0.98in}  \quad \quad    \quad \quad  \, \, \,  \quad \quad   \quad   \quad  
    \, \, +{{ 1 \, 9\,x \, -486\,x^2 } \over { (1 \, -24\,x )^2}}
    \cdot \,  _2F_1\Bigl( [{{ 1} \over { 6}}, \, {{5 } \over {6 }} ], \, [1], \, \, 27 \, x\Bigr), 
\end{eqnarray}
we can rewrite (\ref{christolabissol21}) in terms of the modular form (\ref{christolabissol2new}). 
One can thus write the solution of $\, M_2$ as:
\begin{eqnarray}
 \label{HeunG}
 &&  \hspace{-0.98in}  \, 
  {{ 2 \, +27\, x^4 } \over { 1 \, -27\, x^4 }} \cdot \,
  x \cdot \, {{d } \over {dx}}
  \, _2F_1\Bigl( [{{ 1} \over { 6}}, {{5 } \over {6 }} ], \, [1], \, \, 27 \, x^4\Bigr)
\, \,  \, 
+   {{  1 \, +18\, x^4} \over { 1 \, -27\, x^4 }} \cdot \, x^4 \cdot \,
_2F_1\Bigl( [{{ 1} \over { 6}}, {{5 } \over {6 }}  ], \, [1], \, \, 27 \, x^4\Bigr)
\nonumber \\
&&  \hspace{-0.98in}  \quad  \quad \quad \quad 
\, \, = \, \, \,\, \, \,   1 \,  \, \, \,  +{{315} \over {4}} \, x^4 \,  \, +{{225225} \over {64}} \, x^8
\, \, +{{33948915} \over {256}} \, x^{12} \,\,\,  +{{75293843625} \over {16384}} \, x^{16}
\nonumber \\
&&  \hspace{-0.98in}  \quad  \quad \quad \quad \quad \quad \quad \quad \, 
\,  +{{ 9927744261435} \over {65536}}\, x^{20}  \, \, \, \,  \, + \, \, \, \cdots 
\end{eqnarray}
The order-three linear differential operator $\, L_3$ has the
hypergeometric solution
\begin{eqnarray}
  \label{christolabissol13}
  &&  \hspace{-0.98in}  \quad \quad  \quad \quad  \quad \quad  \quad \quad  \quad
  _3F_2\Bigl( [{{ 7} \over { 12}}, {{11 } \over {12 }}, {{15 } \over {12 }} ],
  \, [{{ 3} \over { 4}}, \, 1], \, \, 27 \, x^4\Bigr),
\end{eqnarray}
while the order-three linear differential operator $\, M_3$  has the hypergeometric solution:
\begin{eqnarray}
  \label{christolabissol12}
  &&  \hspace{-0.98in}  \quad \quad  \quad \quad  \quad \quad  \quad \quad  \quad
  _3F_2\Bigl( [{{ 13} \over { 12}}, {{17 } \over {12 }}, {{21 } \over {12 }} ],
  \, [{{ 1} \over { 4}}, \, 1], \, \, 27 \, x^4\Bigr). 
\end{eqnarray}
These two linear differential  operators are such that  $\, L_3$
is actually homomorphic to the adjoint of $\, M_3$, and, of course,  $\, M_3$
is homomorphic to the adjoint of $\, L_3$,
but  $\, L_3$ is {\em not} homomorphic to the adjoint of $\, L_3$ (and
 $\, M_3$ is not homomorphic to the adjoint of $\, M_3$).  We have again,
a pair of dual linear differential operators.

Since these calculations are in the $\, \alpha \, \rightarrow \, 0$ limit, let us
expand in $\, \alpha$  the rational function (\ref{christolater}):
\begin{eqnarray}
  \label{christolaterexpand}
  &&  \hspace{-0.98in}  \quad \quad 
   {{1} \over { 1 \, \, -  x^4 - y^4  - z^4  \, \, - \alpha \cdot  \, x}}
  \,  \, \,\, = \,  \,\,  \, \, {{1} \over { 1 \, \, -  x^4 - y^4  - z^4}}
  \, \, \,\,  +{{x} \over { (1 \, \, -  x^4 - y^4  - z^4)^2 }} \cdot \, \alpha
  \nonumber \\
  &&  \hspace{-0.98in}  \quad \quad \quad \quad \quad \quad 
  \, \, +{{x^2} \over { (1 \, \, -  x^4 - y^4  - z^4)^3 }} \cdot \, \alpha^2 \, \, \, \,
  +{{x^3} \over { (1 \, \, -  x^4 - y^4  - z^4)^4 }} \cdot \, \alpha^3
  \nonumber \\
  &&  \hspace{-0.98in}  \quad \quad \quad \quad \quad \quad \quad \quad  \quad\, \,
  +{{x^4} \over { (1 \, \, -  x^4 - y^4  - z^4)^5 }} \cdot \, \alpha^4 \,  \, \,
   \, \, \,+ \, \, \cdots 
\end{eqnarray}
Since the telescoper of a sum of rational functions is the direct sum (LCLM) of the telescopers
of these rational functions, let us consider the telescopers of the first five terms in the RHS of
(\ref{christolaterexpand}).  The telescoper of the first term is of course the order-two linear differential operator
$\, L_2$ annihilating the diagonal of this rational function. The telescoper of the second term (in $\, \alpha^{1}$),
is  the order-three linear differential operator $\, L_3$.The telescoper of  the third term  (in $\, \alpha^{2}$),
is the order-two linear differential operator $\, M_2$. The telescoper of the fourth term (in $\, \alpha^{3}$),
is exactly $\, M_3$. 
The telescoper of the sum of the first orders in $\alpha$ in the expansion (\ref{christolaterexpand})
\begin{eqnarray}
  \label{truncchristolaterexpand}
  &&  \hspace{-0.98in}  \quad \quad  \quad  \quad 
  {{1} \over { 1 \, \, -  x^4 - y^4  - z^4}}
  \, \, \, \,  \,\,  +{{x} \over { (1 \, \, -  x^4 - y^4  - z^4)^2 }} \cdot \, \alpha
  \nonumber \\
  &&  \hspace{-0.98in}  \quad \quad \quad\quad  \quad  \quad 
  \, \, +{{x^2} \over { (1 \, \, -  x^4 - y^4  - z^4)^3 }} \cdot \, \alpha^2 \, \, \, \,\, 
  +{{x^3} \over { (1 \, \, -  x^4 - y^4  - z^4)^4 }} \cdot \, \alpha^3, 
 \end{eqnarray}
is actually the LCLM of the four telescopers $\, L_2$, $\, M_2$, $\, L_3$ and $\, M_3$
{\em which is precisely the $\, \alpha \, \rightarrow \, 0 \, $
  limit of the order-ten linear differential operator !}

Let us now consider the telescopers of the next  $\alpha$ orders
in  the expansion (\ref{christolaterexpand}).
The telescoper of the last rational function in (\ref{christolaterexpand}), namely
$\, x^4/(1 \, \, -  x^4 - y^4  - z^4)^5$, is an order-two linear differential operator $\, N_2$.
One can thus write the solution of $\, N_2$ as:
\begin{eqnarray}
 \label{alpha4}
 &&  \hspace{-0.98in}  \quad \quad \quad \quad  {\cal D}_1 \, \, = \, \, \, 
{{3} \over { 48}} \cdot \,  {{ 1 \, +540 \, x^4 \, +4374 \, x^8 } \over { (1 \, -27\, x^4)^3 }} \cdot \,
x \cdot \, {{d } \over {dx}} \,   _2F_1\Bigl( [{{ 1} \over { 3}}, {{2 } \over {3 }} ], \, [1], \, \, 27 \, x^4\Bigr)
\nonumber \\
&&  \hspace{-0.98in}  \quad \quad \quad \quad \quad \quad \quad 
\, \, 
+ {{3} \over { 2}} \cdot \, {{ (19 \, +216\, x^4)} \over { (1 \, -27\, x^4)^3 }}
\cdot \, x^4 \cdot \,  _2F_1\Bigl( [{{ 1} \over { 3}}, {{2 } \over {3 }} ], \, [1], \, \, 27 \, x^4\Bigr)
 \\
&&  \hspace{-0.98in} \quad  \quad \quad  \quad
\, \, = \, \, \,\,\, 30\, x^4 \,\,\, +3780\, x^8 \, +277200\, x^{12} \,
+15765750\, x^{16} \, +771891120\, x^{20} \,  \, \, + \, \, \, \cdots
\nonumber 
\end{eqnarray}
 The telescoper of
\begin{eqnarray}
  \label{alpha8} 
  &&  \hspace{-0.98in}  \quad \quad  \quad\quad  \quad \quad  \quad \quad  \quad \quad  \quad
      \,  {{x^8} \over { (1 \, \, -  x^4 - y^4  - z^4)^9}}, 
\end{eqnarray}
is an order-two linear differential operator whose analytic solution reads:
\begin{eqnarray}
\label{alpha8}
&&  \hspace{-0.98in}  \quad  \quad \quad  \quad {\cal D}_2 \, \, = \, \, \, \, 
-{{3} \over { 672}} \cdot \,
{{ p_1} \over { (1 \, -27\, x^4)^7 }} \cdot \,
x \cdot \, {{d } \over {dx}} \,   _2F_1\Bigl( [{{ 1} \over { 3}}, {{2 } \over {3 }} ], \, [1], \, \, 27 \, x^4\Bigr)
\nonumber \\
&&  \hspace{-0.98in}  \quad   \quad   \quad \quad  \quad \quad   \quad \quad  
\, \, 
+ {{3} \over { 28}} \cdot \, {{ p_2 } \over { (1 \, -27\, x^4)^7 }}
\cdot \, x^4 \cdot \,  _2F_1\Bigl( [{{ 1} \over { 3}}, {{2 } \over {3 }} ], \, [1], \, \, 27 \, x^4\Bigr)
 \\
&&  \hspace{-0.98in}  \quad \quad \quad  \quad  \quad
= \, \, \, \,\, 2970\, x^8 \,\, +900900\, x^{12} \, +137837700\, x^{16} \, +14665931280\, x^{20} \, 
\nonumber \\
&&  \hspace{-0.98in}  \quad \quad \quad  \quad  \quad  \quad  \quad \quad  \quad
\, +1236826871280\, x^{24} \, +88597190167200\, x^{28} \, \,  \, \, \, + \, \, \, \cdots
\nonumber
\end{eqnarray}
where:
\begin{eqnarray}
\label{p1p2}
&&  \hspace{-0.98in}  \quad
p_1 \, \,  \, = \, \,  1 \,\, -714\, x^4 \, -924372\, x^8 \,
    -54587520\, x^{12} \, -530141922\, x^{16} \, -554824404\, x^{20},
\nonumber \\
&&  \hspace{-0.98in}  \quad 
p_2 \, \,  \, = \, \,   1 \,\, +27030\, x^4 \, +2062098\, x^8\, +23960772\,x^{12} \, +29170206\, x^{16}.
\end{eqnarray}

\vskip .3cm

If we consider, instead of the telescoper, the diagonal of the rational function (\ref{christolaterexpand}),
only the terms in $\, \alpha^{4\,n}$ $\, \, n \, = \, 0, \, 1, \, 2, \, \, \cdots\, $
will contribute, the other ones, corresponding to
{\em non-vanishing cycles}~\cite{Deligne}, give zero contributions. Consequently we get
for the diagonal
of the rational function (\ref{christolaterexpand}):
\begin{eqnarray}
\label{series}
&&  \hspace{-0.98in}  \quad \quad \quad 
      {\rm Diag}\Bigl(  {{1} \over { 1 \, \, -  x^4 - y^4  - z^4  \, \, - \alpha \cdot  \, x}}  \Bigr)
\nonumber \\
&&  \hspace{-0.98in}  \quad \quad \quad \quad 
      \, \,  = \, \, \,\, \,
      _2F_1\Bigl( [{{ 1} \over { 3}}, {{2 } \over {3 }} ], \, [1], \, \, 27 \, x^4\Bigr)
      \, \, \,\, + {\cal D}_1 \cdot \, \alpha^4
\, \,\, \, + {\cal D}_2 \cdot \, \alpha^8 \,\, \, \, + \cdots
\end{eqnarray}
\begin{eqnarray}
\label{series}
&&  \hspace{-0.98in}  
\, = \, \, \,
1 \, \,\, +(30 \, \alpha^4  \, +6)  \cdot \, x^{4} \,
\, \, +(2970 \, \alpha^8 \, +3780 \, \alpha^4 \, +90) \cdot \, x^{8}
\nonumber \\
&&  \hspace{-0.98in}  \quad 
\, \, +(371280 \, \alpha^{12}   +900900 \, \alpha^8 +277200 \, \alpha^4  +1680) \cdot \, x^{12}
\nonumber \\
&&  \hspace{-0.98in}  \quad \, \,  
+(51482970 \, \alpha^{16} +185175900 \, \alpha^{12} +137837700 \, \alpha^8
+15765750 \, \alpha^4 +34650)\cdot \, x^{16}
\nonumber \\
&&  \hspace{-0.98in}  \quad 
\, \, 
+(7571343780 \, \alpha^{20} \, +36141044940\, \alpha^{16} \, +44975522592\, \alpha^{12}
\nonumber \\
&&  \hspace{-0.98in}  \quad   \quad \quad \quad \, \, 
\, +14665931280\, \alpha^8 \, +771891120 \, \alpha^4 \, +756756) \cdot \, x^{20} \, \, \, + \cdots
\nonumber \\
&&  \hspace{-0.98in}     \, \,
 = \, \, \, \, 1 \, \, +6\, x^4 \, +90\, x^8 \, +1680\, x^{12} \, +34650\, x^{16}
   \, +756756\, x^{20} \,  \, \, + \, \, \, \cdots 
 \nonumber \\
 &&  \hspace{-0.98in}   \quad  \, \,
 +\Bigl( 30\, x^4 \, +3780\, x^8 \, +277200\, x^{12} \,
+15765750\, x^{16} \, +771891120\, x^{20} \, + \, \, \, \cdots  \Bigr) \cdot \, \alpha^4
  \nonumber \\
 &&  \hspace{-0.98in}  \quad  \, \,
  +\Bigl(2970\, x^8  +900900\, x^{12}  +137837700\, x^{16}  +14665931280\, x^{20}
  \, + \, \, \, \cdots   \Bigr)
  \cdot \, \alpha^8
 \, \, \, + \, \, \, \cdots
 \nonumber 
\end{eqnarray}

\vskip .3cm

\section{Birational symmetries from collineations\\}
\label{colline}

\subsection{Birational symmetries from collineations: a first example\\}
\label{collinefirst}

Let us consider  a collineation transformation
not preserving $\, (x, \, y) \, = \, \, (0, \, 0)$:
\begin{eqnarray}
\label{collinBIS}
\hspace{-0.96in}&&  \quad \quad  \quad \quad  \quad \quad
(x, \, y)  \, \quad  \longrightarrow \, \quad  \quad
\Bigl({{2 \, +x  \, +3\, y} \over{ 1\, -\, x \, +2 \, y}}, \, \, \, \,
     {{ 1 \, +5\, x \, +7\, y} \over{ 1\, -\, x \, +2\, y}} \Bigr), 
\end{eqnarray}
and let us now introduce the following
{\em birational transformation} associated with the collineation (\ref{collinBIS}):
\begin{eqnarray}
\label{collinxyzBIS}
\hspace{-0.96in}&&  \quad  \quad 
(x,  \, y, \, z)  \, \,  \quad  \longrightarrow \, \quad  \quad
\nonumber \\
\hspace{-0.96in}&&  \quad  \quad   \, \quad  \quad 
\Bigl({{2 \, +x  \, +3\, y} \over{ 1\, -\, x \, +2 \, y}}, \, \, \, \,
     {{ 1 \, +5\, x \, +7\, y} \over{ 1\, -\, x \, +2\, y}}, \, \, \, 
     {{  x \, y \, z \cdot \, (1 \, -x \, +2\,y)^2 } \over {
         (2 \, + x \, + 3 \,y) \cdot \, (1 \, + 5\, x \, + 7\, y) }}\Bigr), 
\end{eqnarray}
which preserves the product $\, p \, = \, \, x \, y \, z$.

Let us transform the simple rational function (\ref{simplest}) with
the birational transformation (\ref{collinxyzBIS}).
It becomes the rational function:
\begin{eqnarray}
\label{NewRatcolli}
\hspace{-0.96in}&&  \quad  \quad \quad \quad \quad \quad  
       {\cal R} \, \, = \, \, \,
       {{ (1\, -\, x \, +2 \, y) \cdot \, (2 \, + x \, + 3 \,y) \cdot \, (1 \, + 5\, x \, + 7\, y)
         } \over {   {\cal D} }}, 
\end{eqnarray}
where the denominator $\, {\cal D}$ reads:
\begin{eqnarray}
\label{NewRatcolliDenom}
\hspace{-0.96in}&&  \quad  
{\cal D} \, \, = \, \, \, 
x^4\, y\, z \, -6\, x^3\, y^2\, z \, +12\, x^2\, y^3\, z \, -8\, x\, y^4\, z
\, -3\, x^3\, y\, z \, +12\, x^2\, y^2\, z \,
-12\, x\, y^3\, z
\nonumber \\
\hspace{-0.96in}&&  \quad  \quad   \, 
\, +3\, x^2\, y\, z \, -6\, x\, y^2\, z \,
-35\, x^3 \, -194\, x^2\, y \, -323\, x\, y^2 \, -x\, y\, z \, -168\, y^3 \, -87\, x^2
\nonumber \\
\hspace{-0.96in}&&  \quad  \quad  \quad  \quad  \, 
\, -251\, x\, y \, -178\, y^2 \, -36\, x \, -50\, y \, -4.
\end{eqnarray}
The intersection of the algebraic surface $\,\, {\cal D} \, = \, \, 0\, $ with the
algebraic surface $\, p \, = \, \, x\, y \, z$, is an elliptic curve. One gets, almost
instantaneously\footnote[1]{Using the $\, j$\_$invariant$
  command in Maple with(algcurves). }, the Hauptmodul of this  elliptic curve:
\begin{eqnarray}
\label{HauptNewRatcolli}
\hspace{-0.96in}&&  \quad  \quad \quad \quad  \quad \quad \quad  \quad \quad
{\cal H } \, \, = \, \, \, {{1728\, p^3\cdot \, (1 \, -27\, p) } \over {(1 \, -24 \, p)^3 }}. 
\end{eqnarray}
This  Hauptmodul must be the same\footnote[2]{If one expects an {\em algebraic geometry interpretation} of the
  calculation of the diagonal of rational functions or telescopers~\cite{MDPI}.} as the Hauptmodul (\ref{Hauptsimplest}) of
the elliptic curve (\ref{diagsimplest}), since the two algebraic curves
are birationaly equivalent, being related by a birational transformation namely (\ref{collinBIS}).
The calculation of the telescoper of (\ref{NewRatcolli}) is really massive: it gives, after one month
of computation, an order-eleven linear differential
operator\footnote[5]{We thank C. Koutschan for performing
these slightly ``extreme'' computations.}. The result being too massive, let us consider
other examples of birational transformations associated with collineations
simpler than (\ref{collinxyzBIS}). 

\vskip .3cm

{ \bf Remark B 1.1:} The  diagonal of the rational function (\ref{NewRatcolli})
is a very simple series:
\begin{eqnarray}
\label{diagtoosimple} 
\hspace{-0.96in}&&  \quad \quad \quad 
Diag\Bigl({\cal R}\Bigr)  
\,   = \, \,   - {{1} \over {2}} \cdot \, {{ 1} \over {1 \, +x/4 }}
 \nonumber \\
 \hspace{-0.96in}&& \quad \quad \quad \quad 
 \,   = \, \, 
- {{1} \over {2}} \, + {{1} \over {8}} \cdot \, x \,  - {{1} \over {32}} \cdot \, x^2 \,
+ {{1} \over {128}} \cdot \, x^3 \, - {{1} \over {512}} \cdot \,x^4 \, \, + \, \cdots
\end{eqnarray}

\vskip .3cm

{ \bf Remark B 1.2:} If one considers, instead of (\ref{NewRatcolli})  the rational function with the
same denominator(\ref{NewRatcolliDenom}) but where the numerator
is normalised to $\, 1$, 
\begin{eqnarray}
\label{NewRatcollinorm}
\hspace{-0.96in}&&  \quad  \quad \quad \quad \quad \quad  \quad \quad  \quad \quad \quad \quad
       {\cal R} \, \, = \, \, \,
       {{ 1
         } \over {   {\cal D} }}. 
\end{eqnarray}
The diagonal of (\ref{NewRatcollinorm}) is the same as (\ref{diagtoosimple}) up to a factor two:
\begin{eqnarray}
\label{diagtoosimple2} 
\hspace{-0.96in}&&  \quad \quad \quad  \quad \quad \quad  \quad \quad
Diag\Bigl({\cal R}\Bigr)  
\,   = \, \,   - {{1} \over {4}} \cdot \, {{ 1} \over {1 \, +x/4 }}.
\end{eqnarray}
The telescoper of (\ref{NewRatcollinorm}) is an order-seven linear differential
operator which factorises as follows:
\begin{eqnarray}
\label{finalremationLL7}
\hspace{-0.96in}&&  \quad  \quad \, \, \,   
\, L_{7} \, = \, \, F_2 \cdot \, G_2 \cdot \, H_2 \cdot \, H_1
\quad \quad  \, \,\, \, \, \hbox{with:} \quad \quad \quad \, \, \, 
H_1 \, = \, \, D_x \, + {{1 } \over { 4\, +x}}, 
\end{eqnarray}
where the order-two linear differential operator $\, F_2$
is quite large and is (non-trivially) homomorphic to the order-two  linear differential
operator $\, L_2$ which is the telescoper
of the rational function (\ref{simplest}), and where the order-two
linear differential   operators $\, G_2$ and $\, H_2$ 
have algebraic solutions. The diagonal (\ref{diagtoosimple2})
is solution of the order-one operator $\, H_1$.
 The homomorphism between $\, F_2$ and $\, L_2$ gives 
\begin{eqnarray}
\label{HauptNewRatcolliintertwinAA}
\hspace{-0.96in}&&  \quad  \quad \quad 
F_2 \cdot \, X_1 \, \, = \, \, \, Y_1 \cdot \, L_2
\quad  \quad \, \, \,\hbox{where:}  \quad \quad \, \, \,
X_1 \, \, = \, \, \, \,   A(x) \cdot \, D_x \,  \, + B(x), 
\end{eqnarray}
where $\, A(x)$ and $\, B(x)$ are rational functions.
Consequently a solution $\, {\cal S}$ of the telescoper $\, L_7$ (but not of the product
$\, G_2 \, H_2\, H_1$ in (\ref{finalremationLL7})) will be related to the
hypergeometric solution $\, _2F_1([1/3,2/3],[1],\, 27 \, x)$
of  the order-two linear differential operator $\, L_2$, 
as follows:
\begin{eqnarray}
\label{finalremationA} 
\hspace{-0.96in}&&  \quad  \quad  \quad  \quad  \quad 
X_1\Bigl( \, _2F_1\Bigl([{{1} \over {3}}, \, {{2} \over {3}}],[1],\, 27 \, x\Bigr) \Bigr)
\, \, \,  = \, \, \,  \,  G_2 \cdot \, H_2 \cdot \, H_1 \,  \cdot \, {\cal S}.
\end{eqnarray}

\vskip .3cm

{\bf Remark B 1.3:}  Note that the diagonal of the rational function (\ref{NewRatcolli})
is a very simple series (\ref{diagtoosimple}). 
Therefore the solution $\, {\cal S}$ of the telescoper, associated with
 an elliptic curve of Hauptmodul (\ref{HauptNewRatcolli})
(see equation (\ref{finalremationA})) corresponds to a ``period'', an integral over a non-vanishing cycle,
 and is different from the integral over a vanishing cycle, namely the
diagonal (\ref{diagtoosimple}).

\vskip .3cm 

{\bf Remark B 1.4:} The factorisation  (\ref{finalremationLL7}) is far from being unique. The product of the
last three factors can be seen to be a direct sum:
\begin{eqnarray}
\label{finalremationLL7directsum}
\hspace{-0.96in}&&  \quad  \quad \quad  \quad  \quad  \quad  \quad \quad \, \, \,  \,   
 G_2 \cdot \, H_2 \cdot \, H_1 \,\, = \, \, \,   \,  \tilde{G}_2 \oplus \, \tilde{H}_2 \oplus \, H_1,
\end{eqnarray}
where the two new order-two operators $\, \tilde{G}_2$ and $\, \tilde{H}_2$ are simpler,
with, again, algebraic function solutions.

\vskip .3cm 

\subsection{Birational symmetries from collineations. A simpler example \\}
\label{collinesimpler}

Let us consider the following
birational transformation associated with a collineation:
\begin{eqnarray}
\label{collinxyzBIS2}
\hspace{-0.96in}&&  \quad  \quad   \, \, 
(x,  \, y, \, z)  \, \,  \quad  \longrightarrow \, \quad  \quad
\nonumber \\
\hspace{-0.96in}&&  \quad  \quad  \quad   \, \quad  \quad 
\Bigl({{x  \, +3\, y} \over{ 1\, -\, x \, +2 \, y}}, \, \, \, \,
     {{ 1 \, +5\, x \, +y} \over{ 1\, -\, x \, +2\, y}}, \, \, \, 
     {{  x \, y \, z \cdot \, (1 \, -x \, +2\,y)^2 } \over {
         (x \, + 3 \,y) \cdot \, (1 \, + 5\, x \, + 7\, y) }}\Bigr), 
\end{eqnarray}
which preserves the product $\, p \, = \, \, x \, y \, z$.
Again, if one transforms the simple rational function (\ref{simplest}) with
the birational transformation (\ref{collinxyzBIS2}), one gets
the rational function of the form
\begin{eqnarray}
\label{NewRatcolli2}
\hspace{-0.96in}&&  \quad  \quad \quad \quad \quad \quad  \, \, \,
       {\cal R} \, \, = \, \, \,
       {{ (1\, -\, x \, +2 \, y) \cdot \, (x \, + 3 \,y) \cdot \, (1 \, + 5\, x \, +  y)
         } \over {   {\cal D} }}, 
\end{eqnarray}
and, again, the  intersection of the algebraic surface $\,\, {\cal D} \, = \, \, 0\, $ with the
algebraic surface $\, p \, = \, \, x\, y \, z$, is an elliptic curve, corresponding to eliminate
$\, z \, = \, p/x/y \, $ in $\,\, {\cal D} \, = \, \, 0$. One gets immediatly the
same Hauptmodul (\ref{HauptNewRatcolli})  for this new elliptic curve.

The telescoper of the rational function (\ref{NewRatcolli2}) is an order-ten linear
differential operator\footnote[1]{We thank C. Koutschan for providing this order-ten linear
differential operator.}.
This telescoper is obtained using about nine days of computation time. 
It uses 286 evaluation points (in contrast with the 462 evaluation points required
for (\ref{NewRatcolliDenom})), and one uses in total 38 primes (of size $\, 9 \cdot \, 22\cdot \, 10^{18}$)
to reconstruct the solution with Chinese remaindering.
The telescoper of the rational function (\ref{NewRatcolli2}) factors as follows:
\begin{eqnarray}
\label{finalremationL10bc}
\hspace{-0.96in}&&  \quad  \quad \quad \quad \quad \quad \quad  \quad \, 
\, L_{10} \, \, = \, \, \,
F_2 \cdot \, G_2 \cdot \, H_1 ,\cdot \, I_1 \cdot \, J_2 \cdot \, K_2,  
\end{eqnarray}
The order-two linear differential operator $\, F_2$ in (\ref{finalremationL10bc}) is homomorphic to the
order-two  linear differential operator $\, L_2$ which is the telescoper
of the rational function (\ref{simplest}), and  the order-two
linear differential   operators $\, G_2$, $\, J_2$ and $\, K_2$ 
have algebraic solutions.

{\bf Remark B 2.1:} The factorisation of (\ref{finalremationL10bc}) is far from being unique. As usual we have
a mix between product and direct-sum of factors. The order-ten operator being quite large it is difficult
to get the direct-sum factorisation of $\, L_{10}$ in (\ref{finalremationL10bc}). One finds, however, quite easily
that  $\, L_{10}$ has two simple rational function solutions
\begin{eqnarray}
\label{factorsL1M1bc}
\hspace{-0.96in}&&  \quad \quad \quad \quad \quad \quad \quad
{{1} \over {(x-35) \cdot \, (4\, x\, +3)}}, \quad \quad  \quad {{x} \over { (x-35) \cdot \, (4\, x\, +3)}},
\end{eqnarray}
corresponding to  two order-one operators
$\, L_1 \, = \, D_x \, + (8\, x-137)/(4\,x+3)/(x-35)\, $ and $\, M_1\, = \, D_x \, +(4\,x+3)/(x+21)/(x-35) \, -1/x \,$
and, thus, can be rightdivided by the LCLM of $\, L_1$ and $\, M_1$. 
In fact the product\footnote[2]{Note that the product $\,G_2 \cdot \, H_1$, or the  product $\,G_2 \cdot \, H_1 \,\cdot \, I_1$,
  or the product $\, H_1 \,\cdot \, I_1 \cdot \, J_2 \cdot \, K_2$,
  are also  direct sums. In contrast the product $\, F_2 \cdot \, G_2$ is {\em not} a direct sum.}
of the last factors at the right of the factorization of $\, L_{10}$
can be seen to be a direct sum:
\begin{eqnarray}
\label{finalremationL10bcdirect}
\hspace{-0.96in}&&  \quad  \quad \quad \quad \quad 
G_2 \cdot \, H_1 \,\cdot \, I_1 \cdot \, J_2 \cdot \, K_2 \, \, = \, \, \,
L_1 \, \oplus \, M_1 \, \oplus  \, \tilde{G}_2 \, \oplus  \, \tilde{J}_2  \, \oplus  \, K_2, 
\end{eqnarray}
where $\, \tilde{G}_2$ and  $ \, \tilde{J}_2$  are (much) simpler order-two operators than
$\, G_2$ and  $\, J_2$, again with algebraic function solutions.

The result remaining still too large, let us consider
another example of birational transformation associated with collineations,
simpler than (\ref{collinxyzBIS}) or (\ref{collinxyzBIS2}). 

\vskip .3cm 

{\bf Remark B 2.2:} If one considers, instead of (\ref{NewRatcolli2})  the rational function with the
same denominator $\,{\cal D}$  but where the numerator
is normalised to $\, 1$, 
\begin{eqnarray}
\label{NewRatcolli2norm}
\hspace{-0.96in}&&  \quad  \quad \quad \quad \quad \quad  \quad \quad \quad \quad \quad
       {\cal R} \, \, = \, \, \,
       {{ 1
         } \over {   {\cal D} }}. 
\end{eqnarray}
The telescoper of the rational function (\ref{NewRatcolli2norm}) is an order-seven linear differential operator
\begin{eqnarray}
\label{finalremationL7c}
\hspace{-0.96in}&&  \quad  \quad \quad \quad \quad \quad  \quad \quad \quad \quad
\, L_{7} \, = \, \, F_2 \cdot \, G_1 \cdot \, G_2  \cdot \, H_2, 
\end{eqnarray}
where the order-two linear differential operator $\, F_2$
is (non-trivially) homomorphic to the order-two  linear differential
operator $\, L_2$ which is the telescoper
of the rational function (\ref{simplest}), and
where the order-two linear differential operators $\, G_2$ and $\, H_2$
have simple algebraic solutions.
This factorisation (\ref{finalremationL7c}) is not unique.  Introducing the order-one
operator $ \, \tilde{G}_1 \, = \, D_x \, +1/x$, one can see that $ \, \tilde{G}_1$ rightdivides
$\, L_{7}$ and that the product of the three factors, at the right of the decomposition (\ref{finalremationL7c}),
can be written as a direct sum
\begin{eqnarray}
\label{finalremationL7cdirect}
\hspace{-0.96in}&&  \quad  \quad \quad  \quad \quad \quad \quad  \quad
G_1 \cdot \, G_2  \cdot \, H_2 \, \, = \, \,  \, \tilde{G}_1 \,  \oplus \, \tilde{G}_2  \oplus \, \, H_2, 
\end{eqnarray}
where the solutions of $\, \tilde{G}_2$ are algebraic.

\vskip .3cm 

{\bf Remark B 2.3:} In \ref{colline} we encounter many order-two linear differential operators
with algebraic solutions\footnote[9]{We thank A.Bostan and S. Yurkevich for revisiting most of our
  order-two linear differential operators with algebraic solutions. We thank J-A. Weil
  for showing us that all these  order-two linear differential operators
  have a twelve elements {\em dihedral}
  differential Galois group: they all have algebraic solutions of degree 12.
}. Even for large order-two linear differential operators
one can see quite easily\footnote[1]{ Using hypergeometricsols in DEtools of Maple.} that the {\em log-derivative}
of these solutions are algebraic functions, but finding that the algebraic expression (minimal polynomial)
of the solutions is much harder\footnote[5]{Showing that the solutions are algebraic without having their
exact expressions, can be achieved by showing that their $\,p$-curvatures are zero,  recalling the Andr\'e-Christol
conjecture that one must have a basis of globally bounded solutions, or looking for rational solutions
of symmetric powers of the operators.}. In principle these algebraic functions solutions of order-two linear
differential operators can be written as pullbacked $\, _2F_1$ hypergeometriic functions, but again it is
a difficult task~\cite{Weil}. 

\vskip .3cm 

\subsection{Birational symmetries from collineations. An even  simpler example \\}
\label{collinesimpler}

Let us consider the following
birational transformation associated with a collineation:
\begin{eqnarray}
\label{collinxyzBIS22}
\hspace{-0.96in}&&  \quad  \quad 
(x,  \, y, \, z)  \, \,  \quad  \longrightarrow \, \quad  \quad
\nonumber \\
\hspace{-0.96in}&&  \quad  \quad \,\, \, \, \quad  \quad 
\Bigl({{x  \, +3\, y} \over{ 1\, -\, x \, +2 \, y}}, \, \, \, \,
     {{ 5\, x \, +7\, y} \over{ 1\, -\, x \, +2\, y}}, \, \, \, 
     {{  x \, y \, z \cdot \, (1 \, -x \, +2\,y)^2 } \over {
         (x \, + 3 \,y) \cdot \, (5\, x \, + 7\, y) }}\Bigr), 
\end{eqnarray}
which preserves the product $\, p \, = \, \, x \, y \, z$,
and also preserves the origin $\, (x, \, y, \, z) \, = \, \, (0, \, 0, \, 0)$.
Again, if one transform the simple rational function (\ref{simplest}) with
the birational transformation (\ref{collinxyzBIS22}), one gets
the rational function of the form:
\begin{eqnarray}
\label{NewRatcolli22}
\hspace{-0.96in}&&  \quad  \quad \quad \quad \quad \quad   \quad
       {\cal R} \, \, = \, \, \,
       {{ (1\, -\, x \, +2 \, y) \cdot \, (x \, + 3 \,y) \cdot \, (5\, x \, + 7\, y)
         } \over {   {\cal D} }}, 
\end{eqnarray}
and, again, the  intersection of the algebraic surface $\,\, {\cal D} \, = \, \, 0\, $ with the
algebraic surface $\, p \, = \, \, x\, y \, z$, is an elliptic curve, corresponding to eliminate
$\, z \, = \, p/x/y \, $ in $\,\, {\cal D} \, = \, \, 0$. One gets immediatly the
same Hauptmodul (\ref{HauptNewRatcolli}) for this new elliptic curve.
The telescoper of the rational function (\ref{NewRatcolli22}) is an
order-ten  linear differential operator
\begin{eqnarray}
\label{finalremationL10b}
\hspace{-0.96in}&&  \quad  \quad \quad \quad \quad \quad \quad 
\, \,\,\,  L_{10} \, = \,\, \, F_2 \cdot \, G_2 \cdot \, H_1 \, \cdot \, I_1 \cdot \, J_2 \cdot \, K_2,  
\end{eqnarray}
where the order-two linear differential operator $\, F_2$
is\footnote[4]{The order-two linear differential operator $\, F_2$
  is a quite ``massive'' operator: 30391 characters. }
(non-trivially) homomorphic to the order-two  linear differential
operator $\, L_2$ which is the telescoper
of the rational function (\ref{simplest}) and where the solutions of
$\, G_2$, $\, J_2$ and $\, K_2$ are two {\em algebraic functions}.
The order-two linear differential operator $\, F_2$ is of the form
\begin{eqnarray}
\label{oftheformAD}
\hspace{-0.96in}&&  \quad  \quad \quad \quad \quad \quad \quad 
\, \,\,\,
F_2  \, \, = \, \, \, D_x^2 \,\, +{{A_1(x)} \over {D_1(x)}} \cdot \, D_x \,\, +{{A_0(x)} \over {D_0(x)}}, 
\end{eqnarray}
where $\, A_1(x)$ and $\, A_0(x)$ are polynomials of degree 41 and 55 respectively,
where $\, D_1(x)$ and $\, D_0(x)$ read
\begin{eqnarray}
\label{factors}
\hspace{-0.96in}&&  \quad \quad 
\, D_1(x) \, \, = \, \, 
\lambda(x) \cdot  P_{14}(x) \cdot P_{20}(x),  \quad   
D_0(x) \, \, = \, \,  x \cdot  \lambda(x) \cdot P_{14}(x) \cdot  P_{20}(x)^2, 
\end{eqnarray}
with: 
\begin{eqnarray}
\label{factors}
\hspace{-0.96in}&&  \quad \quad \quad 
\lambda(x) \, \, = \, \, \, (219024 -6916931\, x -23604075\, x^2)\cdot \,
(7 \, -225\,x) \cdot \, (5 \, -243\, x)
\nonumber \\
\hspace{-0.96in}&&  \quad  \quad \quad \quad \quad \quad  \quad 
\times \, (1 -27 \, x )\cdot \, (35 \, -x)\cdot \, (21 \, +x) \cdot \, x, 
\end{eqnarray}
where $\, P_{14}(x) \, $ and  $\, P_{20}(x) \, $ are polynomials of degree 14 and 20 respectively. The order-two
operator linear differential $\, G_2$ yielding algebraic solutions is also a quite ``large'' linear differential operator.

\vskip .3cm 

{\bf Remark B 3.1:} The factorisation of (\ref{finalremationL10b}) is far from being unique. As usual we have
a mix between product and direct-sum of factors. The order-ten linear differential  operator being quite large it is difficult
to get the direct-sum factorisation of $\, L_{10}$ in (\ref{finalremationL10b}). One finds, however, quite easily
that  $\, L_{10}$ has two simple rational function solutions
\begin{eqnarray}
\label{factorsL1M1}
\hspace{-0.96in}&&  \quad \quad \quad \quad \quad \quad 
{{1} \over {(x-35) \cdot \, (x\, +21)}}, \quad \quad \quad \quad {{x} \over {(x-35) \cdot \, (x\, +21)}},
\end{eqnarray}
corresponding to  two order-one operators
$\, L_1 \, = \, D_x \, +2 \, (x-7)/(x+21)/(x-35)$ and $\, M_1\, = \, D_x \, +2 \, (x-7)/(x+21)/(x-35) \, -1/x$
and, thus, can be rightdivided by the LCLM of $\, L_1$ and $\, M_1$. More interestingly,
the product\footnote[1]{In contrast note that the product $\, F_2 \cdot \, G_2$
  in  the decomposition (\ref{finalremationL10b})
  is {\em not} a direct-sum.} $\, H_1 \cdot \, I_1 \cdot \, J_2 \cdot \, K_2$
in the decomposition (\ref{finalremationL10b}) of $\, L_{10}$, can be seen as the direct sum of
$\, L_1$, $\, M_1$, $\, K_2$ and two new (and simpler !) order-two
linear differential operators  $\, \tilde{G}_2$ and $\, \tilde{J}_2$:
  \begin{eqnarray}
\label{factorsL1M1J2K2}
\hspace{-0.96in}&&  \quad \quad \quad \quad  \quad
G_2 \cdot \, H_1 \,\cdot \, I_1 \cdot \, J_2 \cdot \, K_2
\,\, \, = \, \, \, \,  L_1 \, \oplus \, M_1 \, \oplus \,  \tilde{G}_2  \, \oplus \, \tilde{J}_2 \, \oplus \, K_2.
\end{eqnarray}
It was easy to see that the log-derivative of the solutions of the order-two operator $\, J_2$
were algebraic functions, but harder to see that these solutions were actually algebraic. One now finds 
immediately that the solutions of $\, \tilde{J}_2$ are algebraic functions.

\vskip .3cm 

{\bf Remark B 3.2:}  If one considers, instead of (\ref{NewRatcolli22})  the rational function with the
same denominator $\,{\cal D}$  but where the numerator
is normalised to $\, 1$, 
\begin{eqnarray}
\label{NewRatcolli222norm}
\hspace{-0.96in}&&  \quad  \quad \quad \quad \quad \quad \quad \quad \quad \quad \quad \quad\quad 
       {\cal R} \, \, = \, \, \,
       {{ 1
         } \over {   {\cal D} }}. 
\end{eqnarray}
Its telescoper is an order-seven linear differential operator
\begin{eqnarray}
\label{finalremationL7}
\hspace{-0.96in}&&  \quad  \quad \quad \quad \quad \quad  \quad \quad \quad\quad
\, L_{7} \,\, = \, \,\, F_2 \cdot \, G_1 \cdot \, G_2  \cdot \, H_2, 
\end{eqnarray}
where the order-two linear differential operator $\, F_2$
is (non-trivially) homomorphic to the order-two  linear differential
operator $\, L_2$ which is the telescoper
of the rational function (\ref{simplest}), and
where the order-two linear differential operators $\, G_2$ and $\, H_2$
have simple algebraic solutions.

\vskip .3cm 

\subsection{Birational symmetries from collineations. Another  example \\}
\label{collineevenanothersimplerfirst}

Let us consider the following
birational transformation associated with a collineation:
\begin{eqnarray}
\label{collinxyzBIS2222}
\hspace{-0.96in}&&  \quad  \quad \quad \quad 
(x,  \, y, \, z)  \, \,  \quad  \longrightarrow \, \quad  \quad
\nonumber \\
\hspace{-0.96in}&&  \quad \quad \quad   \quad   \, \quad  \quad 
\Bigl({{x } \over{ 1\, -\, x \, +2 \, y}}, \, \, \, \,
     {{  y} \over{ 1\, -\, x \, +2\, y}}, \, \, \, 
       z  \cdot \, (1 \, -x \, +2\,y)^2\Bigr), 
\end{eqnarray}
which preserves the product $\, p \, = \, \, x \, y \, z$,
and also preserves the origin $\, (x, \, y, \, z) \, = \, \, (0, \, 0, \, 0)$.
Again, if one transforms the simple rational function (\ref{simplest}) with
the birational transformation (\ref{collinxyzBIS2222}), one gets
the rational function of the form:
\begin{eqnarray}
\label{NewRatcolli2222}
\hspace{-0.96in}&&  \quad  \quad \quad \quad \quad \quad  \quad \quad \quad \quad 
       {\cal R} \, \, = \, \, \,     {{ 1\, -\, x \, +2 \, y
         } \over {   {\cal D} }}, 
\end{eqnarray}
and again the  intersection of the algebraic surface $\,\, {\cal D} \, = \, \, 0\, $ with the
algebraic surface $\, p \, = \, \, x\, y \, z$, is an elliptic curve, corresponding to eliminate
$\, z \, = \, p/x/y \, $ in $\,\, {\cal D} \, = \, \, 0$. One gets immediatly the
same Hauptmodul (\ref{HauptNewRatcolli}) for this new elliptic curve.
The telescoper of the rational function (\ref{NewRatcolli2222}) is an
order-four linear differential operator
\begin{eqnarray}
\label{finalremationL4bb}
\hspace{-0.96in}&&  \quad  \quad \quad \quad \quad \quad \quad \quad \quad  \quad \quad
\, L_{4} \, = \, \, F_2 \cdot \, G_2, 
\end{eqnarray}
where the order-two linear differential operator $\, F_2$
is (non-trivially) homomorphic to the order-two  linear differential
operator $\, L_2$ which is the telescoper
of the rational function (\ref{simplest}) and where the solutions of
$\, G_2$ are two {\em algebraic functions} of series expansion:
\begin{eqnarray}
\label{s0s1}
\hspace{-0.98in}&&  
s_0 \, \, = \, \, \,
1 \,\,  +{{105} \over {4}} \cdot \, x \, +{{12753} \over {16 }} \cdot  \, x^2
\, +{{876225} \over {32}} \cdot \, x^3 \, +{{251403765} \over { 256}} \cdot \, x^4 \,\,\, + \, \, \, \cdots 
\nonumber \\
\hspace{-0.98in}&&  
s_1 \, \, = \, \, \, x \,\, +{{105} \over {4}} \cdot \,x^2 \, +{{7385} \over {8 }} \cdot  \,x^3
\, \, +{{2111725} \over {64}} \cdot \, x^4 \, \,
+{{155849463} \over {128}} \cdot \, x^5
\, \,\, \, + \, \, \, \cdots 
\end{eqnarray}
The series $\, s \, = \, \, s_1 \, \, $ is, for instance,
solution of the polynomial equation $\, P(s, x) \, = \, 0$,
where $\, P(s, x)$ reads:
\begin{eqnarray}
\label{s2pol}
\hspace{-0.98in}&& \quad \quad 
P(s, x) \, \, = \, \, \,
2847312 \cdot \, p(x)^3 \cdot \, s^6 \, \,\,
+158184 \cdot\, p(x)^2 \cdot \, s^4 \, \,\,
+5040 \cdot \, p(x)^2 \cdot \, s^3
\nonumber \\
\hspace{-0.98in}&&\, \, \quad \quad  \quad 
+2197 \cdot \, p(x) \cdot \, s^2 \, \,\,
+140 \cdot \, p(x) \cdot \, s \, \,\,
+4\, x \cdot \, (243\, x +35), 
\end{eqnarray}
with $\,\, p(x)\, = \, \, 243\, x^2\,+35\, x\, -1$.
The  series expansions of the algebraic solutions of $\, P(s, x) \, = \, 0 \, \, $ read:
\begin{eqnarray}
\label{s2polseries}
\hspace{-0.98in}&& 
{\cal S}(u) \, = \,  \,  \,
 u   \, \,   \, \, +{{448451640\,u^4 -38438712\,u^3  -20761650\,u^2 +1377667\,u  +221830 } \over { 17710 }} \cdot \, x
\nonumber \\
\hspace{-0.98in}&& \, \,  \, \,  
 + 3 \cdot \, {{448451640\,u^4 -38438712\,u^3 -20761650\,u^2 +1450531\,u +221830 } \over {2024 }} \cdot \, x^2
 \, \,  \, + \, \, \, \cdots
 \nonumber 
\end{eqnarray}
where $\, u \, = \, 0, \,  -1/6, \,  1/6, \,  5/26,  \,  -4/39,  \, -7/78$. One finds that
\begin{eqnarray}
\label{s2polseriesfinds}
\hspace{-0.98in}&& \quad  \quad  \quad 
15 \cdot \, {\cal S}\Bigl({{1} \over {6}} \Bigr) \, +8 \cdot \,  {\cal S}\Bigl(-{{1} \over {6}} \Bigr)
\, +13\cdot \, {\cal S}\Bigl(-{{7} \over {78}} \Bigr) \, = \,  \, 0,
\nonumber \\
\hspace{-0.98in}&& \quad  \quad  \quad 
13 \cdot \, {\cal S}\Bigl({{1} \over {6}} \Bigr) \, +8\cdot \,  {\cal S}\Bigl(-{{4} \over {39}} \Bigr)
\, +15\cdot \, {\cal S}\Bigl(-{{7} \over {78}} \Bigr) \, = \,  \, 0,
\nonumber \\
\hspace{-0.98in}&& \quad  \quad  \quad 
15825411\cdot \, {\cal S}\Bigl({{1} \over {6}} \Bigr) \, \, -1771 \cdot \, {\cal S}\Bigl({{5} \over {6}} \Bigr)
\, +29373604 \cdot \, {\cal S}\Bigl(-{{7} \over {78}} \Bigr) \, = \,  \, 0,
\end{eqnarray}
and that the two solutions (\ref{s0s1}) of $\, G_2$ read:
\begin{eqnarray}
\label{s2polseriesfinds2}
\hspace{-0.98in}&& \quad \quad \quad  \quad  \quad  \quad 
s_0 \, \, = \, \, \, {\cal S}(0), \quad  \quad \quad 
s_1 \, \, = \, \, \,
{{521} \over {32}} \cdot \, {\cal S}\Bigl({{1} \over {6}} \Bigr)
\,\, +{{611} \over {32}} \cdot \,  {\cal S}\Bigl(-{{7} \over {78}} \Bigr). 
\end{eqnarray}
The homomorphism between $\, F_2$ and $\, L_2$ gives 
\begin{eqnarray}
\label{HauptNewRatcolliintertwinnew}
\hspace{-0.96in}&&  \quad  \quad  
F_2 \cdot \, X_1 \, \, = \, \, \, Y_1 \cdot \, L_2, 
\quad  \quad \quad \quad  \quad \, \, \,\hbox{where:}  
\nonumber \\
\hspace{-0.96in}&&  \quad  \quad 
X_1 \, \, = \, \, \, \,
 \,  \alpha(x) 
 \cdot \,   \Bigl((3240\,x^2 \,+6\,x \,+1) \cdot \, D_x \,  \, +1080\,x \, -6\Bigr),
 \quad  \quad \quad  \, \, \quad  \,\hbox{with:}
 \nonumber \\
 \hspace{-0.96in}&&  \quad \quad 
  \alpha(x) \, \, = \, \, {{81 } \over {10 \cdot \, (1 \, -35\, x \, -243\, x^2) \cdot \, (1 \, -27\, x)}}.
\end{eqnarray}
Consequently a solution $\, {\cal S}$ of the telescoper $\, L_4$
(but not of $\, G_2$ in (\ref{finalremationL4bb}))
will be related to the
hypergeometric solution $\, _2F_1([1/3,2/3],[1],\, 27 \, x)$
of  the order-two linear differential operator $\, L_2$, 
as follows:
\begin{eqnarray}
\label{finalremation2s} 
\hspace{-0.96in}&&  \quad  \quad  \quad  \quad  \quad  \quad  \quad 
X_1\Bigl( \, _2F_1\Bigl([{{1} \over {3}}, \, {{2} \over {3}}],[1],\, 27 \, x\Bigr) \Bigr)
\, \, \,  = \, \, \,  \,  G_2  \cdot \, {\cal S}.
\end{eqnarray}
The formal series solutions of the order-four linear differential operator (\ref{finalremationL4bb})
are (of course ...) the two (algebraic) solutions (\ref{s0s1}) of  $\, G_2$, together with a  solution
with a $\, \ln(x)^{1}$, and a series $\, s_2$, analytic at $\, x \, = \, 0$:
\begin{eqnarray}
\label{finalremation2exp} 
\hspace{-0.96in}&& 
s_2 \, \, = \, \, \, x^2 \, +{{93} \over {2}} \cdot \, x^3 \,
+{{31185} \over {16}} \cdot \, x^4 \, +{{2488035} \over {32}} \cdot \, x^5
\, +{{1953542437} \over {640}} \cdot \, x^6 \,\, \, +  \, \, \, \cdots 
\end{eqnarray}
Relation (\ref{finalremation2s}) is actually satisfied with $\,\,\, {\cal S} = \, \,  5103 \cdot \, s_2$.
Note that the series for (\ref{finalremation2s}) is a series with {\em integer} coefficients:
\begin{eqnarray}
\label{finalremation2} 
\hspace{-0.96in}&&  \, \, 
{{1} \over {2}} \cdot \, {{1} \over {5103 }} \cdot \,
       X_1\Bigl( \, _2F_1\Bigl([{{1} \over {3}}, \, {{2} \over {3}}],[1],\, 27 \, x\Bigr) \Bigr)
\, \, \,  = \, \, \,  \,1 \,\, +87\, x \, +5358\, x^2 \, +282459\, x^3 
\nonumber \\ 
\hspace{-0.96in}&&  \quad \, 
 \, +13662531\, x^4 \, +626640714\, x^5 \, +27758265651\, x^6 \, +1200939383487\, x^7 
\, \, \, + \, \, \, \cdots \nonumber
\end{eqnarray}

\vskip .2cm 

{\bf Remark B 4.1:} Note that the diagonal $\, \delta$
of the rational function (\ref{NewRatcolli2222}) reads:
\begin{eqnarray}
\label{finalremation2expdiag} 
\hspace{-0.96in}&& \quad \quad \,
\delta \, \, = \, \, \,
1 \,\, +4\, x \,\, +108\, x^2 \, +1960\, x^3 \, +43240\, x^4 \, +965664\, x^5 \, +22377600\, x^6
\nonumber \\
\hspace{-0.96in}&& \quad  \quad \quad \quad  \,
\, +528712272\, x^7 \, +12698698320\, x^8 \, +308814134200\, x^9 \, + \, \, \cdots 
\end{eqnarray}
We expect this diagonal to be a solution of the order-four telescoper (\ref{finalremationL4bb}).
This series is actually a linear combination of the three series $\, s_0$, $\, s_1$ and $\, s_2$,
analytic at $\, x \, = \, 0$:
\begin{eqnarray}
\label{finalremation2expdiag} 
\hspace{-0.96in}&& \quad \quad \quad \quad \quad   \quad  \quad  \quad   \quad 
\delta \, \, \,= \, \, \, \,  s_0  \, \, -{{89} \over {4}} \cdot \, s_1 \,\, -105 \cdot \, s_2. 
\end{eqnarray}
It is interesting to see how the three globally bounded series  $\, s_0$, $\, s_1$ and $\, s_2$,
conspire to give a series with integer coefficients, the diagonal (\ref{finalremation2expdiag}). 

\vskip .3cm 

{\bf Remark B 4.2:} These results must be compared with the calculations for the rational function
\begin{eqnarray}
\label{NewRatcolli2222num}
\hspace{-0.96in}&&  \quad  \quad \quad \quad \quad \quad   \quad \quad \quad    \quad  \quad \quad  
       {\cal R} \, \, = \, \, \,     {{ 1 } \over {   {\cal D} }}, 
\end{eqnarray}
where the denominator $\, {\cal D}$ is the same as the one in (\ref{NewRatcolli2222}).
In this case where the numerator has been normalised to $\, 1$,
the diagonal is the same as the diagonal of $\, 1/(1\, -x -y -z)$, namely
$\, _2F_1([1/3, 2/3],[1],\, 27\, x)$, and the telescoper is the same telescoper
as the one for $\, 1/(1\, -x -y -z)$.

\vskip .3cm 

\subsection{Birational symmetries from collineations. Another  example \\}
\label{collineevenanothersimpler}

Let us consider the following
birational transformation associated with a collineation:
\begin{eqnarray}
\label{collinxyzBIS222}
\hspace{-0.96in}&&  \quad  \quad 
(x,  \, y, \, z)  \, \,  \quad  \longrightarrow \, \quad  \quad
\nonumber \\
\hspace{-0.96in}&&  \quad  \quad   \,  \,  \, \quad  \quad 
\Bigl({{x  \, +3\, y} \over{ 1\, -\, x \, +2 \, y}}, \, \, \, \,
     {{  y} \over{ 1\, -\, x \, +2\, y}}, \, \, \, 
     {{  x \,z  \cdot \, (1 \, -x \, +2\,y)^2 } \over {
         x \, + 3 \,y}}\Bigr), 
\end{eqnarray}
which preserves the product $\, p \, = \, \, x \, y \, z$,
and also preserves the origin $\, (x, \, y, \, z) \, = \, \, (0, \, 0, \, 0)$.
Again, if one transform the simple rational function (\ref{simplest}) with
the birational transformation (\ref{collinxyzBIS222}), one gets
the rational function of the form:
\begin{eqnarray}
\label{NewRatcolli222}
\hspace{-0.96in}&&  \quad  \quad \quad \quad \quad \quad   \quad \quad   \quad  
       {\cal R} \, \, = \, \, \,
       {{ (1\, -\, x \, +2 \, y) \cdot \, (x \, + 3 \,y) 
         } \over {   {\cal D} }}, 
\end{eqnarray}
and again the  intersection of the algebraic surface $\,\, {\cal D} \, = \, \, 0\, $ with the
algebraic surface $\, p \, = \, \, x\, y \, z$, is an elliptic curve, corresponding to eliminate
$\, z \, = \, p/x/y \, $ in $\,\, {\cal D} \, = \, \, 0$. One gets immediatly the
same Hauptmodul (\ref{HauptNewRatcolli}) for this new elliptic curve.
The telescoper of the rational function (\ref{NewRatcolli222}) is an
order-seven linear differential operator
\begin{eqnarray}
\label{finalremationL7f1}
\hspace{-0.96in}&&  \quad  \quad \quad \quad \quad \quad  \quad   \quad \quad  
\, L_{7} \,  \, = \,  \, \, F_2 \cdot \, G_2 \cdot \, H_1 \cdot \, H_2, 
\end{eqnarray}
where the order-two linear differential operator $\, F_2$
is (non-trivially) homomorphic to the order-two  linear differential
operator $\, L_2$ which is the telescoper
of the rational function (\ref{simplest}), and where the order-two
linear differential   operators $\, G_2$ and $\, H_2$ 
have algebraic solutions\footnote[5]{In fact one finds
  easily that the solutions of $\, G_2$ are Liouvillian: the log-derivative
of these  solutions  are algebraic functions. Finding that these
  Liouvillian solutions are algebraic functions is much harder. In contrast one finds
  easily that the order-two  linear differential operator $\, H_2$ has algebraic solutions.}
and where $\, H_1$ is an order-one linear differential operator.
 This homomorphism between $\, F_2$ and $\, L_2$ gives 
\begin{eqnarray}
\label{HauptNewRatcolliintertwinA}
\hspace{-0.96in}&&  \quad  \quad \quad 
F_2 \cdot \, X_1 \, \, = \, \, \, Y_1 \cdot \, L_2
\quad  \quad \, \, \,\, \hbox{where:}  \quad \quad \, \, \,\,
X_1 \, \, = \, \, \, \,   A(x) \cdot \, D_x \,  \, + B(x), 
\end{eqnarray}
where $\, A(x)$ and $\, B(x)$ are rational functions.
Consequently a solution $\, {\cal S}$ of the telescoper $\, L_7$
(but not of the product $\, G_2 \cdot \, H_1 \cdot \, H_2$ in (\ref{finalremationL7f1}))
will be related to the
hypergeometric solution $\, _2F_1([1/3,2/3],[1],\, 27 \, x)$
of  the order-two linear differential operator $\, L_2$, 
as follows:
\begin{eqnarray}
\label{finalremation0} 
\hspace{-0.96in}&&  \quad  \quad  \quad  \quad 
X_1\Bigl( \, _2F_1\Bigl([{{1} \over {3}}, \, {{2} \over {3}}],[1],\, 27 \, x\Bigr) \Bigr)
\, \, \,  = \, \, \,  \,\,
G_2 \cdot \, H_1 \cdot \, H_2 \,  \cdot \, {\cal S}.
\end{eqnarray}
In that case the solution of $\,  {\cal S}$ of the telescoper $\, L_7$ reads
\begin{eqnarray}
\label{finalremationexpans} 
\hspace{-0.96in}&& \quad \quad \quad 
 {\cal S} \, \, = \, \,\, \, x^4 \,\,\, \, +{{13316825310791} \over {231428221515}} \cdot \, x^5 \,\, \,
 +{{30360140830595651} \over {11108554632720}} \cdot \,x^6 \, \,\, \, + \, \, \, \cdots 
\end{eqnarray}
and the expansion of (\ref{finalremation0}) reads:
\begin{eqnarray}
\label{finalremationexpans2} 
\hspace{-0.96in}&& \quad \quad \quad 
X_1\Bigl( \, _2F_1\Bigl([{{1} \over {3}}, \, {{2} \over {3}}],[1],\, 27 \, x\Bigr) \Bigr) 
\, \, \,  = \, \, \,  \,  \, {{1} \over {x}}  \,\, \, \, +{{85390121841387522079} \over {629841285410317908}}
\nonumber \\
\hspace{-0.96in}&& \quad \quad \quad \quad \quad  \quad \quad  
\, +{{906492811433323772155053002605} \over {77136236451492696817854192}} \cdot \, x
\, \,\,\, + \, \, \, \cdots
\end{eqnarray}

\vskip .3cm

{ \bf Remark B 5.1:} The factorisation (\ref{finalremationL7f1})  is
far from being unique. Introducing the order-one
linear differential operator $\, L_1  \, = \,\, D_x \, +4/(3+4\,x)$,
one has the following direct-sum decomposition: 
\begin{eqnarray}
\label{finalremationL7}
\hspace{-0.96in}&&  \quad  \quad \quad \quad \quad \quad  \quad \quad
L_{7} \, = \, \,  L_1 \, \oplus \, L_6, 
\\
\hspace{-0.96in}&&  \quad  \quad \quad \quad \quad \quad  \quad \quad
 G_2 \cdot \, H_1 \cdot \, H_2 \,\,  = \, \, \,    L_1 \, \oplus \, \tilde{G}_2 \, \oplus \, H_2,
\end{eqnarray}
where $\, L_6$ is an order-six linear differential operator, and where
the order-two linear differential operator operator
$\, \tilde{G}_2$ is slightly simpler than $\, G_2$.

\vskip .3cm

{ \bf Remark B 5.2:} If one considers, instead of (\ref{NewRatcolli222}),
the rational function with the same denominator $\,{\cal D}$  but where
the numerator is normalised to $\, 1$, 
\begin{eqnarray}
\label{NewRatcolli222norm}
\hspace{-0.96in}&&  \quad  \quad \quad \quad \quad \quad  \quad \quad \quad  \quad \quad \quad\quad
       {\cal R} \, \, = \, \, \,
       {{ 1
         } \over {   {\cal D} }}. 
\end{eqnarray}
its telescoper is an order-four linear differential operator
\begin{eqnarray}
\label{finalremationL7}
\hspace{-0.96in}&&  \quad  \quad \quad \quad \quad \quad \quad  \quad \quad  \quad\quad
\, L_{4} \, = \, \, F_2 \cdot \, G_2. 
\end{eqnarray}
The  order-two linear differential operator $\, F_2$
is (non-trivially) homomorphic to the order-two  linear differential
operator $\, L_2$ which is the telescoper
of the rational function (\ref{simplest}), and
the order-two linear differential operator $\, G_2$ has simple algebraic solutions.

\vskip .3cm 

\subsection{Birational symmetries from collineations. Another simpler example \\}
\label{collineevensimpler}

Let us consider the following
birational transformation associated with a collineation:
\begin{eqnarray}
\label{collinxyzBIS3}
\hspace{-0.96in}&&  \quad  \quad \quad \quad \,
(x,  \, y, \, z)  \, \,  \quad  \longrightarrow \, \quad  \quad
\nonumber \\
\hspace{-0.96in}&&  \quad  \quad \quad  \quad  \,  \,\, \quad  \quad 
\Bigl({{x  \, +3\, y} \over{ 1\, -\, x \, +2 \, y}}, \, \, \, \,
     {{ 1 \, + y} \over{ 1\, -\, x \, +2\, y}}, \, 
     {{  x \, y \, z \cdot \, (1 \, -x \, +2\,y)^2 } \over {
         (x \, + 3 \,y) \cdot \, (1 \, \, +  y) }}\Bigr), 
\end{eqnarray}
which preserves the product $\, p \, = \, \, x \, y \, z$.
Again, if one transform the simple rational function (\ref{simplest}) with
the birational transformation (\ref{collinxyzBIS3}), one gets
the rational function of the form:
\begin{eqnarray}
\label{NewRatcolli3}
\hspace{-0.96in}&&  \quad  \quad \quad \quad \quad \quad  
       {\cal R} \, \, = \, \, \,
       {{ (1\, -\, x \, +2 \, y) \cdot \, (x \, + 3 \,y) \cdot \, (1 \,  \, +  y)
         } \over {   {\cal D} }}, 
\end{eqnarray}
and again the  intersection of the algebraic surface $\,\, {\cal D} \, = \, \, 0\, $ with the
algebraic surface $\, p \, = \, \, x\, y \, z$, is an elliptic curve, corresponding to eliminate
$\, z \, = \, p/x/y \, $ in $\,\, {\cal D} \, = \, \, 0$. One gets immediatly the
same Hauptmodul (\ref{HauptNewRatcolli})  for this new elliptic curve.

The telescoper of the rational function (\ref{NewRatcolli3}) can now be calculated
in only a few hours, and one gets an order-nine linear differential operator of the form  
\begin{eqnarray}
\label{finalremationL9}
\hspace{-0.96in}&&  \quad  \quad \quad \quad \quad \quad  \quad \quad 
\, L_{9} \, \,  = \, \,  \, F_2 \cdot \, G_2 \cdot \, H_1 \cdot \, H_2 \, \cdot \, I_2, 
\end{eqnarray}
where the order-two linear differential operator $\, F_2$
is (non-trivially) homomorphic to the order-two  linear differential
operator $\, L_2$ which is the telescoper
of the rational function (\ref{simplest}), and where the order-two
linear differential   operators $\, G_2$, $\, H_2$ and  $\, I_2$
have algebraic solutions\footnote[2]{In fact one finds
  easily that the solutions of $\, G_2$, $\, H_2$ are Liouvillian: their log-derivative
  are algebraic functions. Finding that these
  Liouvillian solutions are algebraic functions is much harder. In contrast one finds
easily that the order-two  linear differential operator $\, I_2$ has algebraic solutions.}
and where $\, H_1$ is an order-one linear differential operator. This
homomorphism between $\, F_2$ and $\, L_2$ gives 
\begin{eqnarray}
\label{HauptNewRatcolliintertwin}
\hspace{-0.96in}&&  \quad  \quad \quad 
F_2 \cdot \, X_1 \, \, = \, \, \, Y_1 \cdot \, L_2
\quad  \quad \, \, \,\hbox{where:}  \quad \quad \, \, \,
X_1 \, \, = \, \, \, \,   A(x) \cdot \, D_x \,  \, + B(x), 
\end{eqnarray}
where $\, A(x)$ and $\, B(x)$ are quite large rational functions.
Consequently a solution $\, {\cal S}$ of the telescoper $\, L_9$
(but not of the product $\, G_2 \cdot \, H_1 \cdot \, H_2 \, \cdot \, I_2$
in (\ref{finalremationL9}))
will be related to the
hypergeometric solution $\, _2F_1([1/3,2/3],[1],\, 27 \, x)$
of  the order-two linear differential operator $\, L_2$, 
as follows:
\begin{eqnarray}
\label{finalremation} 
\hspace{-0.96in}&&  \quad  \quad  \quad  \quad  \quad 
X_1\Bigl( \, _2F_1\Bigl([{{1} \over {3}}, \, {{2} \over {3}}],[1],\, 27 \, x\Bigr) \Bigr)
\, \, \,  = \, \, \,  \,  G_2 \cdot \, H_1 \cdot \, H_2 \, \cdot \, I_2 \cdot \, {\cal S}.
\end{eqnarray}
If finding the emergence of the hypergeometric function
$\, _2F_1([1/3,2/3],[1],\, 27 \, x)$ is easy to obtain from the (algebraic geometry) 
calculation of the Hauptmodul (\ref {HauptNewRatcolli}), (see (\ref{diagsimplest})),
the telescoper of (\ref{NewRatcolli3}),
or equivalently, the solution $\, {\cal S}$ of that telescoper,
requires to find many linear differential operators, namely the intertwinner $\, X_1$ and
also the right factors $\, G_2$, $\, H_1$, $ \, H_2$ and $ \, I_2$. In contrast
with the birational transformations described in
section \ref{infinite} (see (\ref{simplestbiratQ}), (\ref{simplestbiratQ2}), (\ref{simplestbiratQ3})),
which simply preserve the diagonals of the rational functions, we have here,
with the birational transformation (\ref{collinxyzBIS3}), again
{\em two birationally equivalent underlying elliptic curves}, but
a {\em much more convoluted ``covariance''} requiring to find many linear differential operators.
The ``elliptic curve skeleton'' (the j-invariant or the Hauptmodul) is preserved, but the
right factors dressing $\, G_2$, $\, H_1$, $ \, H_2$ and $ \, I_2$
and the intertwiner $\, X_1$ are quite involved.

\vskip .3cm

{\bf Remark B 6.1:} In fact the order-nine operator (\ref{finalremationL9})
is a direct sum. It can be written in the form 
\begin{eqnarray}
\label{finalremationL9directsum}
\hspace{-0.96in}&&  \quad  \quad \quad \quad \quad \quad  \quad \quad 
\, L_{9} \, \, = \, \, \, L_8   \, \oplus \, L_1,
\\
\hspace{-0.96in}&&  \quad  \quad \quad \quad \quad \quad  \quad \quad 
G_2 \cdot \, H_1 \cdot \, H_2 \, \cdot \, I_2
\,  \, \, = \, \, \, \, L_1 \, \oplus \,  \tilde{G}_2 \, \oplus \,  \tilde{H}_2 \, \oplus \,  I_2, 
\end{eqnarray}
where the order-one operator reads:
\begin{eqnarray}
\label{finalremationL9directsumorderone}
\hspace{-0.96in}&&  \quad  \quad \quad \quad \quad \quad \quad \quad 
 L_1 \, \, = \, \, \, D_x \,\, + {{4} \over { 3  \, +4 \, x }},
\end{eqnarray}
where $\, L_8$ is an order-eight operator, 
and where the operators with a tilde
are much simpler than the operators without a tilde.

\vskip .2cm

{\bf Remark B 6.2:} Again if one considers, instead of (\ref{NewRatcolli3}),
the rational function with the same
denominator $\, {\cal D}$, but where the numerator has been normalised to $\, 1$,
\begin{eqnarray}
\label{NewRatcolli3norm}
\hspace{-0.96in}&&  \quad  \quad \quad \quad \quad \quad  \quad \quad \quad \quad    \quad \quad   
       {\cal R} \, \, = \, \, \,
       {{ 1 } \over {   {\cal D} }}, 
\end{eqnarray}
one finds an order-seven telescoper which factorises as follows:
\begin{eqnarray}
\label{finalremationL77q}
\hspace{-0.96in}&&  \quad  \quad \quad \quad \quad \quad \quad \quad  \quad \, \,  
\, L_{7} \,\, = \,\, \, F_2 \cdot \, G_1 \cdot \, H_2 \, \cdot \, I_2, 
\end{eqnarray}
where the order-two linear differential operator $\, F_2$
is (non-trivially) homomorphic to the order-two  linear differential
operator $\, L_2$ which is the telescoper
of the rational function (\ref{simplest}), and where the order-two
linear differential   operators  $\, H_2$ and  $\, I_2$
have algebraic solutions.

\vskip .2cm

{\bf Remark B 6.3:} Again the factorisation (\ref{finalremationL77q})
is far from being unique. Introducing the order-one
linear differential operator  $\, L_1 \, = \, D_x \, +1/x$, 
one has the two following direct-sum decompositions
\begin{eqnarray}
\label{finalremationL77}
\hspace{-0.96in}&&  \quad  \quad \quad \quad \quad  \quad  \quad 
\, L_{7} \, = \, \,  L_{6} \, \oplus \, L_1,
\\
\hspace{-0.96in}&&  \quad  \quad \quad \quad \quad  \quad  \quad 
 G_1 \cdot \, H_2 \, \cdot \, I_2  \, \, = \, \, \,  L_1 \, \oplus \, \tilde{H}_2 \, \oplus \, I_2, 
\end{eqnarray}
where the order-two linear differential operator $\, \tilde{H}_2$ 
is slightly simpler than $\,H_2$.

\vskip .2cm

{\bf Remark B 6.4:} As far as an {\em algebraic geometry approach}
of diagonals and telescopers is concerned (see~\cite{MDPI}),
we see that the concept of telescopers, which describes {\em all} the periods, can be
more interesting than the concept of diagonals which often
yields to diagonals that can be  almost trivial functions (being
simple rational functions, or being simply equal to zero).  The examples of \ref{colline}
show that the differential algebra approach of creative telescoping cannot be
totally replaced by an algebraic geometry approach~\cite{MDPI}. The algebraic geometry approach
provides very quickly some  precious information on the telescoper (the Hauptmodul),
but not the telescoper itself.
In fact one might consider the opposite point of view:
creative telescoping could be seen as a tool to get  effective algebraic geometry
results. 

\vskip .2cm

{\bf Remark B 6.5:} The examples displayed in this appendix
can be seen as an illustration of the ``dialogue of the deaf'' between mathematicians
and physicists. Some mathematicians will point out the fact that the
calculation of the Hauptmodul (\ref{HauptNewRatcolli}) 
underlines the essence of the problem, namely the existence
of an underlying elliptic curve, and will see the explicit calculation of the
telescoper, and all its periods, as a laborious and slightly
useless piece of work. In particular they will consider the ``dressing''
right-factors occurring in the decompositions (\ref{finalremationL10bc}), (\ref{finalremationL10b}), ...  
as a totally and utterly spurious information, and they will also probably see the explicit expression of the
large order-two operators $\, F_2$ as superfluous, retaining only the order-two linear differential operator $\, L_2$,
prefering to ignore, or forget, the intertwiner $\, X_1$ in (\ref{HauptNewRatcolliintertwinA})
or (\ref{HauptNewRatcolliintertwin}). Along this line they may consider the other solutions of the telescoper,
namely the ``periods'' (associated with non-vanishing cycles) that are not diagonals, as irrelevant. In contrast
for a physicist, getting all the periods, and the explicit expression of the
telescoper will be seen as essential\footnote[1]{In contrast with mathematicians a physicist will not be
  interested in the certificates in the creative tespocoping equation, but only in the telescopers.}. Recalling
the $\, \chi^{(n)}$ components of the susceptibility of the Ising model,
it is essential to get the explicit expression of the linear differential operators (telescopers)
annihilating these $\, \chi^{(n)}$'s even if these (large)
linear differential operators~\cite{2009-chi5,High} are products (and direct sums)
of a large set of factors. In the framework of integrable models, beyond diagonals, a physicist will always
seek for a linear differential operator corresponding to an elliptic curve (resp. K3 surface, Calabi-Yau manifold, ...)
even if it is ``buried'' as a left factor of a large telescoper,
like the $\, F_2$'s in (\ref{finalremationL10bc}) or (\ref{finalremationL10b}).

\end{document}